\documentclass[cernpreprint, atlasdraft=false, UKenglish, texmf, orcidlogo]{atlasdoc}

\usepackage[subcaption]{atlaspackage}

\usepackage{atlaspackage}
\usepackage{comment}
\usepackage[T1]{fontenc}
\usepackage[utf8]{inputenc}  %
\usepackage{graphicx} %
\usepackage{siunitx} %
\usepackage{multirow}   %
\usepackage{booktabs}   %
\usepackage[backref=false]{atlasbiblatex}

\usepackage{atlasphysics}
\graphicspath{{logos/}{figures/}}
\AtlasTitle{Measurements of $ZZ \rightarrow \ell\ell\nu\nu$ and $ZZjj \rightarrow \ell\ell\nu\nu jj$ productions in $pp$ collisions at $\sqrt{s}=13$ TeV with the ATLAS detector}

\AtlasAbstract{
This article presents measurements of $Z$ boson pair production cross-sections at a center-of-mass energy of $\sqrt{s} = 13$ TeV, using 140 fb$^{-1}$ of proton--proton data collected with the ATLAS detector at the Large Hadron Collider. The analysis includes both inclusive $ZZ$ production and $ZZ$ production in association with two jets ($ZZjj$), where one $Z$ boson decays into a pair of charged leptons $\ell^+\ell^-$ ($\ell = e, \mu$), and the other decays into a neutrino pair.
The measured fiducial cross-section of the inclusive $ZZ$ production is $21.0\pm1.0$ fb, while for the $ZZjj$ production it is $0.96^{+0.18}_{-0.16}$ fb.
Extrapolation of the $ZZ$ cross-section to the full phase space, for $Z$ bosons within the mass range $66-116$~GeV, yields $\sigma (pp \rightarrow ZZ) = 15.38\pm0.81$ pb.
Differential cross-sections are presented for an extended set of kinematic observables.
The results are found to be in good agreement with Standard Model predictions.
Constraints on anomalous gauge boson self-interactions are derived using both the vertex function formalism and the effective field theory framework.
}
\AtlasRefCode{STDM-2019-27}

\PreprintIdNumber{CERN-EP-2025-243}

\AtlasJournal{EPJ C}
\AtlasCoverEgroupAnalysisTeam{atlas-stdm-2019-27-analysis-team@cern.ch}

\hypersetup{pdftitle={ATLAS document},pdfauthor={The ATLAS Collaboration}}

\begin{document}

\maketitle

\tableofcontents

%
%

%

\section{Introduction}
\label{sec:intro}

Precise measurements of total and differential cross-sections for diboson production at the Large Hadron Collider (LHC)~\cite{Evans:2008zzb} are essential to test the predictions of the Standard Model (SM) and probe for potential physics beyond it. %

This article presents measurements of integrated and differential cross‑sections for $Z$ boson pair ($ZZ$) production at $\sqrt{s} = 13$~\TeV using ATLAS proton--proton ($pp$) collision data recorded in 2015–2018, corresponding to an integrated luminosity of $140$ fb$^{-1}$. The measurements include inclusive $ZZ$ production and $ZZ$ production with two associated jets ($ZZjj$), where one $Z$ boson decays into a pair of charged leptons $\ell^+\ell^-$ ($\ell = e, \mu$) and the other decays into neutrinos. Decays of \( Z \) bosons via \(\tau\)-leptons, where the \(\tau\) later decays leptonically, are included in the signal definition, but their contribution is negligible after event selection. The measurement targets final states with two on-shell $Z$ bosons.

\begin{figure}[!hbtp]
\centering
\subfloat[]{
\includegraphics[width=0.2\linewidth]{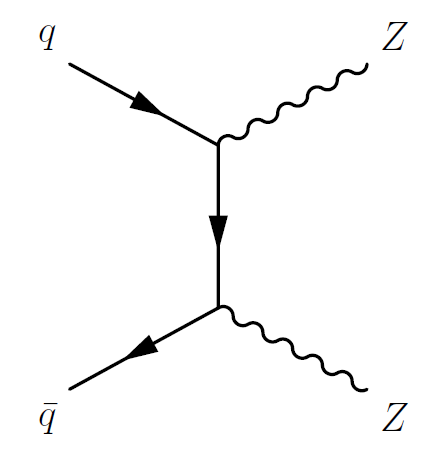}\hspace{4.5 mm}
}
\subfloat[]{
\includegraphics[width=0.2\linewidth]{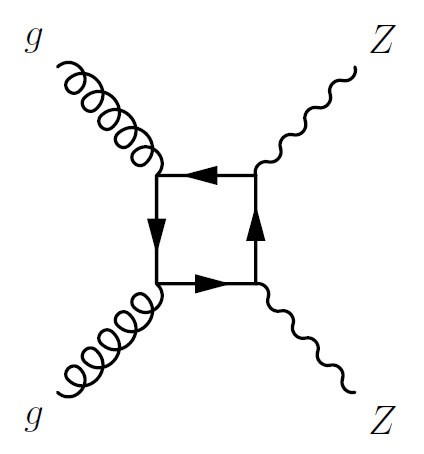}\hspace{4.5 mm}
}
\subfloat[]{
\includegraphics[width=0.2\linewidth]{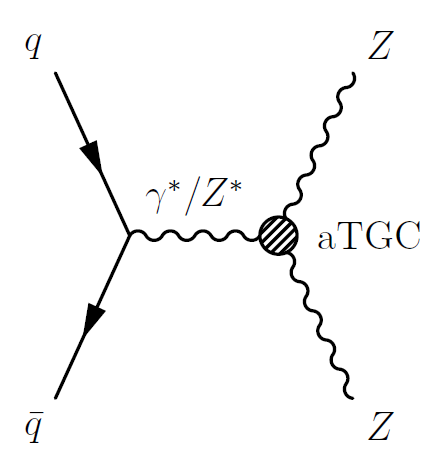}
}
\caption{Example of Feynman diagrams for the production of a pair of $Z$ bosons from the  (a) $q\bar q$ and (b) $gg$ initial states at the LHC.
The s-channel diagram (c) is not allowed at LO in the SM; it represents an anomalous triple gauge coupling (aTGC) of a $ZZZ$ or $ZZ\gamma$ vertex.
}
\label{fig:feynmanZZinclusive}
\end{figure}

In the SM, $Z$ boson pairs are produced at the LHC primarily through quark--antiquark ($q\bar{q}$) annihilation, as illustrated in Figure~\ref{fig:feynmanZZinclusive}(a), which shows a leading-order (LO) Feynman diagram.
An additional contribution arises from gluon--gluon ($gg$) fusion via a quark loop, depicted in Figure~\ref{fig:feynmanZZinclusive}(b), which accounts for approximately 10\% of the total $ZZ$ production cross-section for $\sqrt(s)=13$~\TeV $pp$ collisions.
Production via the $s$-channel, shown in Figure~\ref{fig:feynmanZZinclusive}(c), is forbidden at LO in the SM, as it would involve an anomalous triple gauge coupling (aTGC) through a $ZZZ$ or $ZZ\gamma$ vertex. Consequently,
$ZZ$ production serves as a sensitive probe of neutral anomalous triple gauge couplings (aTGCs) predicted by physics theories beyond the SM (BSM)~\cite{Gounaris:2000tb_1}.

Various measurements of $ZZ$ production at the LHC were carried out by the ATLAS Collaboration~\cite{STDM-2012-02,STDM-2014-16,STDM-2016-15,STDM-2017-03,STDM-2022-17} and the
CMS Collaboration ~\cite{CMS-SMP-12-007, CMS-SMP-13-005, CMS-SMP-16-001, CMS-SMP-19-001}
using data
taken at $\sqrt{s}$ = $7$, $8$, and $13$ ~\TeV. Two complementary final states originating from the decays of the
$Z$ boson pairs are used in those measurements, one with four charged leptons (4$\ell$) and the other with two charged leptons and two neutrinos ($\ell\ell\nu\nu$).
At 13~\TeV, the ATLAS Collaboration measured the inclusive fiducial cross-section in the $\ell\ell\nu\nu$ channel with uncertainty of 7\%~\cite{STDM-2017-03}.

\begin{figure}[!hbtp]
\centering
\subfloat[]{
\includegraphics[width=0.21\linewidth]{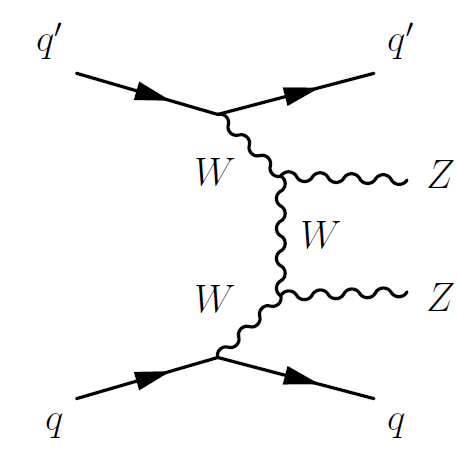}\hspace{4.5 mm}
}
\subfloat[]{
\includegraphics[width=0.2\linewidth]{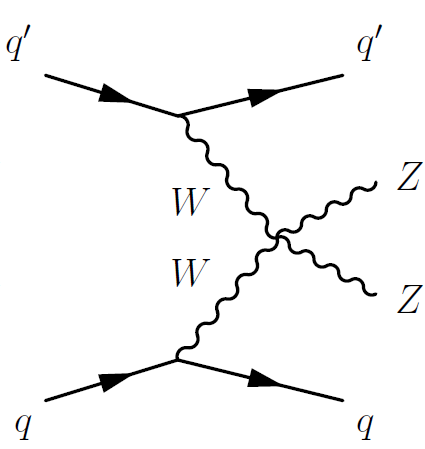}\hspace{4.5 mm}
}
\subfloat[]{
\includegraphics[width=0.21\linewidth]{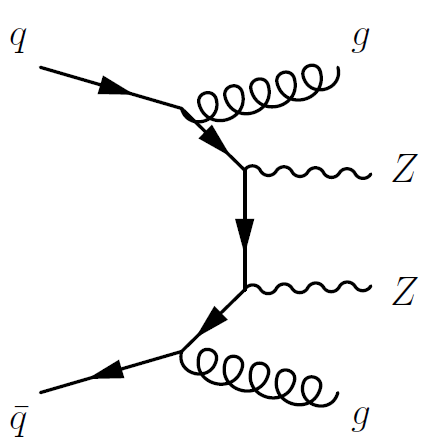}
}
\caption{
Typical diagrams for the production of $ZZjj$ events.  The electroweak production, containing triple and quartic        interaction vertices, is shown in (a) and (b), while (c) shows the QCD production with gluons in the final state.
}
\label{fig:feynmanZZvbs}
\end{figure}

The analysis also targets inclusive $ZZjj$ production, with contributions from both QCD- and electroweak-induced processes.
The electroweak component is dominated by the vector boson scattering (VBS) topology.
Representative Feynman diagrams for these processes are shown in Figure~\ref{fig:feynmanZZvbs}.
The EW production of $ZZjj$ events is particularly sensitive to the $WWZ$ and $WWZZ$ weak-boson self-interactions, which arise from the non-Abelian structure of the electroweak interactions.
These are depicted in Figure~\ref{fig:feynmanZZvbs}(a) and Figure~\ref{fig:feynmanZZvbs}(b), respectively.

As a result, $ZZjj$ production serves as a powerful channel for investigating
if these triple and quartic gauge couplings are in agreement with the SM predictions and for setting constraints on aTGCs and anomalous quartic gauge couplings (aQGCs).

The EW $ZZjj$ production was observed using a combination of the $ZZjj \to 4\ell jj$ and $ZZjj \to \ell\ell \nu\nu jj$ channels  ~\cite{STDM-2017-19}.
The ATLAS Collaboration has also measured the differential cross-section for $ZZjj \rightarrow 4\ell jj$ %
production~\cite{STDM-2020-02}, from which limits on aQGCs
were derived at the 95\% confidence level (CL).
In the present analysis, the $ZZ jj \rightarrow \ell\ell\nu\nu jj$ final state is studied, providing greater sensitivity to aQGCs as the branching ratio is approximately six times larger than that of the $4\ell jj $ channel.

Using nearly four times the integrated luminosity of the %
previous statistically limited measurement~\cite{STDM-2017-03},
this article presents cross-section measurements of $\ell\ell\nu\nu$ and $\ell\ell\nu\nu jj$ production in fiducial phase spaces.
The cross-sections are extracted by fitting signal and background models to data in dedicated regions. The total $pp \rightarrow ZZ$ production cross-section is obtained by extrapolating the fiducial measurement to the full phase space, defined in terms of the invariant masses of the $Z$ bosons as reconstructed from their decay leptons.
Differential cross-sections are measured by correcting for detector effects (unfolding) in terms of eight and seven kinematic variables for the inclusive $ZZ$ and $ZZjj$ production processes, respectively.

The inclusive $ZZ$ measurement is used to search for aTGCs by fitting the transverse momentum  ($p_\text{T}$) distribution of the dilepton system from the Z boson decay.
The $ZZjj$ measurement is used to probe aQGCs by fitting the transverse mass ($m_{\text{T}}$) distribution of the $ZZ$ system, defined in Section~\ref{sec:xs-measurement}.
Limits on aTGCs and aQGCs are derived using a simplified model within the SM Effective Field Theory (SMEFT) framework~\cite{Falkowski2017,Degrande2014,Ellis2023}. For comparison with previous results, the anomalous Lagrangian vertex function (VF) formalism~\cite{Baur2000} is also employed to set limits on aTGC parameters.

The structure of this article is as follows. Section~\ref{sec:atlas_detector} provides an overview of the ATLAS detector subsystems. Section~\ref{sec:dataMC} describes the data samples and Monte Carlo (MC) simulations used for both signal and background processes.
The event selection criteria are detailed in Section~\ref{sec:object_event_selection}, while
the definition of the experimental fiducial phase space is presented in Sections~\ref{sec:FiducialPhaseSpace}.
Section~\ref{sec:Systematics} discusses the estimate of systematic uncertainties in the measurements. The SM cross-section predictions are presented in Section~\ref{sec:theorypred}.
The results of the integrated and differential cross-section measurements are reported in Section~\ref{sec:xs-measurement}. The interpretation of the results, including the study of anomalous gauge boson self-interactions and the constraints on aTGC and aQGC parameters, is presented in Section~\ref{sec:aTGC-aQGC}. Finally, Section~\ref{sec:conclusion} summarises the key results.


%
%

%

\section{The ATLAS detector}
\label{sec:atlas_detector}

The ATLAS experiment~\cite{PERF-2007-01} at the LHC is a multipurpose particle detector
with a forward--backward symmetric cylindrical geometry and a near \(4\pi\) coverage in
solid angle.\footnote{ATLAS employs a right-handed coordinate system with its origin at the nominal interaction point (IP) at the detector's centre. The \(z\)-axis aligns with the beam direction, while the \(x\)-axis points from the IP to the LHC ring's centre, and the \(y\)-axis points upwards. The transverse plane $(x, y)$ is described using cylindrical coordinates $(r, \phi)$, where $\phi$ denotes the azimuthal angle around the beam direction. The pseudorapidity $\eta$ is defined by the polar angle $\theta$ as \(\eta = -\ln \tan(\theta/2)\).}

The detector consists of an inner tracking detector (ID) surrounded by a thin superconducting solenoid
providing a \qty{2}{\tesla} axial magnetic field, electromagnetic and hadronic calorimeters, and a muon spectrometer (MS).
The inner tracking detector covers the pseudorapidity range \(|\eta| < 2.5\).
It consists of silicon pixel, silicon microstrip, and transition radiation tracking detectors.
Lead/liquid-argon (LAr) sampling calorimeters provide electromagnetic (EM) energy measurements with high granularity within the region \(|\eta|< 3.2\).
A steel/scintillator-tile hadronic calorimeter covers the central pseudorapidity range (\(|\eta| < 1.7\)).
The endcap and forward regions are instrumented with LAr calorimeters
for EM and hadronic energy measurements up to \(|\eta| = 4.9\).
The muon spectrometer surrounds the calorimeters and is based on
three large superconducting air-core toroidal magnets with eight coils each.
The field integral of the toroids ranges between \num{2.0} and \qty{6.0}{\tesla\metre}
across most of the detector.
The muon spectrometer includes a system of precision tracking chambers up to \(|\eta| = 2.7\) and fast detectors for triggering up to \(|\eta| = 2.4\).
The luminosity is measured by the LUCID--2~\cite{LUCID2} detector, which is located close to the beampipe.
A two-level trigger system is used to select events~\cite{TRIG-2016-01}.
The first-level trigger is implemented in hardware and uses a subset of the detector information
to accept events at a rate close to \qty{100}{\kHz}.
This is followed by a software-based trigger that
reduces the accepted rate of complete events to \qty{1.25}{\kHz} on average
depending on the data-taking conditions.
A software suite~\cite{SOFT-2022-02} is used for simulation, reconstruction, and analysis of real data and simulated events, and for detector operations, trigger and data acquisition.

%

%

%

%

%

%

%


%

\section{Data and Monte Carlo simulations}
\label{sec:dataMC}

The data used for this analysis were recorded using single-lepton ($e,~\mu$) and multi-lepton triggers, corresponding %
to an integrated luminosity of  $140$ ~\ifb~\cite{DAPR-2021-01}
after the application of data quality requirements~\cite{DAPR-2018-01}.
For the single-lepton trigger, the transverse momentum thresholds vary from $20$ to $26$ ~\GeV, while for the multi-lepton triggers, they vary from $10$ to $14$ GeV, depending on the data-taking periods~\cite{TRIG-2018-01,TRIG-2018-05}.
The overall trigger efficiency for the $ZZ\rightarrow \ell\ell\nu\nu$ events selected in this analysis is higher than 99\%{}.

Simulated events for the signal and background processes are produced using MC event generators that model the underlying physics processes in $pp$ collisions. After the initial event generation, the final-state particles are passed through a detector simulation ~\cite{SOFT-2010-01} based on the \textsc{Geant4} framework~\cite{GEANT4:2002zbu}.
The simulated events are processed using the same reconstruction algorithms and trigger requirements as those used in the data.
The event samples also include simulations of additional $pp$ interactions from the same and neighbouring bunch crossings, commonly referred to as pile-up, which are simulated using \PYTHIA[8.1]~\cite{Sjostrand:2007gs} with the A3 tune~\cite{ATL-PHYS-PUB-2016-017}. The simulated events are reweighted to reproduce the pile-up conditions observed in the data.

Scale factors, derived from data control samples (such as $Z \rightarrow \ell^+\ell^- ~\textrm{and} ~J/\psi\rightarrow \ell^+\ell^-$ decays), are applied to simulated events to correct for discrepancies between data and MC simulations. These corrections account for differences in the trigger efficiencies, reconstruction, identification, isolation, and interaction vertex reconstruction for electrons and muons. The electron and jet energies and the muon momenta in the simulated events are smeared to account for the differences in resolution between data and MC simulations.

The signal processes $pp\rightarrow ZZ\rightarrow \ell\ell\nu\nu$ and $pp\rightarrow ZZjj \rightarrow \ell\ell\nu\nu jj$ can occur via either QCD or EW production mechanisms.
For QCD production, both quark-initiated ($q\bar q \rightarrow ZZ$) and gluon-initiated ($gg \rightarrow ZZ$) processes were modelled using \textsc{Sherpa} 2.2.2~\cite{Bothmann:2019yzt}, including off-shell effects and Higgs boson ($H$)  contributions, where appropriate.
For the quark-initiated process,
the events were simulated using matrix elements at NLO accuracy in QCD for up to one additional parton and at LO accuracy for up to three additional parton
emissions.
Samples for the loop-induced processes $gg\rightarrow ZZ$
were simulated using LO-accurate matrix elements in QCD for up to one
additional parton emission.
A k-factor of 1.7 was applied to normalise the $gg$-induced cross-section to NLO accuracy~\cite{Caola2015}. The NLO EW corrections ~\cite{Accomando:2004de, Kallweit:2017khh} were also included in the $q\bar q \rightarrow \ell\ell\nu\nu$ process with a $p_{\text{T}}(Z)$ dependent k-factor, where $p_{\mathrm{T}}(Z)$ is the transverse momentum of the $Z$ boson decaying into neutrinos.

The matrix element calculations were matched
and merged with the \SHERPA parton shower based on Catani--Seymour
dipole factorisation~\cite{Gleisberg:2008fv,Schumann:2007mg} using the MEPS@NLO
prescription~\cite{Hoeche:2011fd,Hoeche:2012yf,Catani:2001cc,Hoeche:2009rj}.
The virtual QCD corrections were provided by the
\OPENLOOPS library~\cite{Buccioni:2019sur,Cascioli:2011va,Denner:2016kdg}. The
\NNPDF[3.0nnlo] set of PDFs was used~\cite{Ball:2014uwa}, along with the
dedicated set of tuned parton-shower parameters developed by the
\SHERPA authors.

An alternative sample for the quark-initiated QCD production of the $pp\rightarrow ZZ\rightarrow \ell\ell\nu\nu$ process was produced with \POWHEGBOX[v2]~\cite{Nason:2004rx,Frixione:2007vw,Alioli:2010xd} at NLO accuracy in QCD.
Events were interfaced to \PYTHIA[8.210]~\cite{Sjostrand:2014zea}
for the modelling of the parton shower, hadronisation, and underlying
event, with parameters set according to the \AZNLO
tune~\cite{STDM-2012-23}. The \CT[10] PDF set~\cite{Lai:2010vv} was used
for the hard-scattering processes, whereas the \CTEQ[6L1] PDF
set~\cite{Pumplin:2002vw} was used for the parton shower.

\textsc{MadGraph5}~\cite{Alwall:2011uj} in combination with the \NNPDF[2.3lo] PDF set~\cite{Alwall:2014hca} was used to simulate the electroweak contributions to both the
$ZZ$ and $ZZjj$ fiducial cross-sections.
The calculation is accurate to LO in perturbative QCD and uses the \NNPDF[2.3lo] PDF set~\cite{Ball:2012cx}.   The events were interfaced with \PYTHIA[8.2] to provide parton showering, hadronisation, and underlying-event activity using the A14 set of tuned parameters~\cite{ATL-PHYS-PUB-2014-021}.

An alternative sample for the electroweak production of $ZZjj \rightarrow \ell\ell\nu\nu jj$ events was produced using \SHERPA~2.2.2 at LO accuracy. The \NNPDF[3.0nlo] PDF set~\cite{Ball:2014uwa} was used. Events were produced with the default \SHERPA\ parton shower, hadronisation, and underlying-event modelling. Matrix-element calculations included all relevant tree-level electroweak diagrams and were merged with the \SHERPA\ parton shower using the MEPS@LO prescription~\cite{Hoeche:2011fd,Hoeche:2012yf,Catani:2001cc,Hoeche:2009rj}.%

The main backgrounds for the inclusive $ZZ \rightarrow \ell\ell\nu\nu$ and the $ZZjj \rightarrow \ell\ell\nu\nu jj$ signals originate from both resonant and non-resonant $\ell^+ \ell^-$ production along with a significant missing transverse momentum component, either from neutrinos or from mismeasured jets accompanying the charged leptons. In both the $\ell\ell\nu\nu$ and the $\ell\ell\nu\nu jj$ final states, the dominant contribution comes from the $WZ$ production process, particularly in the decay channels $WZ \rightarrow \ell\nu\ell\ell$ (where the third lepton evades detection) and $WZ \rightarrow q\bar{q}\ell\ell$. This process is considered a resonant background, as it results in an $\ell^+ \ell^-$  pair originating from an on-shell $Z$ boson. %
Another important background is the $Z$+jets production process, i.e., the production of single $Z$ bosons in association with jets,  where the $Z$ boson decays into $\ell^+ \ell^-$.
Non-resonant backgrounds arise mainly from processes such as $WW \rightarrow \ell\nu\ell\nu$ and top ($t$) pair production, $t\bar{t} \rightarrow WWbb \rightarrow \ell\nu\ell\nu bb$.
Other backgrounds arise from processes producing top quarks in association with $W$ or $Z$ bosons ($tW$, $t\bar{t}V$ with $V=W,Z$), $ZZ$ without neutrinos in the decay, i.e., $ZZ(\rightarrow 4\ell, 2\ell 2q)$, tribosons $VVV$, $W$ bosons in association with jets ($Wj$), single $Z$ bosons decaying into $\tau\tau$, and $WH$ events.

The diboson background samples were simulated using \SHERPA or \POWHEG as follows. The $WZ \rightarrow \ell\nu\ell\ell$ and $ZZ \rightarrow 4\ell$ samples were simulated using  \SHERPA[2.2.2]. The $WZ$ and $ZZ$ samples with "semileptonic" final states, where one boson decays leptonically and the other hadronically, and the gluon-initiated $WW$ sample, were generated with \SHERPA[2.2.1].
The quark-initiated $WW$ samples were generated with \POWHEGBOX[v2].
Interference effects between \(WW\) and \(ZZ\) for same-flavour charged leptons and neutrinos were not considered.

The production of \ttbar events was modelled using the \POWHEGBOX[v2]\cite{Frixione:2007vw} generator at NLO in QCD, with the \NNPDF[3.0nlo] PDF set. The \hdamp parameter, which regulates the matching between \POWHEG matrix elements and the parton shower, was set to be 1.5 times the top quark mass~\cite{ATL-PHYS-PUB-2016-020}. The events were interfaced with \PYTHIA[8.230] for modelling the parton shower, hadronisation, and underlying event, using the A14 tune and the \NNPDF[2.3lo] PDF set~\cite{Ball:2014uwa}.

For associated production of top quarks with $W$ bosons ($tW$), the \POWHEGBOX[v2] generator was again used at NLO in QCD, employing the five-flavour scheme and the \NNPDF[3.0nlo] set of PDFs. To avoid double counting with the \( t\bar{t} \) production, the diagram-removal scheme~\cite{Frixione:2008yi} was applied. As with \ttbar production, events were interfaced to \PYTHIA[8.230] using the A14 tune and the \NNPDF[2.3lo] PDF set.
The production of $t\bar{t}V$ events was modelled using the
\MGNLO[2.3.3]~\cite{Alwall:2014hca} generator at NLO with the
\NNPDF[3.0nlo] PDF set.
The events were interfaced to \PYTHIA[8.210] using the A14 tune and the
\NNPDF[2.3lo] PDF set. %

The production of $V$+jets ($V$ = $W$, $Z$), with all three lepton flavours,
was simulated with the \SHERPA[2.2.1] generator, using NLO matrix elements for up to two partons and LO matrix elements for up to four partons. These events were matched with the \SHERPA parton shower using the \MEPSatNLO prescription, with the set of tuned parameters developed by the \SHERPA authors. The \NNPDF[3.0nnlo] set of PDFs was used, and the samples were normalised to an NNLO prediction~\cite{Anastasiou:2003ds}.

The triboson backgrounds $ZZZ$, $WZZ$, and $WWZ$ with fully leptonic decays with at least four prompt charged leptons were modelled
using \SHERPA{2.2.2} at LO accuracy in QCD and the \CT[10] PDF set~\cite{Lai:2010vv}. %
The $Z\to\tau\tau$ events were simulated using \textsc{Sherpa}~2.2.1, with up to two partons at NLO and up to four partons in total, merged using the ME+PS@NLO scheme.
The $WH$ samples were simulated using \POWHEGBOX[v2] at NLO accuracy in QCD, interfaced to \PYTHIA[8.186]~\cite{Sjostrand:2007gs} for the modelling of parton showering and hadronisation.

For all simulated samples, except those generated with \SHERPA, the decays of \(b\)- and \(c\)-hadrons are handled by \EVTGEN ~\cite{Lange:2001uf}.

The MC samples used in the interpretation of the analysis in terms of aTGCs and aQGCs are discussed in Section~\ref{sec:aTGC-aQGC}.


\section{Event selection}
\label{sec:object_event_selection}

The event selection in both the $ZZ$ and $ZZjj$ signal regions (SRs) and the background control regions (CRs)
relies on multiple reconstructed objects, including electrons, muons, jets, and the missing transverse momentum  $E_\text{T}^{\text{miss}}$.

Events are first required to have a collision vertex associated with at least two tracks, each with $p_\text{T}$ greater than 0.5 ~\GeV. The vertex with the highest sum of $p^2_\text{T}$  of the associated tracks is identified as the primary vertex.

Electrons are reconstructed from energy clusters in the EM calorimeter that match to tracks in the ID. They are identified using a likelihood function based on EM shower shapes, track properties and track-to-cluster matching. Electrons must meet a ``medium'' likelihood requirement, achieving an overall identification efficiency of approximately 90\%~\cite{EGAM-2021-02}. The electron momentum is calculated from the cluster energy and track direction. The cluster's pseudorapidity must be within $|\eta| < 2.47$, excluding the transition region between the barrel and endcap calorimeter regions $1.37 < |\eta| < 1.52$. The leading (subleading) electron is required to have a transverse momentum ($p_{\mathrm{T}}$) greater than 30~GeV (20~GeV).

Muons are identified by tracks reconstructed in the MS, which are then matched to tracks reconstructed in the ID. A ``medium'' identification selection is applied, based on requirements for the number of hits in both the ID and the MS~\cite{MUON-2018-03}. The muon pseudorapidity must satisfy $|\eta| < 2.5$.  The muon momentum is determined by combining the MS measurement, corrected for energy deposited in the calorimeters, with the ID measurement. The leading (subleading) muon is required to have $p_{\mathrm{T}} > 30$~GeV (20~GeV).

Electron and muon candidates are required to originate from the primary vertex. They are required to satisfy $|d_0/\sigma_{d_0}| < 3(5)$ for muons (electrons), where $d_0$ is the transverse impact parameter calculated relative to the measured beam-line position and $\sigma_{d_0}$ is its uncertainty.
They must also have a longitudinal impact parameter, $z_0$ (the difference
between the value of $z$ of the point on the track at which $d_0$ is defined and
the longitudinal position of the primary vertex), which satisfies $|z_0 \cdot \sin(\theta)| < 0.5$ mm.

To suppress hadronic and non-prompt lepton background, electrons and muons are required to be isolated from other particles by satisfying the particle-flow isolation criteria ~\cite{EGAM-2021-01, MUON-2018-03}, which are based on clusters of energy deposits in the calorimeter and inner detector track information.

Jets are reconstructed using the anti-\(k_{t}\) algorithm~\cite{Cacciari:2008gp} with a radius parameter of \( R = 0.4 \), applied to particle-flow objects that combine information from the inner detector and calorimeter~\cite{Fastjet}. The jet-energy scale is derived from simulation and further refined using in situ techniques based on dijet, \(\gamma+\)jet and \(Z+\)jet events~\cite{JETM-2018-05}. A jet-vertex tagger~\cite{PERF-2014-03} is applied to jets with $p_\text{T} <$ 60 ~\GeV and $|\eta|< $ 2.4 to suppress jets originating from pile-up. In addition, jets with $p_{\text{T}} > 20$~\GeV and $|\eta| < 2.5$ containing $b$-hadrons are identified using the DL1r multivariate $b$-tagging algorithm~\cite{FTAG-2020-08, FTAG-2019-02}. The 85\% efficiency working point is used, which corresponds to rejection factors of 6.3 for $c$-jets and 33 for light-flavour jets.
Jets are required to have transverse momentum $p_{\mathrm{T}} > 30~\GeV$ and pseudorapidity $|\eta| < 4.5$.

Overlaps between reconstructed objects are accounted for with a removal procedure that is mainly based on the angular separation between the different final-state objects (electrons, muons, jets). The procedure is the same as the one described in Ref.~\cite{SUSY-2015-10}.

The neutrinos in the $\ell\ell\nu\nu$ final state do not interact with the ATLAS detector and cannot be reconstructed. Their presence is identified using the $E_\text{T}^{\text{miss}}$ variable, which is computed as the magnitude of the negative vector sum of transverse momenta of all leptons and jets and the tracks originating from the primary vertex but not associated with any of the leptons or jets~\cite{JETM-2020-03}.
Another quantity used in the event selection is the total transverse energy, $H_\text{T}$, which is the scalar sum of the transverse momenta of all selected leptons and jets.

\subsection{Signal regions}
The event selection criteria are designed to maximize the sensitivity to the $ZZ$ and $ZZjj$ signal processes while effectively reducing background contributions. Two signal regions are defined: one targeting inclusive \( ZZ \) production and another targeting the \( ZZjj \) topology.
In both SRs events are required to
contain exactly two oppositely charged electrons or muons, with an invariant mass between 80 and 100~\GeV. %
The azimuthal angle between the missing transverse momentum and the  $Z$ boson decaying into a charged lepton pair must satisfy $\Delta\phi({E}_\text{T}^\text{miss}, Z) > 2.2$. The angular separation between the two leptons, defined as \(\Delta R_{\ell\ell} = \sqrt{(\Delta\eta)^2 + (\Delta\phi)^2}\), must satisfy \(\Delta R_{\ell\ell} < 1.8\).

Events containing $b$-tagged jets, referred to as $b$-jets in the following, with $p_{\mathrm{T}} > 30$ ~\GeV and $|\eta| < 2.5$ are rejected.
The ratio of the missing transverse momentum to the total transverse energy, $E_{\text{T}}^{\text{miss}}/H_\text{T}$, must be greater than 0.65 to ensure a $p_{\mathrm{T}}$ balance between the two $Z$ bosons.

For the $ZZ$ SR, events are required to have $E_{\text{T}}^{\text{miss}} > 110$~\GeV. The $ZZjj$ SR has a stricter $E_{\text{T}}^{\text{miss}}$ requirement of $E_{\text{T}}^{\text{miss}} > 150$~\GeV to minimize the background contribution from $Z$+jets production. %


\subsection{Background control regions}

For the $ZZ\rightarrow \ell\ell\nu\nu$ and $ZZjj \rightarrow \ell\ell\nu\nu jj$ processes, background sources are categorised into "resonant" and "non-resonant" types.
Resonant backgrounds include $WZ$ production, dominated by $W\rightarrow\ell\nu$ decays where the lepton is not reconstructed, and the $Z$+jets process, where two leptons form a $Z$ candidate, with "fake"  $E_\text{T}^\text{miss}$ arising from jet mis-measurement. Non-resonant backgrounds encompass $WW$ and top-quark processes.

In all considered backgrounds, lepton candidates are either prompt or originate from $\tau$-lepton decays following $W$ or $Z$ decays. The contribution from non-prompt or fake leptons, predominantly from $W$+jets production, is negligible in this analysis. It is therefore included in the “other” background category in the tables and estimated from simulation.

Background contributions are estimated using a simultaneous fit across multiple control regions designed to constrain dominant processes contributing to the selected final states. In the inclusive $ZZ$ phase space, three dedicated CRs are defined with an enhanced contribution from the main background processes: $WZ$, $WW$, and $t\bar t$. Additionally, three CRs target the $Z$+jets background: $Z$+0jets, $Z$+1jet, and $Z$+$\geq$2jets.

For the $WZ$ control region, events are required to contain a third lepton with $p_{\mathrm{T}} > 20~\GeV$ that satisfies the medium identification criteria, along with a same-flavour opposite-charge (SFOS) lepton pair whose invariant mass is consistent with the $Z$ boson mass in the region $80 < m_{\ell\ell} < 100~\GeV$. If multiple such pairs are found, the one with the mass closest to the $Z$ boson mass is selected.
To suppress non-$WZ$ contributions, several further selections are applied:$E_\mathrm{T}^{\mathrm{miss}} > 70$~\GeV, $E_{\mathrm{T}}^{\mathrm{miss}}/H_{\mathrm{T}} > 0.3$, a $b$-jet veto, and $m_\mathrm{T}^W > 30$~\GeV.
The transverse mass of the $W$ ($m_\mathrm{T}^W$) is constructed from the third lepton’s transverse momentum ($\vec{p}_\mathrm{T}^\ell$) and the $\vec{E}_\mathrm{T}^\mathrm{miss}$ vector representing the transverse momentum of the neutrino ($\vec{p}_\mathrm{T}^\nu$), using the formula
\[
m_\mathrm{T}^W = \sqrt{2 \cdot p_\mathrm{T}^\nu \cdot p_\mathrm{T}^\ell \cdot \left[1 - \cos \Delta\phi(\ell,\nu)\right]}
\]
where \( \Delta\phi(\ell, \nu) \) is the angle between the lepton and the neutrino in the transverse plane.

The non-resonant background is estimated using two $e^\pm\mu^\mp$ CRs, where exactly one electron and one muon with opposite charges are selected, fulfilling a $Z$ mass requirement of $80 < m_{\ell\ell} < 100$ ~\GeV. Additional selections are  applied:$E_\mathrm{T}^{\mathrm{miss}} > 70$~\GeV and $E_{\mathrm{T}}^{\mathrm{miss}}/H_{\mathrm{T}} > 0.3$. These selections ensure a sample highly populated by $WW$ and $t\bar t$ events, with minimal signal contamination.
To suppress the $t\bar{t}$ contribution and enhance the $WW$ purity in the $WW$ CR, a veto on $b$-jets is imposed, whereas for the Top CR, at least one $b$-jet is required.

Three CRs for the $Z$+jets background are defined by requiring two SFOS leptons with $80 < m_{\ell\ell} < 100$ ~\GeV and selecting events in the plane $\{ E_{\mathrm{T}}^{\mathrm{miss}}, E_{\mathrm{T}}^{\mathrm{miss}}/H_{\mathrm{T}} \}$ chosen to be well separated from the signal region. Events must satisfy either $70 < E_{\mathrm{T}}^{\mathrm{miss}} < 80$~GeV and $E_{\mathrm{T}}^{\mathrm{miss}}/H_{\mathrm{T}} > 0.3$, or $E_{\mathrm{T}}^{\mathrm{miss}} > 80$~GeV and $0.3 < E_{\mathrm{T}}^{\mathrm{miss}}/H_{\mathrm{T}} < 0.45$. A $b$‑jet veto is applied, and events are categorised into exactly 0‑jet, exactly 1‑jet, and at least 2‑jet bins. Each of these jet-multiplicity regions is used as a single-bin control region in the fit, as no significant statistical gain is achieved by further binning.

For the $ZZjj \to \ell\ell \nu \nu jj$ process, two CRs are defined: one for the resonant $WZjj$ background and one for the non-resonant backgrounds ($WWjj$ and $t\bar t$).
The non-resonant background is constrained using the $e^\pm\mu^\mp$ CR, with an additional requirement of at least two jets. The $WZjj$ background is estimated using the three-lepton CR described above, with the requirement of at least two jets present. The $Z+\text{jets}$  background contribution is estimated from MC simulations and found to be negligible in the signal region.

The event yields for both data and MC simulations in the inclusive $ZZ$ signal region and the six control regions are summarised in Table~\ref{tab:prefitYieldsInclu}. Similarly, the event yields in the signal and control regions for the $ZZjj$ selection are detailed in Table~\ref{tab:prefitYieldsZZjj}.
They include uncertainties from both MC statistical and systematic sources, which are described in Section~\ref{sec:Systematics}.

\begin{table}[htbp]
\caption{Pre-fit MC yields for the inclusive $ZZ$ phase-space. The ``Other'' category includes events from the processes $ZZ(\rightarrow 4\ell$, $2\ell 2q)$, $VVV~(V=W,Z)$, $Wj$, $Z(\rightarrow \tau\tau)$, and $WH$. The selected number of data events in the SR and the CRs are also provided.}
\label{tab:prefitYieldsInclu}
\centering
\renewcommand{\arraystretch}{1.2}
\resizebox{\textwidth}{!}{%
\begin{tabular}{l
S[table-format=4.3(4)]
S[table-format=4.3(4)]
S[table-format=4.3(4)]
S[table-format=4.3(4)]
S[table-format=4.3(4)]
S[table-format=4.3(4)]
S[table-format=4.3(4)]}
\hline
& \textbf{SR} & \textbf{WZ CR} & \textbf{WW CR} & \textbf{Top CR}
& \textbf{Z+0j CR} & \textbf{Z+1j CR} & \textbf{Z+$\geq$2j CR} \\
\hline
$ZZ$                   & 1650(36)     & 2.6(0.7)     & 0.58(0.11)   & 0.005(0.011) & 100(12)      & 150(20)       & 160(34)       \\
$WZ$                   & 740(43)      & 2100(140)    & 22(2)        & 0.9(0.3)     & 80(12)       & 250(28)       & 310(60)       \\
$Z$+0j                 & 80(34)       & 40(17)       & 0.6(1.0)     & 0            & 1700(670)    & 0             & 0             \\
$Z$+1j                 & 80(40)       & 40(14)       & 2.2(0.7)     & 0.12(0.08)   & 0            & 3800(1460)    & 0             \\
$Z$+$\geq$2j           & 13(7)        & 20(12)       & 0.4(0.6)     & 0.15(0.16)   & 0            & 0             & 4700(1700)    \\
$WW$                   & 36.2(1.9)    & 0.65(0.12)   & 672(41)      & 13.9(1.6)    & 48(6)        & 46(5)         & 18.5(1.9)     \\
Top                    & 114(20)      & 57(8)        & 1850(290)    & 5700(710)    & 92(25)       & 200(39)       & 250(43)       \\
Other                  & 48(4)        & 91(7)        & 46(13)       & 3.4(1.3)     & 12(3)        & 44(10)        & 48(10)        \\
\hline
Total prediction       & 2760(120)    & 2340(150)    & 2590(310)    & 5720(710)    & 2050(700)    & 4510(1500)    & 5470(1700)    \\
Data                   & 2917         & 2409         & 2903         & 5736         & 2691         & 5733          & 5769          \\
\hline
\end{tabular}}
\end{table}

\newcommand{\YieldTableSize}{\small} %

\begin{table}[htbp]
\caption{Pre-fit MC yields for the $ZZjj$ phase-space. The ``Other'' category includes events from the processes of $ZZ(\rightarrow 4\ell$, $2\ell 2q), VVV (V=W, Z), Wj, Z(\rightarrow \tau\tau), \textrm{and}~ WH$. The selected number of data events in the SR and the CRs are also provided.}
\label{tab:prefitYieldsZZjj}
\centering
\scriptsize
\renewcommand{\arraystretch}{1.2}
\begin{tabular}{l
S[table-format=3.1(2)]
S[table-format=3.1(2)]
S[table-format=3.3(3)]}
\hline
& \textbf{SR} & \textbf{WZ CR} & \textbf{WW and Top CR} \\
\hline
$ZZ$         & 108(12)   & 1.0(0.3)   & 0.006(0.004) \\
$WZ$         & 62(12)    & 280(60)    & 4.8(1.0)     \\
$Z$+jets     & 2.4(1.2)  & 15(8)      & 0.4(0.2)     \\
$WW$ and Top     & 7.2(1.5)  & 8.2(1.8)   & 620(90)      \\
Other        & 3.0(0.7)  & 12(3)      & 12(14)       \\
\hline
Total prediction      & 183(21)   & 320(60)    & 640(100)     \\
Data        & 184       & 317        & 678          \\
\hline
\end{tabular}
\end{table}


\section{Phase space definitions}
\label{sec:FiducialPhaseSpace}

The $ZZ\rightarrow \ell\ell\nu\nu$ and $ZZjj\rightarrow \ell\ell\nu\nu jj$ production cross-sections are measured within experimental fiducial phase spaces, which are defined by the kinematic acceptance at the stable-particle level in the MC simulation (particles with a proper lifetime $\tau > 10$ ps), as summarised in Table~\ref{tab:fid-ps}. The fiducial phase space is defined for final states with electrons or muons.
The particle-level selections are defined to closely mimic the selection applied at detector level,
and with a relaxed minimum $E_{\text{T}}^{\text{miss}}$  by $15-20$ ~\GeV, about one standard deviation from the value used at the reconstruction level and with a more relaxed dilepton mass window around the $Z$ pole.
Electrons and muons are required to be prompt, i.e. not originating from the decay of hadrons or $\tau$-leptons. A dressed lepton is defined as the momentum four-vector sum of the lepton (electron or muon) and all prompt photons within a cone of $\Delta R_{\ell \gamma} < 0.1$. Dressed leptons are used to reconstruct the $Z$ boson candidates, requiring SFOS lepton pairs.
The  $E_{\text{T}}^{\text{miss}}$
is calculated from the transverse momentum of the $\nu \bar\nu$ system, which results from the decay of the other $Z$ boson in the event. Hadronic jets are labelled as "$j$" in the table. In addition to the fiducial phase space, Table~\ref{tab:fid-ps} also includes the definition of the total phase space.

\begin{table}[!htbp]
\caption{Fiducial and total phase space definitions for inclusive $ZZ$ and $ZZjj$ cross-section measurements and for the total ZZ cross-section.}

\centering
\scriptsize

\renewcommand{\arraystretch}{1.2}
\begin{tabular}{l l}
\hline
\textbf{Common fiducial phase space  definition} & \textbf{Kinematic cut}\\
\hline
A lepton is "dressed" with photons
within an angular separation of  &  $\Delta R_{\ell \gamma} < 0.1$\\
Lepton pseudorapidity range  & $|{\eta}^{\ell} | < 2.5$\\
Transverse momentum of leading (sub-leading) lepton  &  $p_{\text{T}}^{\ell} > 30 ~(20)$ ~\GeV \\
Invariant mass of the dilepton system  &  $76  <m_{\ell\ell}<106$ ~\GeV \\
Missing transverse energy  &   $E_{\text{T}}^{\text{miss}}>95$ ~\GeV\\
Angular separation of the two leptons &  $\Delta R_{\ell\ell}<1.8$ \\
Azimuthal angular separation of the dilepton system and  $E_\text{T}^{\text{miss}}$ & $\Delta \phi (E_\text{T}^{\text{miss}}, {p}_\text{T}^{\ell\ell}) > 2.2$\\
$E_\text{T}^{\text{miss}}$ over $H_{\text{T}}$    &   $E_{\text{T}}^{\text{miss}}/H_\text{T} > 0.65$ \\
\hline
\textbf{Additional requirements for $ZZjj$ phase space} &   \\
\hline
At least 2 hadronic jets with & $p_\text{T}^\text{$j$}>30$ ~\GeV and $ |\eta^\text{$j$}|<4.5$\\
$E_{\text{T}}^{\text{miss}}$ requirement     & $E_{\text{T}}^{\text{miss}} > 130$ ~\GeV \\
\hline
\textbf{Total phase space  definition } &   \\   \hline
Invariant masses $m_{\ell\ell}$ and $m_{\nu\nu}$ & 66 $\leq $  $m_{\ell\ell}$, $m_{\nu\nu}$ $\leq   116$ ~\GeV  \\
\hline
\end{tabular}
\label{tab:fid-ps}
\end{table}


\section{Systematic uncertainties}
\label{sec:Systematics}

\textbf{Experimental systematic uncertainties} are assessed to account for residual discrepancies between data and simulation in the modelling of trigger efficiencies, leptons, jets, $E_{\text{T}}^{\text{miss}}$, luminosity, and the average number of proton--proton interactions per bunch crossing.

The total integrated luminosity of the dataset is known to a precision of 0.83\%, determined using a combination of van der Meer beam separation scans in dedicated %
data-taking sessions and calibrated luminosity-sensitive detectors during regular data-taking periods~\cite{DAPR-2021-01}.

The number of proton--proton interactions per bunch crossing varies across the dataset and is modelled in simulation by reweighting events to match the distribution observed in data. An uncertainty arises from the limited precision in the ratio of predicted to measured inelastic cross-sections within the ATLAS acceptance region~\cite{STDM-2015-05}. To account for this, a systematic variation is applied to the average number of interactions in the simulation.

Efficiencies for lepton reconstruction, identification, isolation, and trigger selections are corrected in simulation using scale factors derived from control samples in data, as discussed in Section~\ref{sec:dataMC}. The uncertainties in these corrections are propagated by varying the scale factors within their uncertainties~\cite{MUON-2018-03,EGAM-2018-01,TRIG-2018-01,TRIG-2018-05}. Differences between data and simulation in the lepton momentum scale and resolution are evaluated by applying additional scaling and smearing variations~\cite{MUON-2022-01,EGAM-2018-01}.

Jets are calibrated as discussed in Section~\ref{sec:object_event_selection}.
The jet energy scale (JES) is derived using test-beam data, LHC collision data, and simulation information.
The JES calibration~\cite{JETM-2018-05} includes corrections that account for detector effects, jet fragmentation, the impact of pile-up, and differences between data and MC simulation.
The impact of the uncertainty in the jet energy resolution (JER)~\cite{JETM-2018-05} is evaluated by applying additional smearing to the jet energy in the simulated samples to match the resolution observed in data.
An additional uncertainty arises from the correction applied to simulated events associated with the jet-vertex-tagger selection.
The uncertainty in the $b$-tagging efficiency, and the corresponding uncertainties in the $c$-jet and light-jet rejection scale factors, are also taken into account~\cite{FTAG-2018-01}. Uncertainties in $E_{\mathrm{T}}^{\mathrm{miss}}$ are evaluated by varying the soft term and propagating the lepton and jet energy scale and resolution uncertainties to the $E_{\mathrm{T}}^{\mathrm{miss}}$ calculation.

\textbf{Theory systematic uncertainties} associated with the modelling of $ZZ$ production affect the measurement.
Two main sources are considered:
(i) QCD scale variations, arising from missing higher-order corrections in the simulation and evaluated by varying the factorisation and renormalisation scales,
and (ii) uncertainties from parton distribution functions and the $\alpha_s$ value used in the PDF.

Uncertainties from missing higher orders are evaluated~\cite{Bothmann:2016nao} using independent variations of the factorisation and renormalisation scales by multiplicative factors of 0.5, 1, and 2.
This results in seven scale combinations, including the nominal pair of scales $(1,1)$ and excluding the two extreme cases where one scale is halved and the other is doubled.
The envelope of these variations defines the QCD scale uncertainty for each prediction.
For $ZZjj$, the QCD- and EW-induced components are treated separately with uncorrelated scale uncertainties.

PDF uncertainties are evaluated following the PDF4LHC prescription~\cite{Butterworth:2015oua}, taking the RMS of the 100 replicas of the NNPDF set (excluding $\alpha_s$ variations).
The effect of the strong coupling constant is assessed separately by varying $\alpha_s$ by $\pm 0.001$.

The uncertainty due to generator modelling is assessed by replacing the nominal quark-initiated \SHERPA $ZZ$ sample with the alternative \POWHEGBOX $ZZ$ sample described in Section~\ref{sec:dataMC}.  The resulting difference in the extracted cross-section is taken as a systematic uncertainty. The uncertainty in the parton showering is calculated in \SHERPA by varying the shower recoil scheme.

In the $ZZjj$ phase space, the $Z+\text{jets}$ background is estimated using MC simulation. A conservative normalisation uncertainty of 50\% is assigned to account for potential mismodelling in events with multiple jets. This value is motivated by the observation in the inclusive \(Z+\geq 2\) jets region, where a correction factor of \(1.10^{+0.44}_{-0.31}\) was found (see Sec.~\ref{sec:xs-measurement}).

Theoretical modelling uncertainties for the background samples are treated analogously to those for the signal, and are included as separate nuisance parameters in the fit.

The experimental and theory uncertainties are incorporated into the analysis through variations in relevant nuisance parameters in simultaneous fits to data performed separately for the $ZZ\to\ell\ell\nu\nu$ and $ZZjj\to\ell\ell\nu\nu jj$ channels. In each case, the signal and corresponding control regions are fitted simultaneously, and the uncertainties are constrained by the data.

The systematic uncertainties evaluated for the cross-sections for both fiducial phase spaces, $ZZ$ and $ZZjj$, are categorised and summarised in Table~\ref{tab:uncertainties}.

\begin{table}[!hbtp]
\caption{The impact of systematic uncertainties on the measured fiducial cross-sections $\sigma_{ZZ \to \ell \ell \nu\nu}$ and $\sigma_{ZZjj \to  \ell \ell \nu\nu jj}$, decomposed into
their major categories. For each category, the contribution is obtained by summing in quadrature the impacts of its
individual contributions. The total systematic uncertainty is calculated similarly by combining the impacts of individual contributions in quadrature. The statistical uncertainty is calculated as the square root of the difference of the squares of the total uncertainty and the total systematic uncertainty.}
\centering
\scriptsize
\renewcommand{\arraystretch}{1.2}
\begin{tabular}{lcc}
\toprule
\textbf{Systematic source} & \textbf{Impact on $\sigma_{ZZ \to \ell \ell \nu\nu}$} & \textbf{Impact on $\sigma_{ZZjj \to \ell \ell \nu\nu jj}$} \\
\midrule
Scale variation                          & 0.99\%  & 3.5\%   \\
PDF and $\alpha_s$          & 0.15\%  & 0.42\%  \\
Alternative signal generator & 0.38\%  & 0.93\%  \\
Z+jets modelling             & --      & 0.43\%  \\
\midrule
Jet energy scale and resolution & 1.7\%   & 8.7\%   \\
MC statistics                & 1.6\%   & 2.9\%   \\
Pile-up reweighting        & 0.39\%  & 4.1\%   \\
$E_\mathrm{T}^\text{miss}$  & 1.4\%   & 0.97\%  \\
Electrons     & 0.38\%  & 0.24\%  \\
Muons         & 0.35\%  & 0.42\%  \\
$b$-tagging                 & 0.28\%  & 1.2\%   \\
\midrule
Luminosity                  & 0.87\%  & 0.82\%  \\
\midrule
Total systematics             & 3.1\%   & 11\%   \\
\midrule
Statistics             & 3.6\%   & 14\%   \\
\midrule
Total            &  4.8\%  &  18\%   \\
\hline
\end{tabular}
\label{tab:uncertainties}
\end{table}



\section{Theory predictions}
\label{sec:theorypred}

A precise test of the SM requires higher-order perturbative QCD calculations for detailed comparisons with experimental data. The \textsc{Matrix}~\cite{MATRIX}  framework, a parton-level MC generator, is used with the \textsc{NNPDF3.0nnlo} PDF~\cite{Ball:2014uwa} to compute
integrated total and fiducial cross-sections, and differential fiducial cross-sections up to NNLO and NLO accuracies in QCD and EW calculations, respectively.
The fiducial phase space acceptance $A_{ZZ}$, i.e., the fraction of events inside the fiducial volume relative to the events in the total phase space, is used to extrapolate the $ZZ$ production cross-section measured in the fiducial phase space to the total phase space. It is defined as
\(
A_{ZZ} = \sigma_{\mathrm{fid}} / \sigma_{\mathrm{total}},
\)
where $\sigma_{\mathrm{fid}}$ and $\sigma_{\mathrm{total}}$ denote the fiducial and total inclusive cross-sections, respectively.

The inclusive $ ZZ\rightarrow \ell\ell\nu\nu$ production cross-sections in the fiducial and total phase space, calculated using \textsc{Matrix}, are presented in Table~\ref{tab:zz-xs}.
Like everywhere else in this article, $\ell\ell$ and $\nu\nu$ are the direct $Z$ decay products, with $\ell = e, \mu$ as usual.
The calculation includes NNLO QCD corrections for
$q\bar q \rightarrow ZZ \rightarrow \ell\ell\nu\nu$  and NLO QCD corrections for $gg\rightarrow ZZ\rightarrow \ell\ell\nu\nu$, along with NLO EW corrections. %
Following the prescription given in
\textsc{Matrix}, three different ways of combining QCD and EW corrections are presented. One additive prescription (NNLO QCD + EW) and two multiplicative prescriptions, where the factorisation of the EW corrections is complete (NNLO QCD × EW) or restricted to the $q \bar{q}$ channel. The differences between those prescriptions allow to assess the impact of QCD-EW mixed effects of $\mathcal{O}(\alpha_{s} \alpha$) and related theory uncertainties.
As seen in Table~\ref{tab:zz-xs}, these differences are at the level of the MC statistical uncertainty.
The result of the additive QCD+EW prescription lies between the results of the other two (multiplicative) configurations and is taken as the nominal  \textsc{Matrix} prediction.
The QCD scale uncertainty covers the variations of predictions from changing the QCD renormalisation and factorisation scales, as described in Section \ref{sec:Systematics}.
While the QCD scale uncertainty (see Section \ref{sec:Systematics}) is considered, uncertainties associated to PDFs are not included in the \textsc{Matrix} prediction. The derived fiducial phase space acceptance $A_{ZZ}$ is also listed in Table~\ref{tab:zz-xs}.

The fiducial cross-sections for the processes $ZZ \rightarrow \ell\ell\nu\nu$ and $ZZjj \rightarrow \ell\ell\nu\nu jj$ are calculated using a combination of Sherpa and MadGraph5 MC samples. The QCD-induced production is simulated with \textsc{Sherpa} and the electroweak production is simulated with \textsc{MadGraph5}, as detailed in Section~\ref{sec:dataMC}. The predicted cross-sections, including theory uncertainties, are presented in Table~\ref{tab:sherpa-xs} for both the $ZZ$ and $ZZjj$ fiducial phase spaces.

For the inclusive $ZZ$ production, the fiducial cross-section calculated with \textsc{Matrix} is about 3\% to 6\% higher than the one calculated with \SHERPA, primarily due to the inclusion of higher-order QCD corrections. Despite this difference, the results from both MC generators are consistent within their respective uncertainties. The acceptance correction factor \( C_{ZZ} \), defined as the ratio of generated signal events passing the selection criteria using reconstructed objects to the number passing the fiducial criteria using generator-level objects, is found to be \( C_{ZZ} = 0.59 \).

\begin{table}[htbp]
\caption{Predicted $ZZ \to \ell\ell\nu\nu$ ($\ell = e,\mu$) production cross-sections at $\sqrt{s} = 13$~\TeV from \textsc{Matrix}, calculated in the fiducial and total phase spaces. The first uncertainty reflects the MC statistical uncertainty, while the second reflects the QCD scale variations. In the third column,  the fiducial phase space acceptance, $A_{ZZ}$, is presented along its corresponding theory uncertainty.}
\centering
\scriptsize
\renewcommand{\arraystretch}{1.3}
\begin{tabular}{l c c c}
\hline
\textbf{Higher-order corrections} &
\textbf{$\sigma^{\mathrm{fid.}}_{ZZ\to \ell\ell\nu\nu}$ [fb]} &
\textbf{$\sigma^{\mathrm{total}}_{ZZ\to \ell\ell\nu\nu}$ [fb]} &
\textbf{$A_{ZZ}$ [\%]} \\
\hline
($q\bar{q}$ NNLO + gg NLO) QCD + NLO EW & $21.03 \pm 0.27 \, ^{+0.08}_{-0.13}$ & $413.9\pm 4.7\,{}^{+10}_{-9.5}$ & $5.08 \pm 0.13$ \\
($q\bar{q}$ NNLO + gg NLO) QCD × NLO EW & $20.74 \pm 0.27 \, ^{+0.10}_{-0.10}$ & $407.6\pm 4.5\,{}^{+10}_{-9.0}$ & $5.09 \pm 0.13$ \\
($q\bar{q}$ NNLO) QCD × NLO EW + gg NLO QCD & $21.28 \pm 0.27\,^{+0.08}_{-0.11}$ & $409.7\pm 4.5\,{}^{+10}_{-9.4}$ & $5.14 \pm 0.13$ \\
\hline
\end{tabular}%
\label{tab:zz-xs}
\end{table}

\begin{table}[htbp]
\caption{
Fiducial cross-sections for $ZZ \to \ell\ell\nu\nu$ and $ZZjj \to \ell\ell\nu\nu jj$ productions at $\sqrt{s} = 13$~\TeV. The $ZZ$ and $ZZjj$ cross-sections are obtained with the QCD production simulated using \SHERPA\
and the electroweak production simulated using \textsc{MadGraph5} (NLO EW). Theory uncertainties from QCD scale variations, PDFs, and $\alpha_s$ are quoted.
}
\centering
\scriptsize
\renewcommand{\arraystretch}{1.4}
\begin{tabular}{l c}
\hline
\textbf{Cross-section}  & \textbf{Calculated value [fb] } \\
\hline
$\sigma^{\text{fid}}_{ZZ \to \ell \ell \nu\nu}$ & $20.1 \pm 1.0\,\text{(QCD scale)} \pm 0.3\,\text{(PDF)} \pm 0.2\,(\alpha_s)$ \\
$\sigma^{\text{fid}}_{ZZjj \to \ell \ell \nu\nu jj}$ & $0.94 \pm 0.19\,\text{(QCD scale)} \pm 0.01\,\text{(PDF)} \pm 0.02\,(\alpha_s)$ \\
\hline
\end{tabular}
\label{tab:sherpa-xs}
\end{table}


\section{Cross-section measurements}
\label{sec:xs-measurement}

The signal strengths in the $ZZ$ and $ZZjj$ signal regions are extracted using a binned profile-likelihood fit to the $\Delta\phi(\ell \ell, \vec{E}^{\text{miss}}_{\mathrm{T}})$ distribution, performed simultaneously across the  SRs and the relevant CRs.
The likelihood function is constructed using the \textsc{HistFactory} framework~\cite{Cranmer:2012sba} implemented in \textsc{RooStats}~\cite{Moneta:2010pm}.
Background normalisation factors are included as free parameters in the fit and are primarily determined by data in the CRs, while the systematic uncertainties described in Section~\ref{sec:Systematics} are included as Gaussian-constrained nuisance parameters.
All nuisance parameters are taken as fully correlated across analysis regions, except for theory uncertainties: QCD scale and PDF uncertainties are correlated across regions and uncorrelated between processes, while the alternative generator uncertainties, which apply only to the signal, are treated as uncorrelated.

For each SR, a signal strength and a set of background normalisation factors are obtained from the fit.
The signal strength parameters, denoted as $\mu_\mathrm{s}^{\mathrm{ZZ}}$ and $\mu_\mathrm{s}^{\mathrm{ZZjj}}$, are defined as the ratio of the observed to predicted number of signal events in the \(ZZ\) and \(ZZjj\) SRs, respectively: $\mu_\mathrm{s}^{\mathrm{SR}} = \frac{N^{\mathrm{SR}}_{\mathrm{data}}}{N^{\mathrm{SR}}_{\mathrm{MC}}}$.
The predicted number of events come from the signal simulations by \SHERPA and \textsc{MadGraph5} described in Section~\ref{sec:dataMC}.
For the inclusive $ZZ$ case, the background scale factors are
$\mu_\mathrm{WZ}$, $\mu_{\mathrm{top}}$, $\mu_{\mathrm{WW}}$,
$\mu_\mathrm{Z+0jets}$, $\mu_\mathrm{Z+1 jet}$ and $\mu_\mathrm{Z+\geq 2 jets}$.
For the $ZZjj$ case, the $WW$ and top backgrounds are grouped in the non-resonant category and the normalisation factors used are $\mu_\mathrm{WZ}$ and $\mu_{\text{non-resonant}}$.
The $Z$+jets background in the $ZZjj$ case and all other minor backgrounds are taken from their MC predictions.

The measured signal strength in each SR is later used to derive the cross-section in the corresponding fiducial phase space.
Theory systematic uncertainties are propagated to the measured cross-section via the ratio of reconstructed to fiducial number of MC signal events, in order to avoid double counting, as the signal strength $\mu_\mathrm{s}^{\mathrm{SR}}$ is defined relative to the reconstructed prediction.

In the inclusive $ZZ$ case, the fitted signal strength is $\mu_\mathrm{s}^\mathrm{ZZ} = 1.05 \pm 0.05$. The background scale factors are found to be: $1.00^{+0.06}_{-0.05}$ for $WZ$, $0.98^{+0.13}_{-0.10}$ for top, $1.27^{+0.18}_{-0.17}$ for $WW$,
$1.14^{+0.44}_{-0.30}$ for $Z \mathrm{+0 jets}$,
$1.06^{+0.37}_{-0.28}$ for $Z \mathrm{+1 jet}$,
and $1.10^{+0.44}_{-0.31}$ for $Z \mathrm{+ \geq 2 jets}$.

In the \(ZZjj\) case, the fitted signal strength is $\mu_s^{ZZjj} = 1.02^{+0.19}_{-0.17}$.
The background normalisation factors are found to be $0.97^{+0.24}_{-0.20}$ for $WZ$ and $1.08^{+0.19}_{-0.15}$ for the non-resonant backgrounds.

\begin{figure}[!htbp]
\minipage{1\textwidth}
\centering
\subfloat[]{
\includegraphics[width=0.32\linewidth]{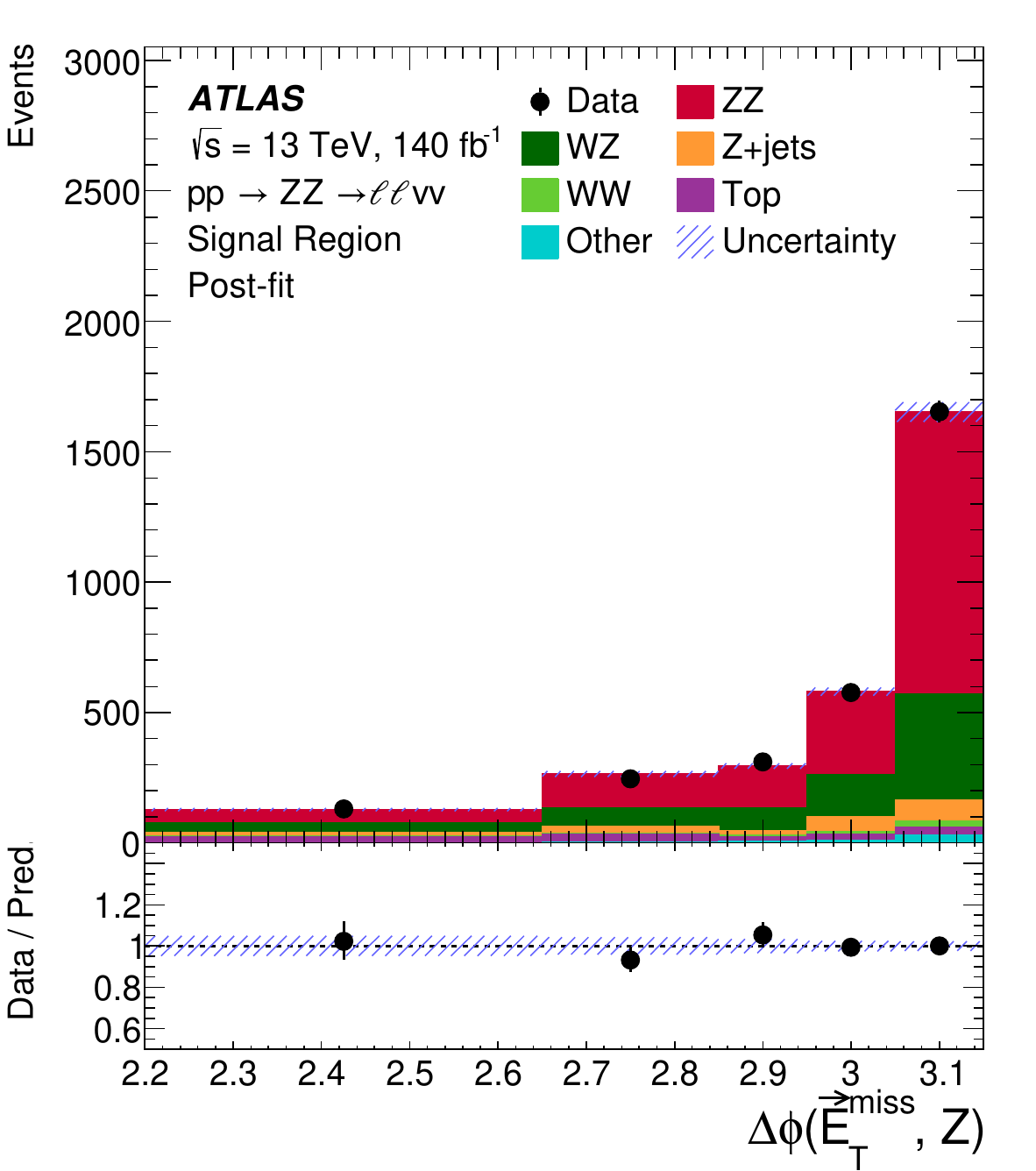}
} \hspace{4 mm}
\subfloat[]{
\includegraphics[width=0.32\linewidth]{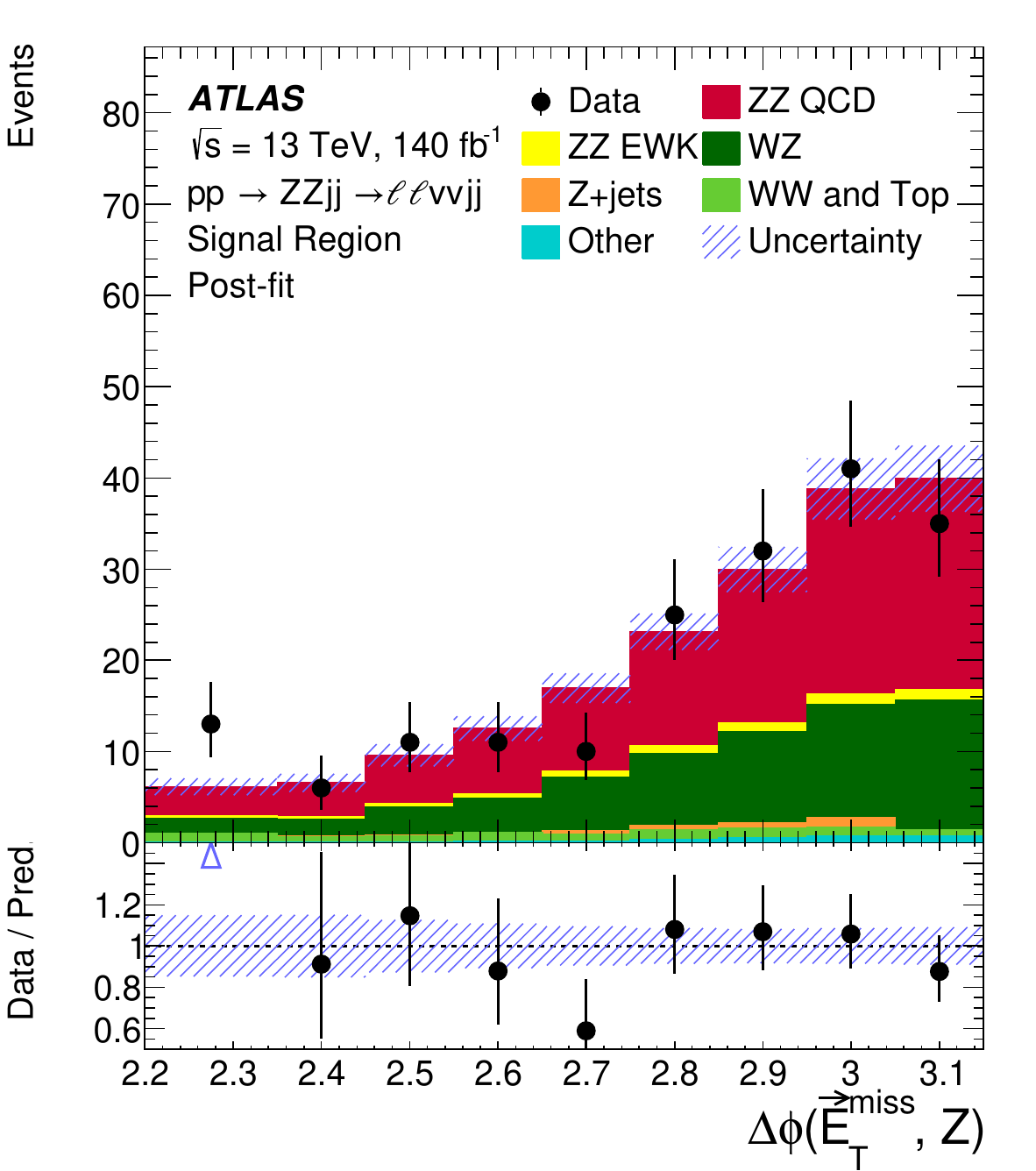}
}\\
\subfloat[]{
\includegraphics[width=0.325\linewidth]{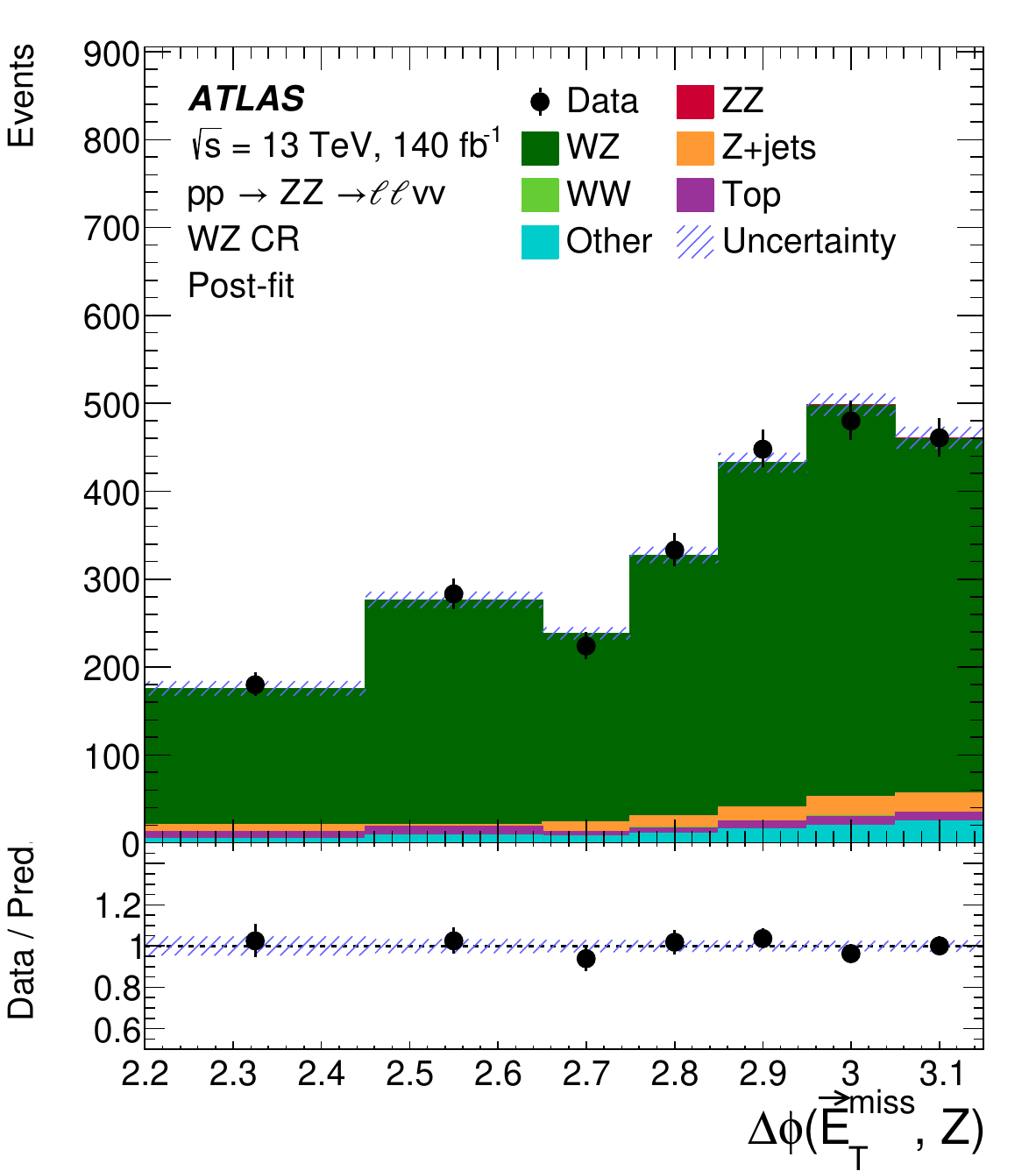}
}
\subfloat[]{
\includegraphics[width=0.32\linewidth]{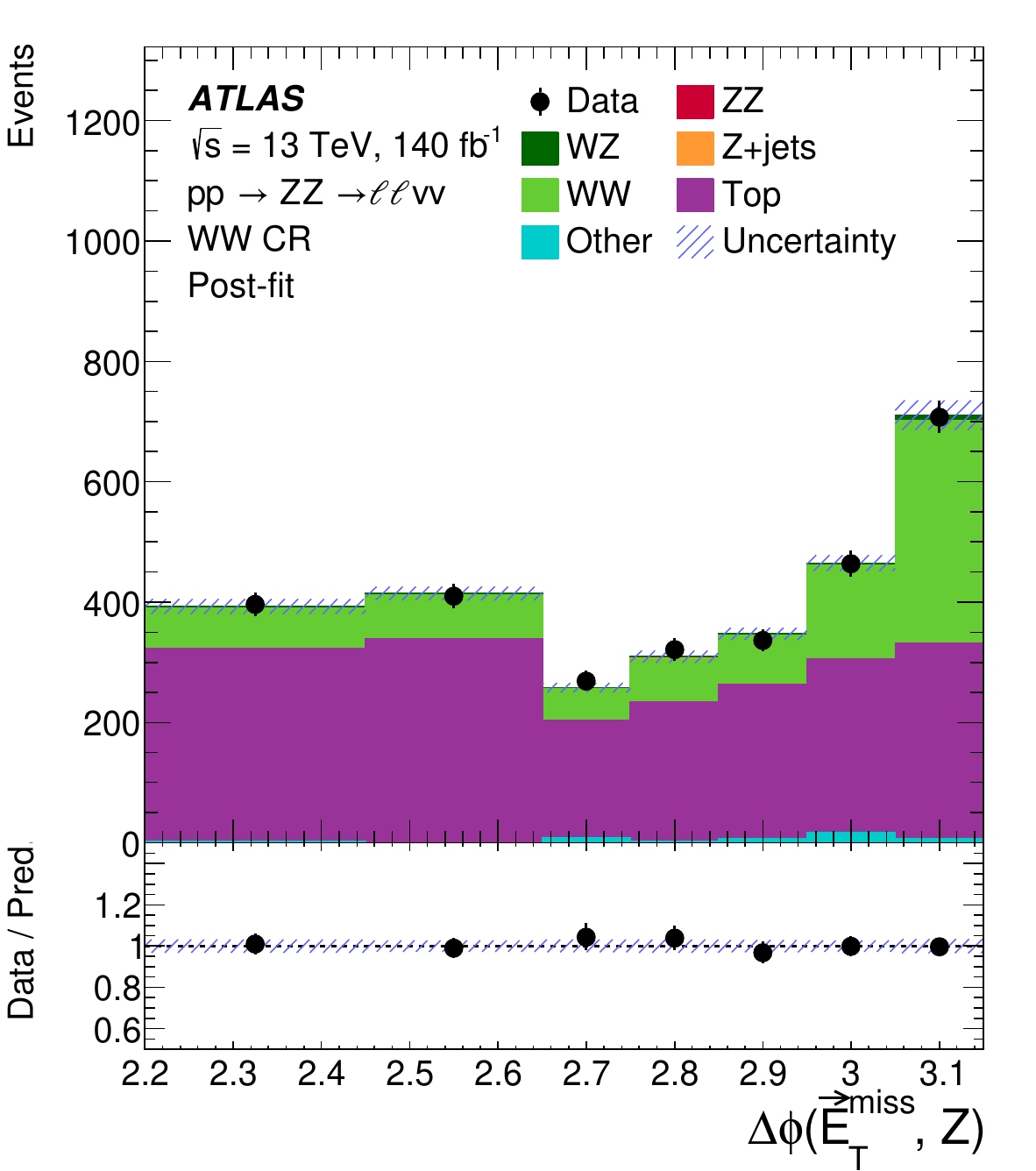}
}
\subfloat[]{
\includegraphics[width=0.32\linewidth]{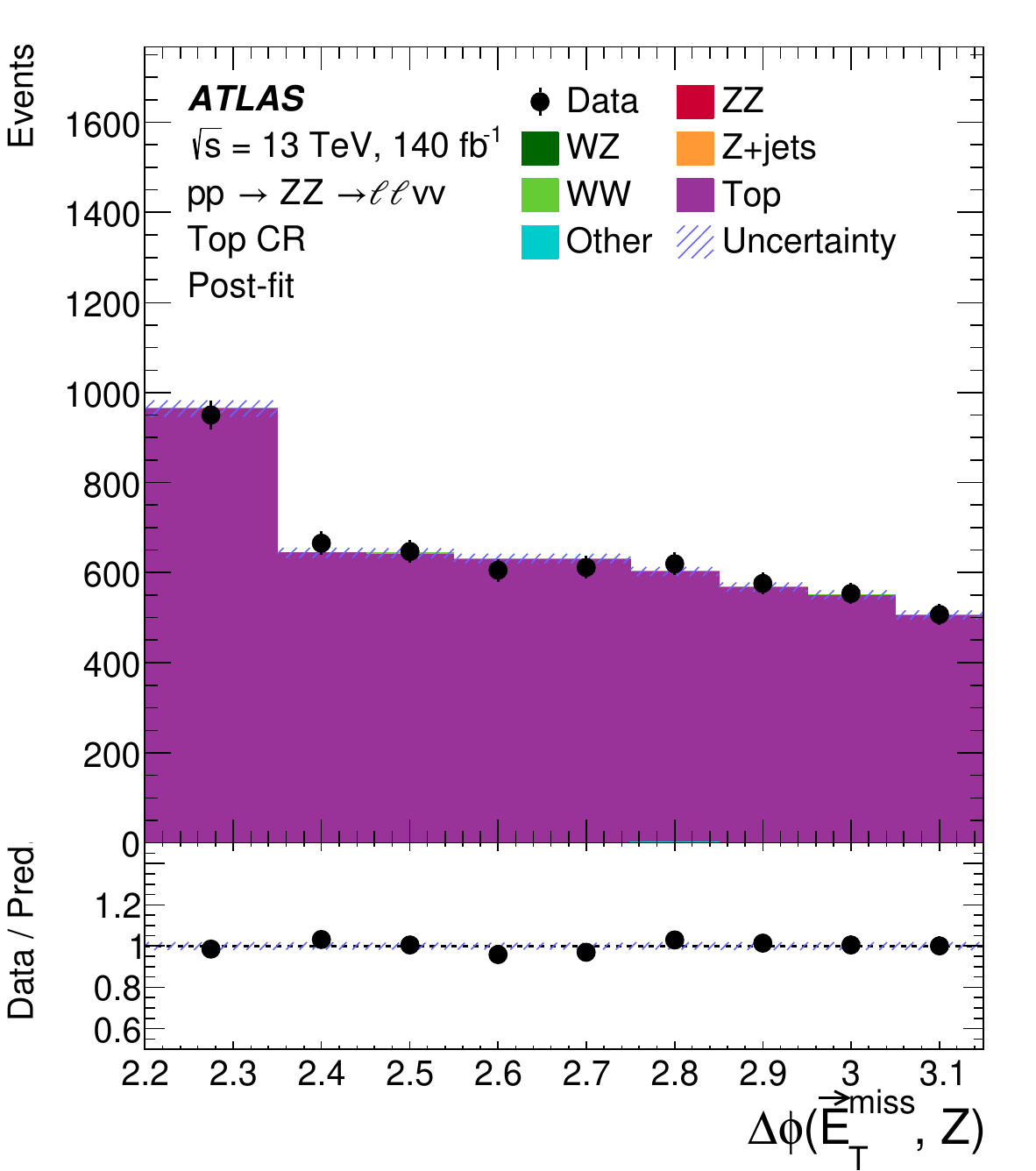}
}\\
\subfloat[]{
\includegraphics[width=0.325\linewidth]{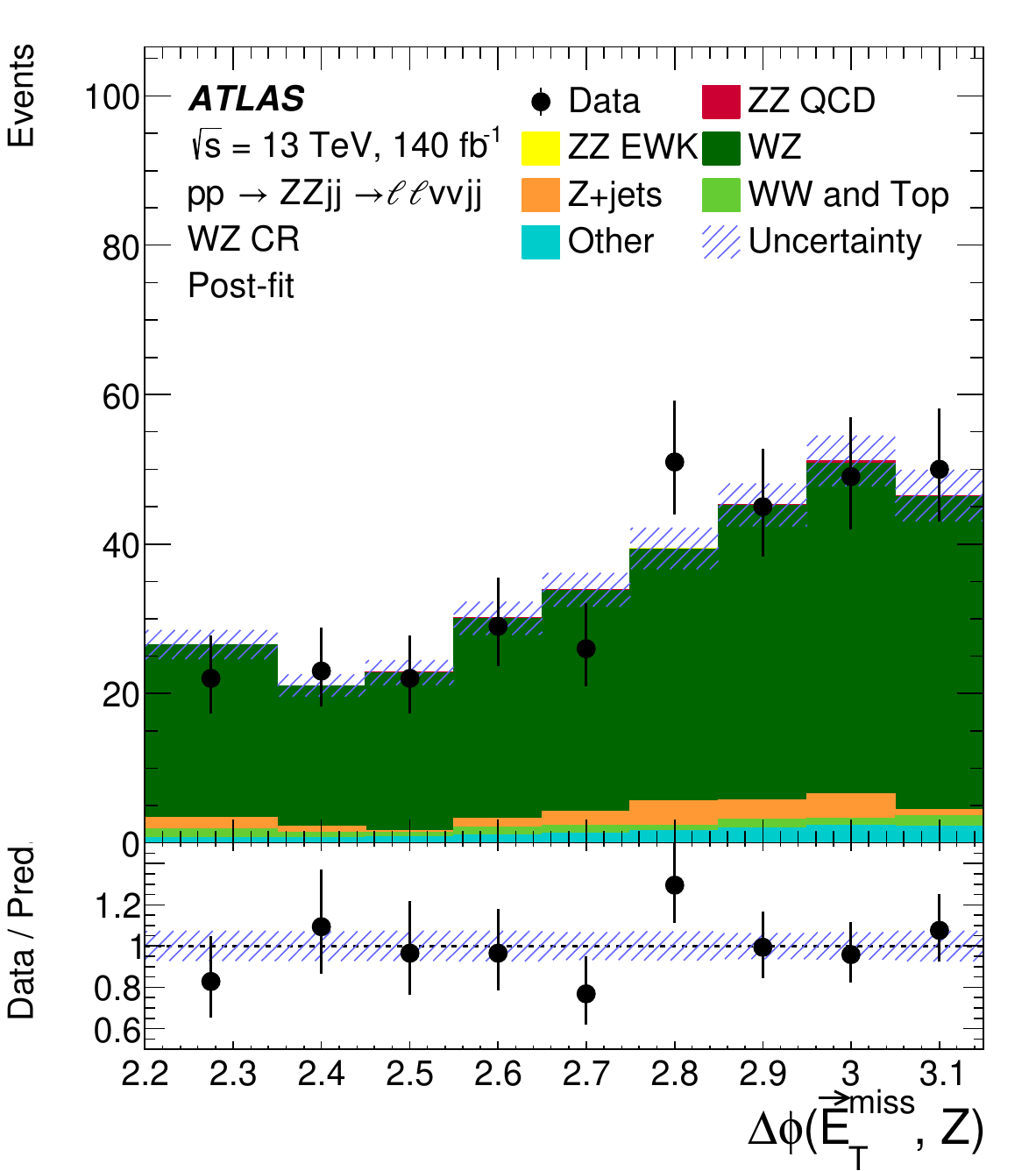}
}
\subfloat[]{
\includegraphics[width=0.325\linewidth]{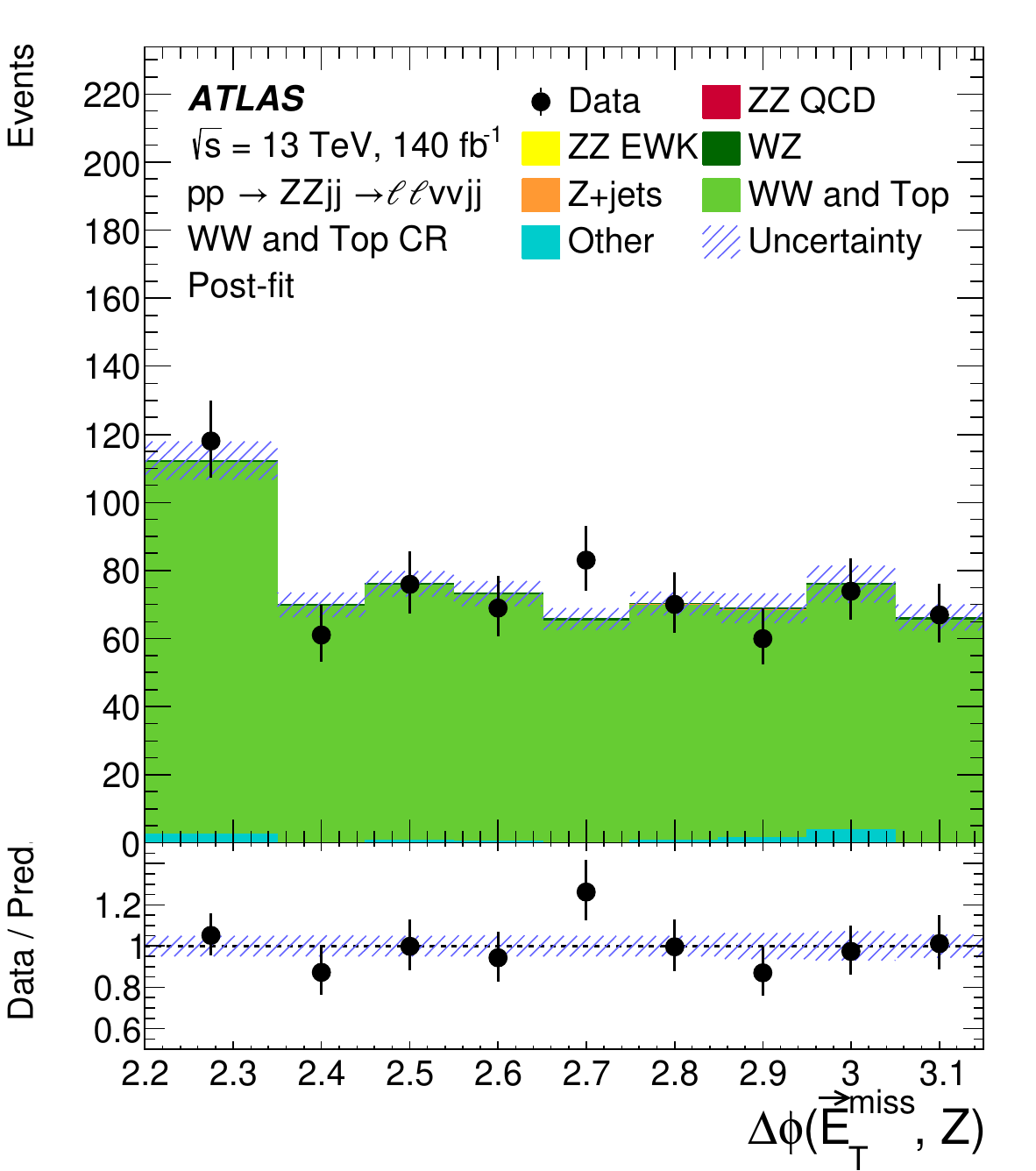}
}
\subfloat[]{
\includegraphics[width=0.325\linewidth]{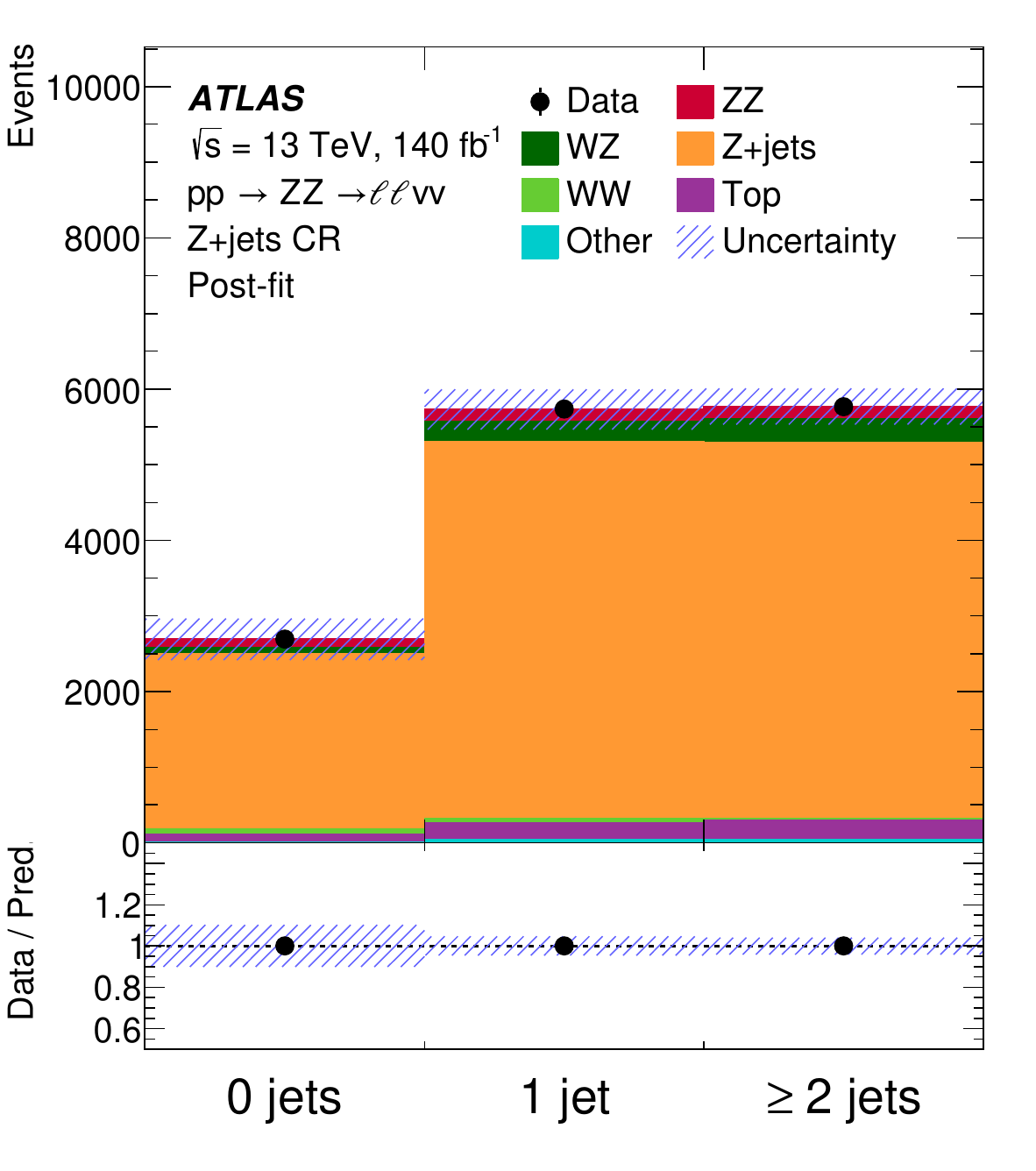}
}
\caption{The post-fit distributions of $\Delta\phi(\vec{E}_\text{T}^\text{miss}, Z)$. Panels (a) and (b) show the signal regions for the inclusive $ZZ$ and $ZZjj$ production, respectively. The "Other" category counts events from processes of $ZZ(\rightarrow 4\ell$, $2\ell 2q), VVV (V=W, Z), Wj, Z(\rightarrow \tau\tau), \textrm{and}~ WH$. Panels (c), (d) and (e) illustrate the background CRs for inclusive $ZZ$ production, while panels (f) and (g) display the $ZZjj$ CRs. Panel (h) presents the post-fit event yields for background events in the $Z$+jets CRs for the inclusive $ZZ$ analysis.  Data points are shown with statistical error bars, while the shaded band represents the total post-fit uncertainty in the prediction, combining statistical and systematic uncertainties. Open markers indicate data points lying outside the vertical range of the plot.}
\label{fig:postfit_observed_ZZCR}
\endminipage\hfill
\end{figure}

The post-fit
$\Delta\phi$($\vec{E}_\text{T}^\text{miss}, Z$) distributions in the $ZZ$ and $ZZjj$ SRs and in the various CRs are shown in Figure~\ref{fig:postfit_observed_ZZCR}.
Representative post-fit kinematic distributions in the signal regions are shown in Figure~\ref{fig:postfit_observed}, where the transverse mass of the $ZZ$ system is defined as
$m_\text{T}^\text{ZZ} = \sqrt{\left[\sqrt{m_\text{Z}^2+(p_\text{T}^{\ell\ell})^2}+\sqrt{m_\text{Z}^2+(E_\text{T}^\text{miss})^2}\right]^2 -|\vec{p}_\text{T}^{\ell\ell}+\vec{E}_\text{T}^\text{miss}|^2}$.

\begin{figure}[!htbp]
\minipage{1\textwidth}
\centering
\subfloat[]{
\includegraphics[width=0.32\linewidth]{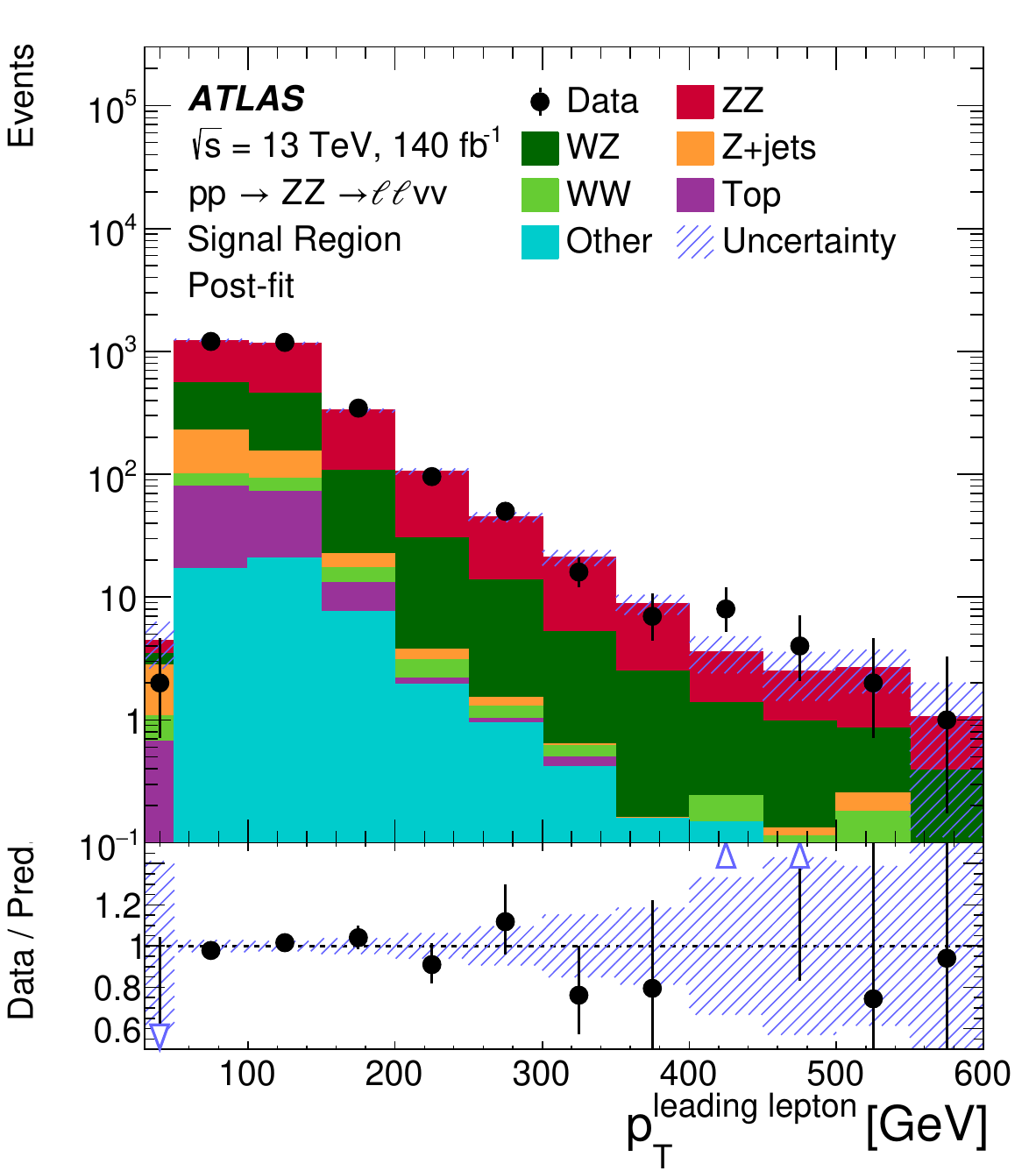}
}
\subfloat[]{
\includegraphics[width=0.32\linewidth]{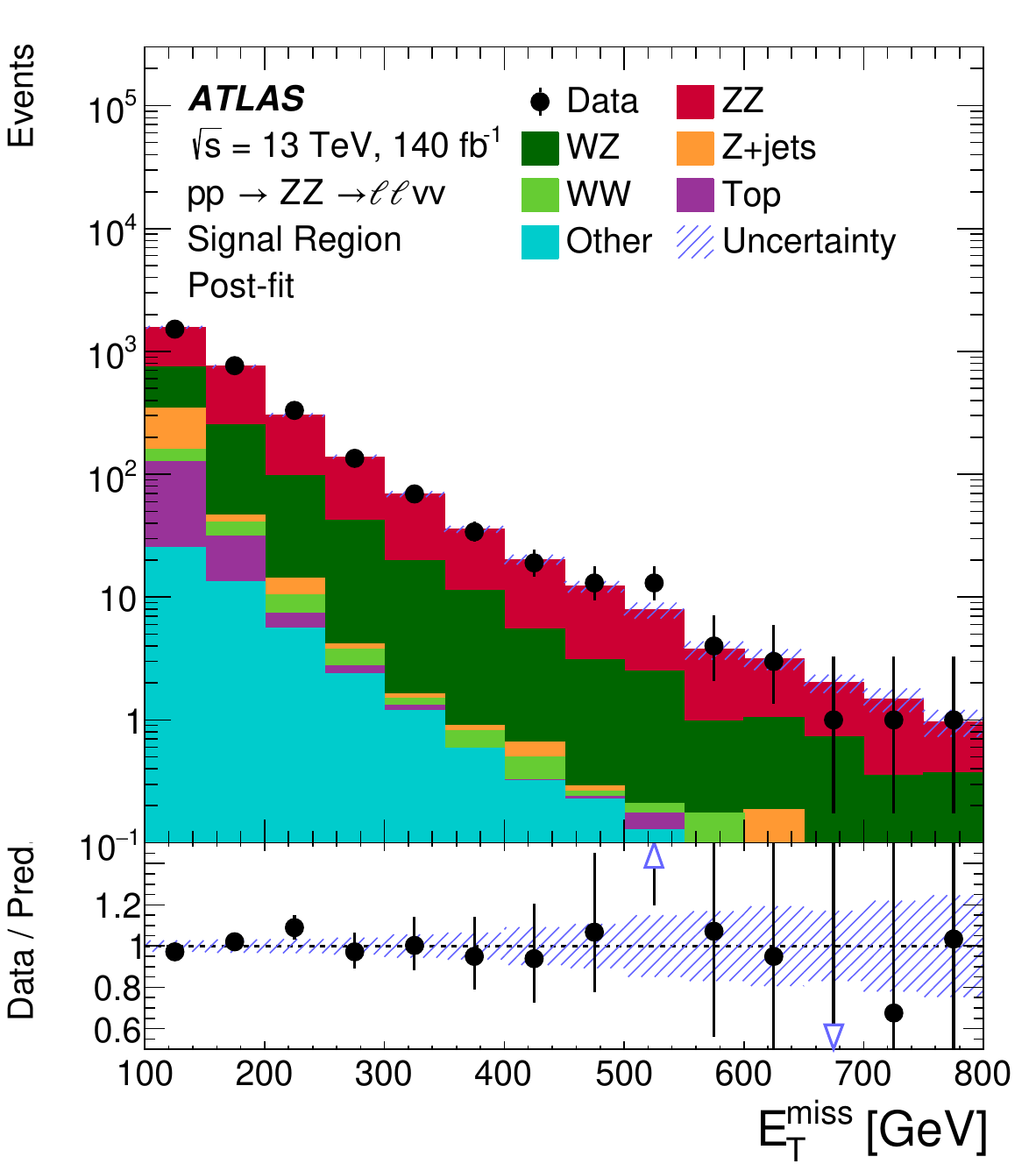}
}
\subfloat[]{
\includegraphics[width=0.32\linewidth]{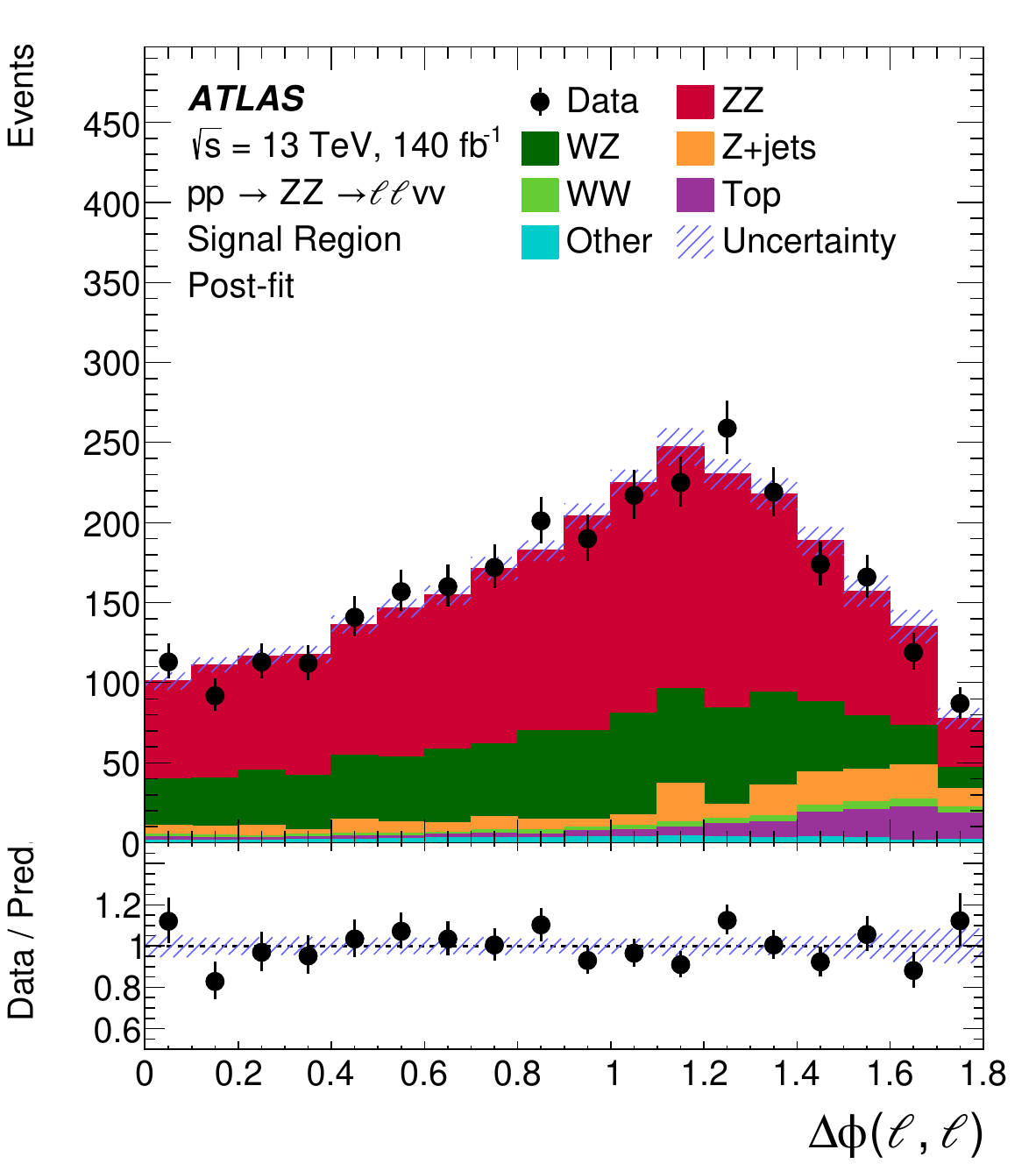}
}\\
\subfloat[]{
\includegraphics[width=0.32\linewidth]{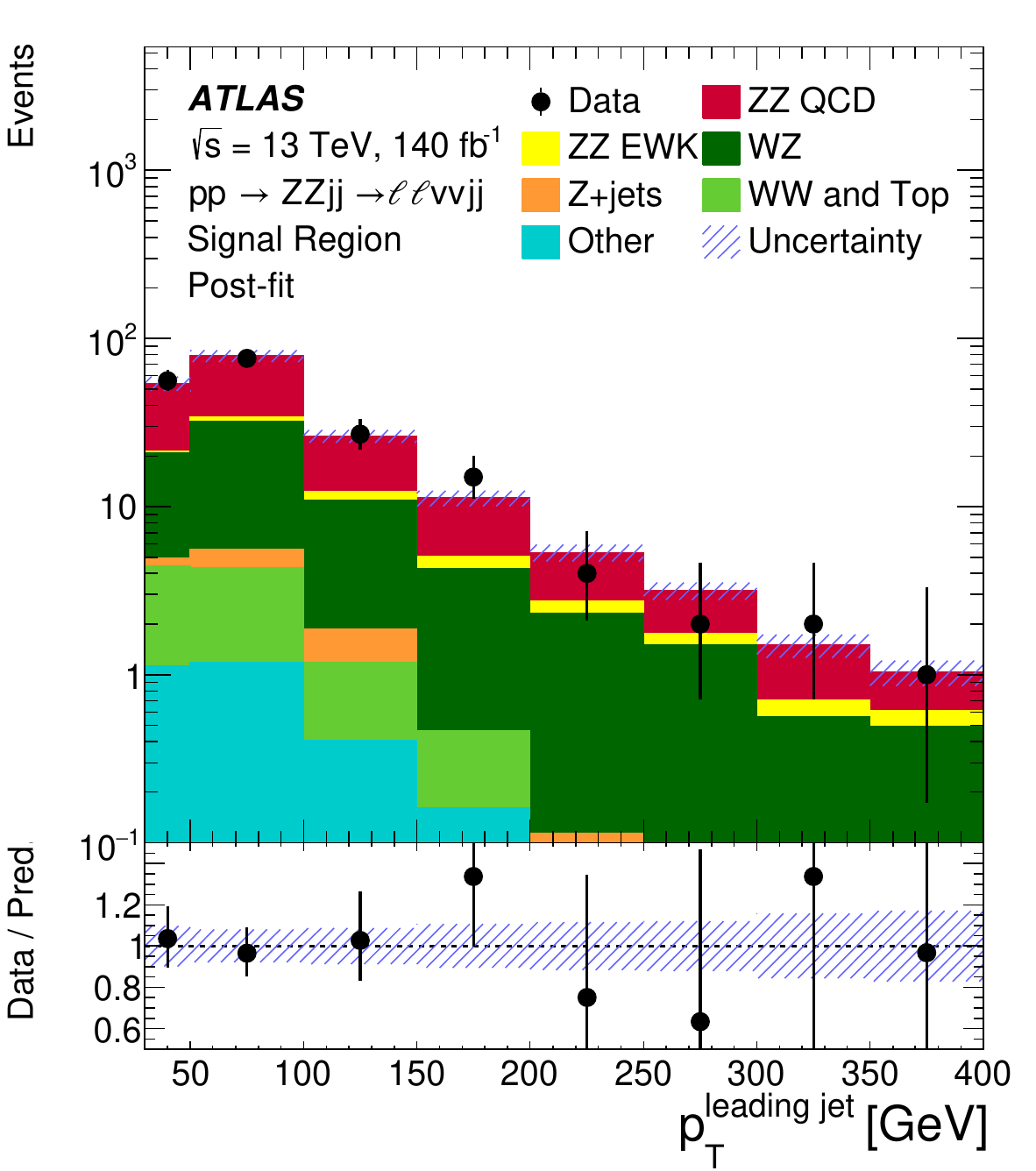}
}
\subfloat[]{
\includegraphics[width=0.32\linewidth]{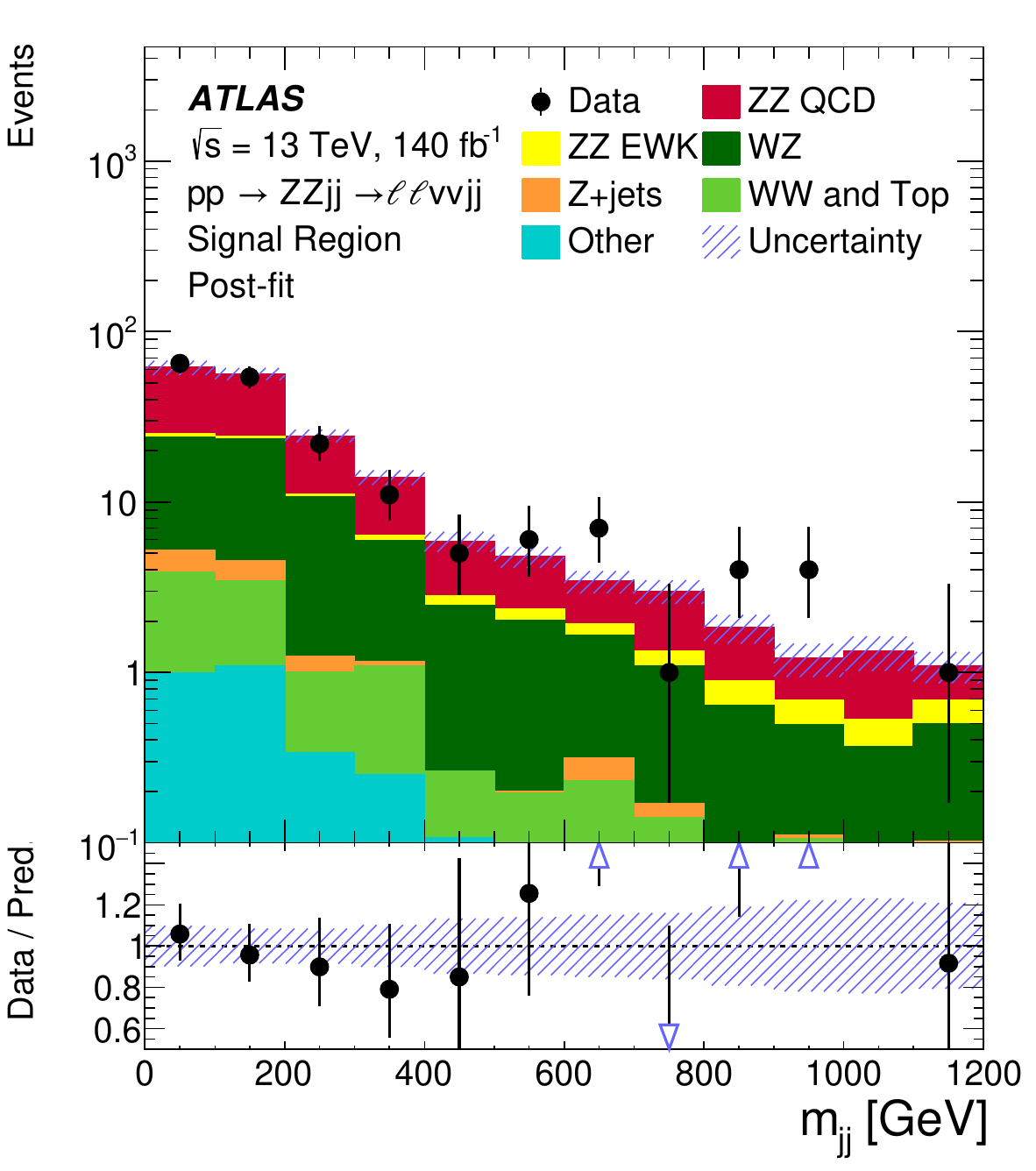}
}
\subfloat[]{
\includegraphics[width=0.32\linewidth]{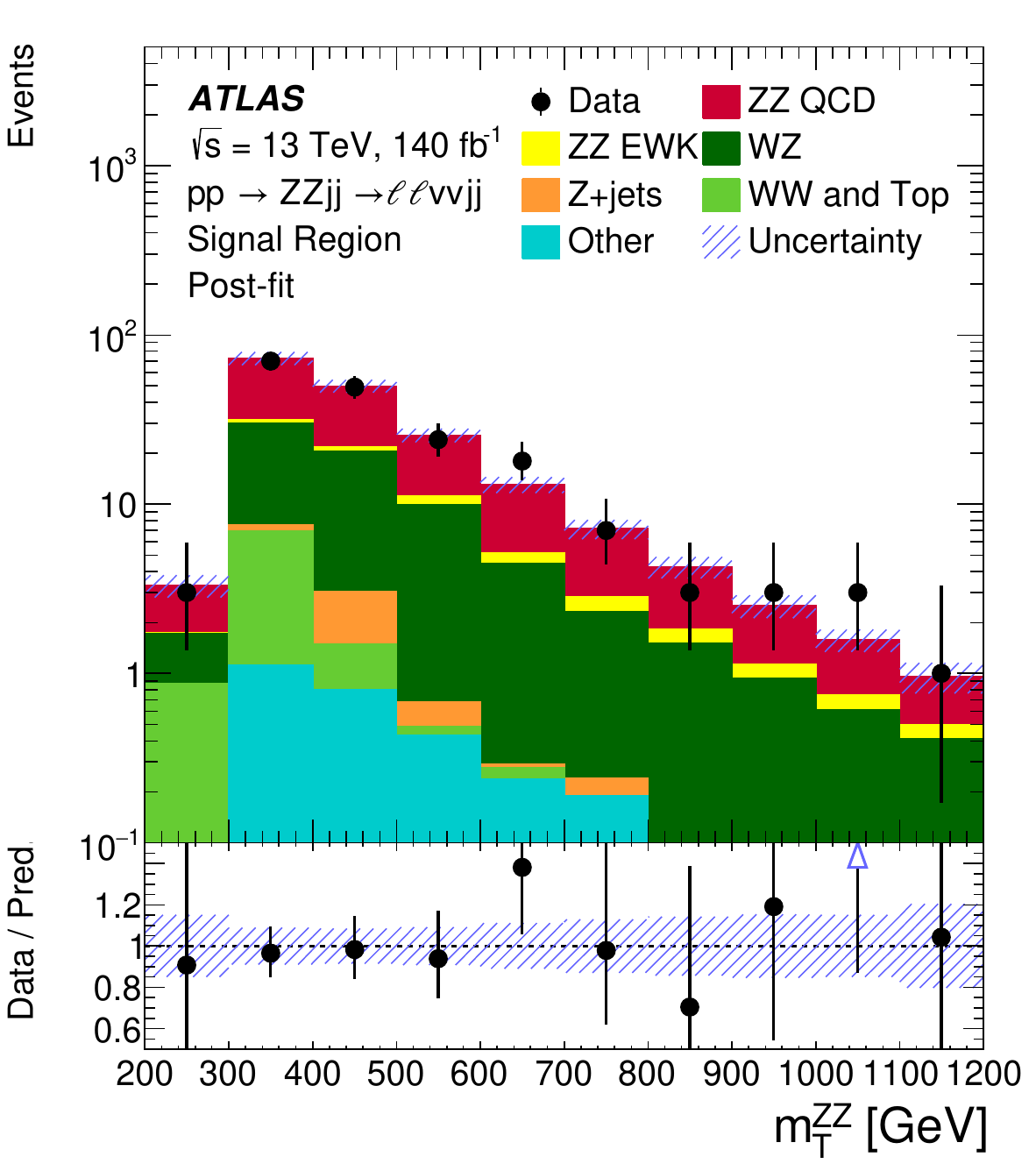}
}
\caption{Post-fit kinematic distributions for the inclusive $ZZ$ SR are shown in the top row: (a) the transverse momentum of the leading lepton $p_\text{T}^\text{leading lepton}$, (b) $E_\text{T}^\text{miss}$ and (c) the azimuthal angle separation between two charged leptons $\Delta\phi(\ell, \ell)$. For the $ZZjj$ SR, the bottom row shows: (d) the leading jet $p_\text{T}^\text{leading jet}$, (e) the invariant mass of the dijet system $m_\text{jj}$,
and (f) the transverse mass of the $ZZ$ system $m_\text{T}^\text{ZZ}$. The data points are shown with statistical error bars, while the shaded band represents the total post-fit uncertainty in the prediction, combining statistical and systematic uncertainties. Open markers indicate data points lying outside the vertical range of the plot.}
\label{fig:postfit_observed}
\endminipage\hfill
\end{figure}

\subsection{Integrated cross-sections}

The integrated cross-sections for the processes $ZZ \rightarrow \ell\ell\nu\nu$ and $ZZjj \rightarrow \ell\ell\nu\nu jj$ in their respective fiducial phase spaces are calculated
by multiplying the measured signal strengths ($\mu_\text{s}^\text{ZZ}$ and $\mu_\text{s}^\text{ZZjj}$) by the SM cross-section predictions ($\sigma^{ZZ}_{MC}$ and $\sigma^{ZZjj}_{MC}$),
calculated with the \textsc{Sherpa} MC simulations, as shown in Table~\ref{tab:sherpa-xs}.

The measured fiducial cross-section for the inclusive $ZZ \to \ell\ell\nu\nu$ production process is found to be

$$\sigma_{ZZ}^{\mathrm{fid}} = 21.03 \pm 0.73\,(\text{stat}) \pm 0.57\,(\text{exp}) \pm 0.18\,(\text{theory}) \pm 0.18\,(\text{lumi}) \text{ fb.} $$

where the uncertainty is separated into the data statistical uncertainty (stat), the uncertainty from experimental systematic uncertainties (exp), the uncertainty from theory systematic uncertainties (theory) and the luminosity uncertainty (lumi).
The measured cross-section is in agreement with the SM prediction of $20.1 \pm 1.1$~fb from \textsc{Sherpa}+\textsc{MadGraph5} and $21.03 \pm 0.30
$~fb from \textsc{Matrix}.

For the $ZZjj \to \ell\ell\nu\nu jj$ production process, the measured fiducial cross-section is

$$\sigma_{ZZjj}^{\mathrm{fid}} = 0.96^{+0.15}_{-0.11}~(\text{stat}) \pm 0.10~(\text{exp}) \pm 0.04~(\text{theory}) \pm 0.01~(\text{lumi}) \text{ fb.}$$

This is compatible with the theory prediction of $0.94 \pm 0.20$~fb from \textsc{Sherpa}+\textsc{MadGraph5}.

The $ZZ\rightarrow \ell\ell\nu \nu$ fiducial production cross-section is extrapolated to the total phase space, using the fiducial acceptance reported in Table~\ref{tab:zz-xs} and the branching ratio~\cite{Zyla:2020zbs} of the $ZZ$ decay mode  $ZZ \rightarrow 2\ell 2\nu $.
The total $pp\rightarrow ZZ$ production is found to be:
$$\sigma^\text{total}_\text{measured}(pp\rightarrow ZZ) = 15.38\pm 0.81 ~\text{pb}.$$
This measurement is in agreement with the SM prediction calculated using \textsc{Matrix}: $$\sigma^\text{total}_\text{SM}(pp\rightarrow ZZ) = 15.40\pm 0.38 ~\text{pb}.$$


\subsection{Differential cross-sections}

Differential cross-sections are measured in the fiducial phase spaces by counting the data events in each bin of the relevant observables, subtracting the expected background contributions and correcting for detector effects. Detector-level distributions are unfolded to stable particle level using an iterative Bayesian method~\cite{DAgostini:1995}, which accounts for resolution effects, reconstruction inefficiencies and bin migrations. The response matrices are derived from simulated signal samples and encode the probability of an event generated in a given particle-level bin being reconstructed in a particular detector-level bin. The number of iterations, between two and five depending on the observable, is optimised to balance statistical uncertainty and unfolding stability. The binning is designed based on purity and efficiency criteria to ensure reliable unfolding and preserve the shape of the distributions. The last bin of each observable is open-ended and extends to the kinematic limit, so that it includes the full tail of the distribution; no separate overflow bin is used.
Validation tests with the nominal and alternative samples are performed in the simulation and no significant bias is observed. The kinematic variables used for the differential cross-section measurements are summarised in Table~\ref{tab:diff-var}.

A CP-sensitive angular variable, $\sin(\varphi) \cos(\theta)$, is constructed in the $ZZ$ signal region following the prescription of Refs.~\cite{Chang:1994cs, STDM-2021-05, Semushin:2025uts}. The polar angle $\theta$ and the azimuthal angle $\varphi$ are defined for the negatively charged lepton in the dilepton rest frame, using a special coordinate system in which the $z$-axis is aligned with the dilepton momentum in the $ZZ$ rest frame. The $x$-axis lies in the reaction plane, sharing the same polar direction as the $z$-axis and the $y$-axis completes a right-handed system. Since neutrinos from $Z\to\nu\bar{\nu}$ decays are not directly detected, the dineutrino momentum is assumed to follow the relation $\vec{p}^{\,\nu\nu}=-\vec{p}^{\,\ell\ell}$, corresponding to the assumption of the $ZZ$ system being at rest in the laboratory frame, and the same assumption is used at particle level to ensure a consistent definition. The differential cross-section as a function of this variable is unfolded to particle level in the fiducial phase space.

Figures~\ref{fig:diff-xs-ZZ} and~\ref{fig:diff-xs-ZZjj} display the measured differential cross-sections. The uncertainties shown in the plots and tables include both statistical and systematic contributions, propagated through the unfolding procedure.
Statistical uncertainties in the data dominate the measurement in most bins; however, systematic uncertainties are significant in specific kinematic regions.

\begin{table}[!hbtp]
\caption{Kinematic variables for the differential cross-section measurements.} %
\centering
\scriptsize
\renewcommand{\arraystretch}{1.2}
\begin{tabular}{l c c}
\hline
\textbf{Observable description} &  \textbf{Observable} &  \textbf{SR} \\
\hline

Azimuthal angular separation of the two leptons & $\Delta\phi(\ell, \ell)$  & $ZZ$ and $ZZjj$ \\
Transverse momentum of the leading lepton  &  $p_\text{T}^{\text{leading lepton}}$ & $ZZ$ only \\
Transverse momentum of the Z boson  & $p_\text{T}^Z$ & $ZZ$ and $ZZjj$ \\
Rapidity of the visible $Z \rightarrow \ell\ell$  &    $|y^Z|$ & $ZZ$ and $ZZjj$ \\
Transverse momentum of the $ZZ$ system  &  $p_\text{T}^{ZZ}$ & $ZZ$ and $ZZjj$ \\
Transverse mass of the $ZZ$ system
&  $m_\text{T}^{ZZ}$ & $ZZ$ and $ZZjj$\\
Number of jets                   &   $N_{\mathrm{jets}}$ & $ZZ$ only\\
CP-sensitive angular variable   &  sin($\varphi$)cos($\theta$) & $ZZ$ only\\
Invariant mass of the dijet system & $m_{\mathrm{jj}}$ & $ZZjj$ only \\
Transverse momentum of the leading jet & $p_\text{T}^{\text{leading jet}}$ & $ZZjj$ only \\
\hline
\end{tabular}
\label{tab:diff-var}
\end{table}

The measured differential cross-sections in the $ZZ$ fiducial phase space show good agreement with SM predictions from the nominal \textsc{Sherpa}+\textsc{MadGraph5}  MC simulation (denoted as \textsc{Sherpa+MG5} in the plots) and from the \textsc{Matrix} calculation. The \textsc{Matrix} predictions exhibit slightly better agreement with the data, which can be attributed to the inclusion of NNLO QCD corrections.

In the $ZZjj$ phase space, the results are compared with SM predictions obtained using \textsc{Sherpa} for strong production and \textsc{MadGraph5} for electroweak production.
The total uncertainties in the $ZZjj$ measurements are larger than in the inclusive $ZZ$ case, reflecting the higher statistical uncertainty and the larger jet-related systematic uncertainties. The \textsc{Sherpa} prediction describes well the data within uncertainties.

\begin{figure}[!hbtp]
\centering
\subfloat[]{
\includegraphics[width=0.32\linewidth]{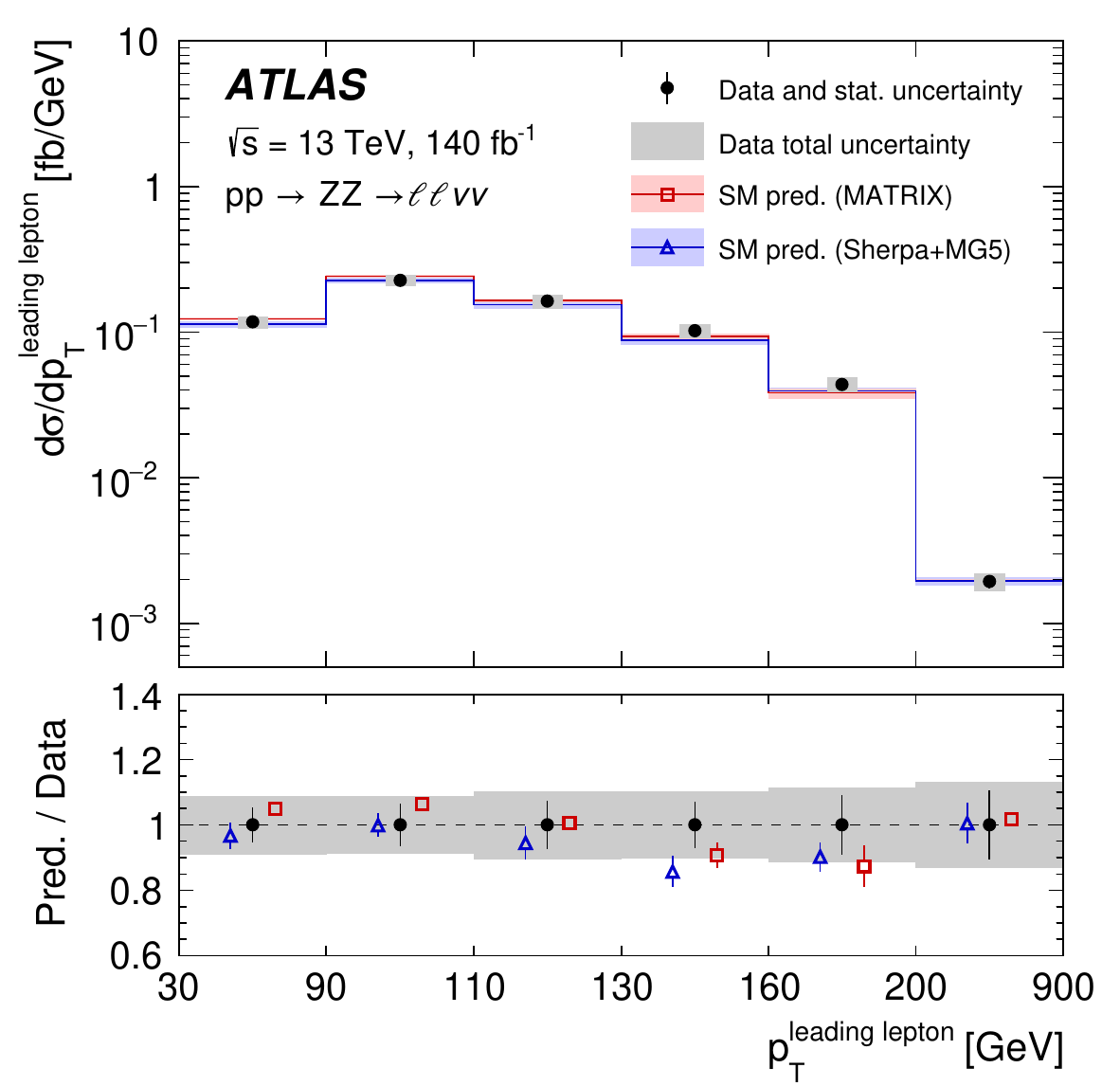}
}
\subfloat[]{
\includegraphics[width=0.32\linewidth]{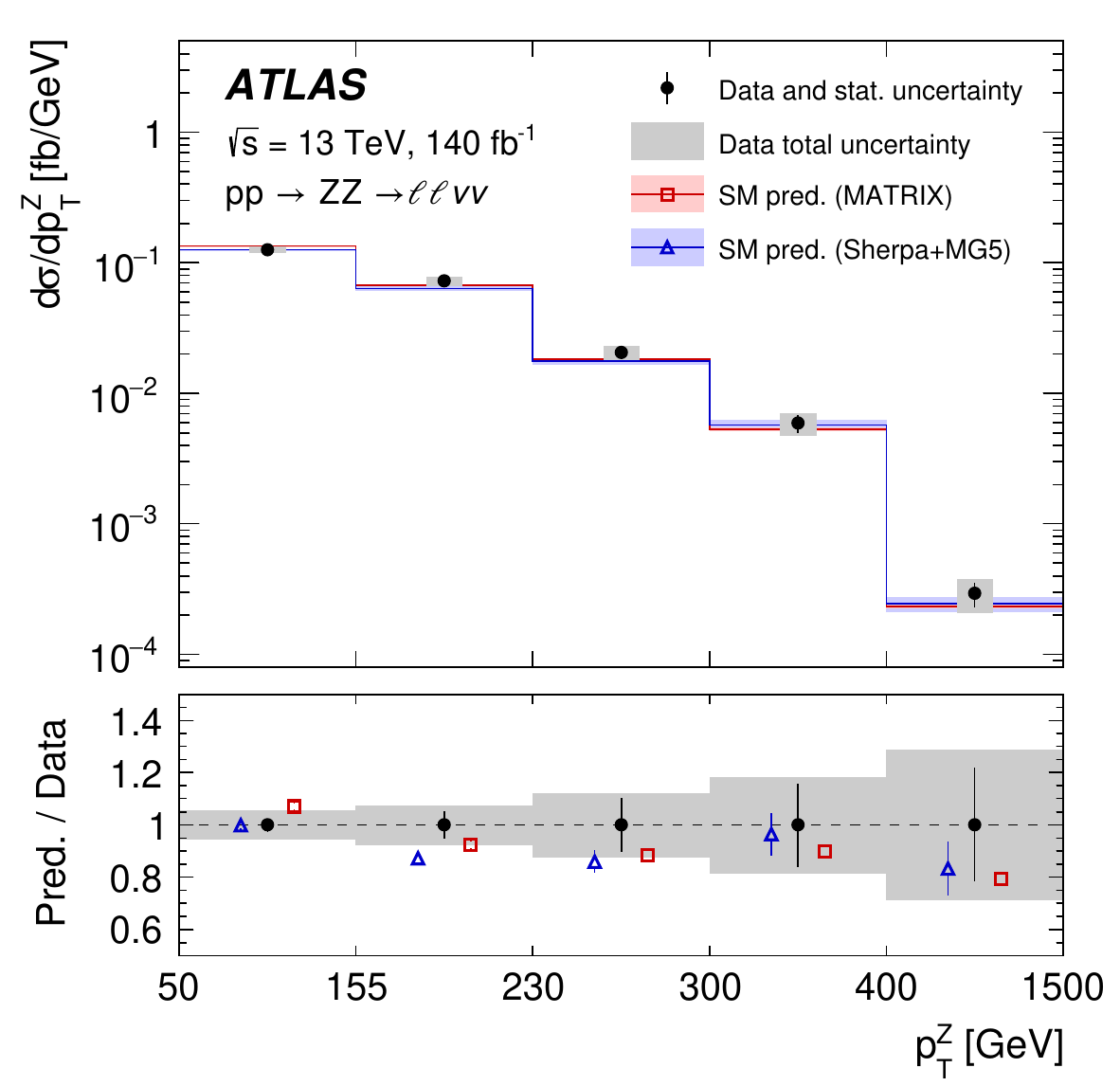}
}
\subfloat[]{
\includegraphics[width=0.32\linewidth]{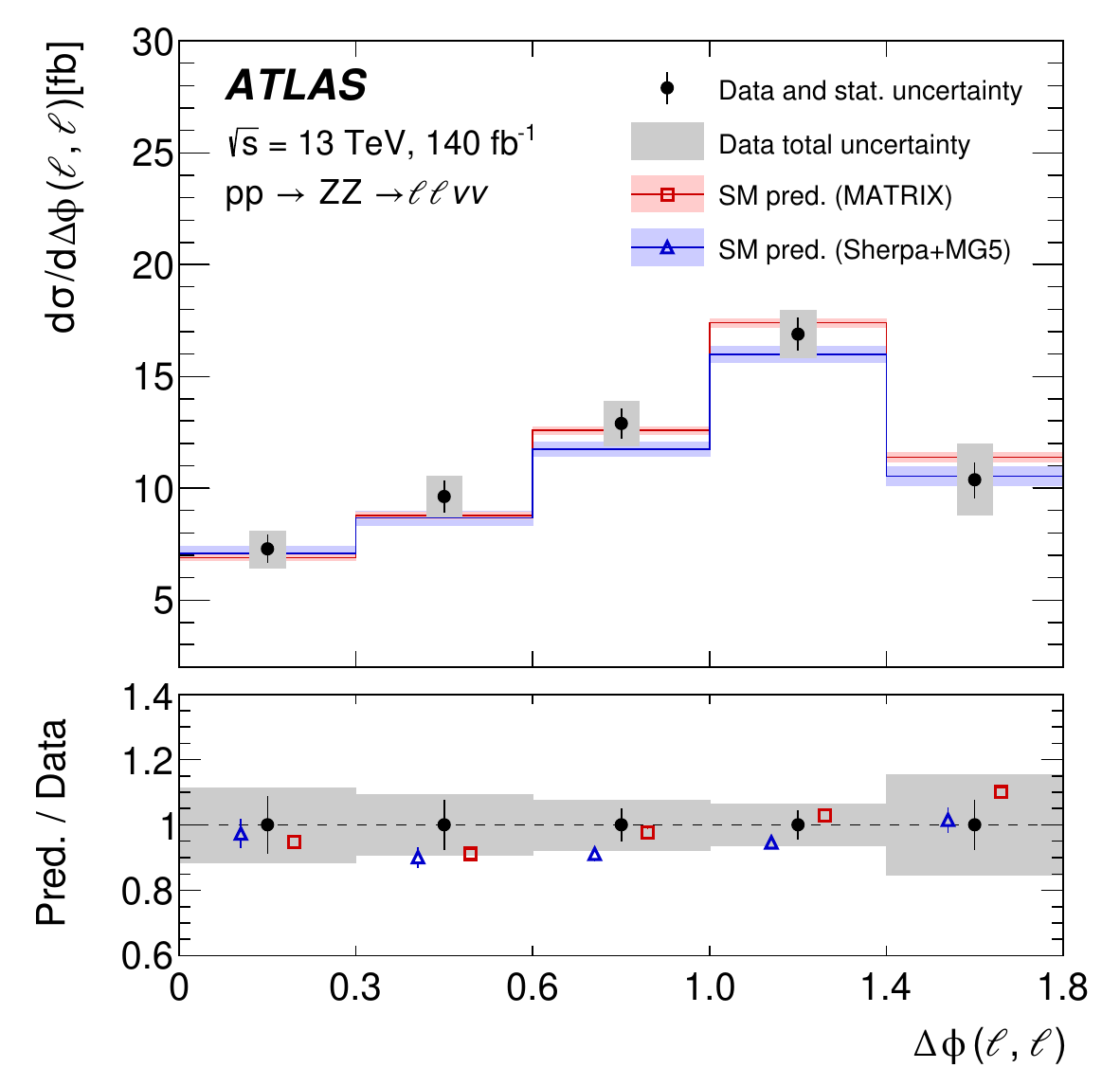}
}\\
\subfloat[]{
\includegraphics[width=0.32\linewidth]{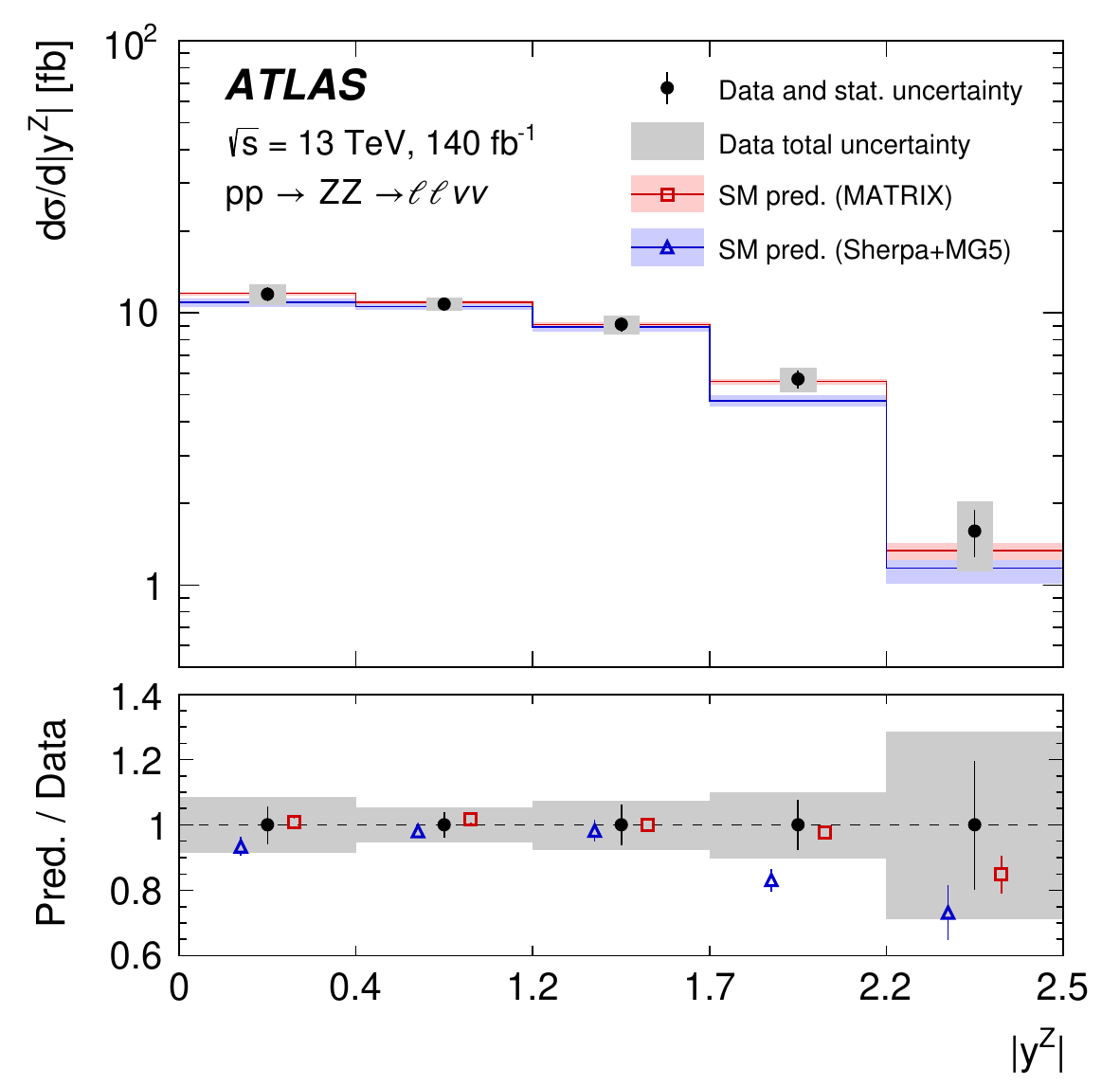}
}
\subfloat[]{
\includegraphics[width=0.32\linewidth]{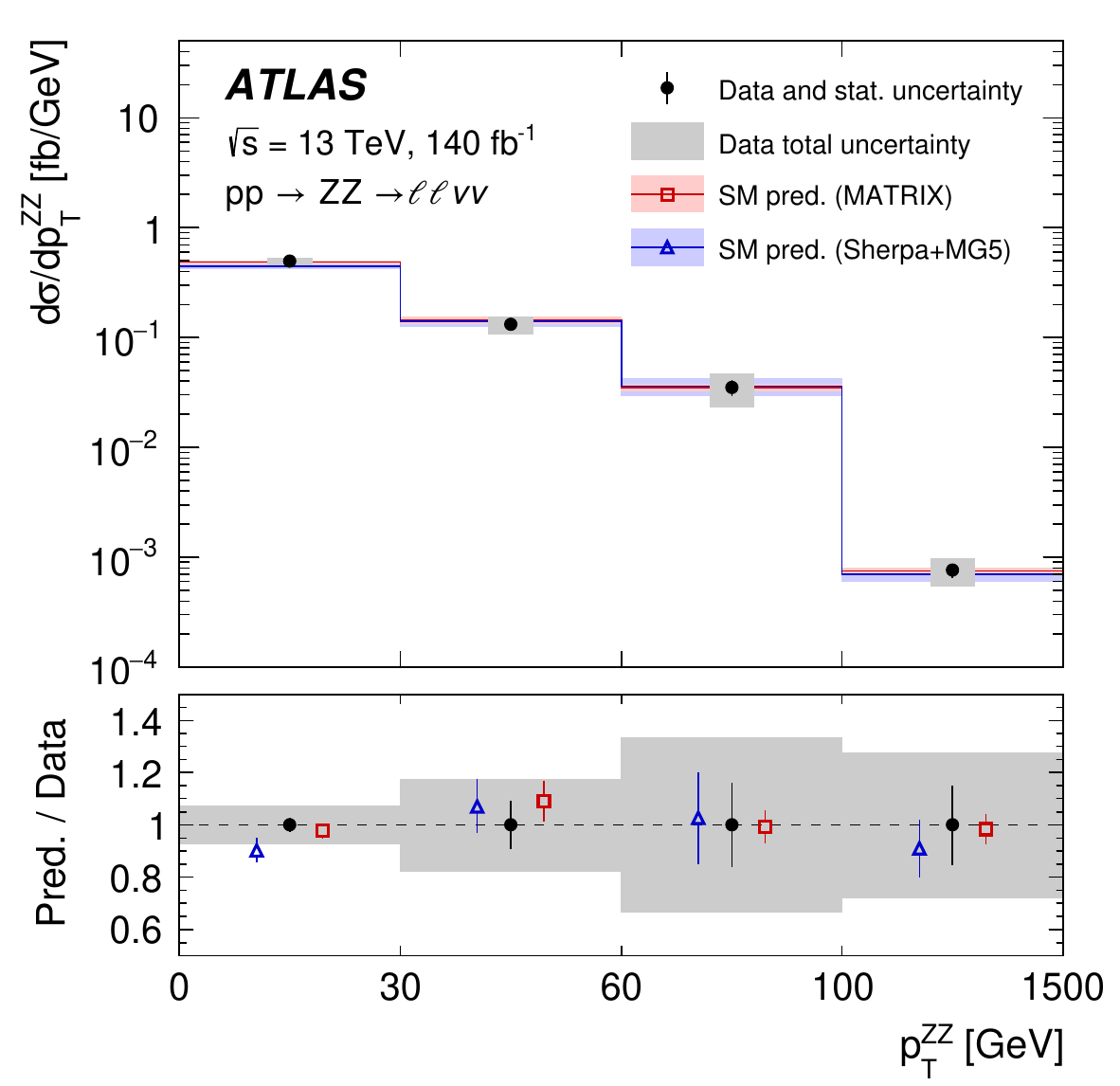}
}
\subfloat[]{
\includegraphics[width=0.32\linewidth]{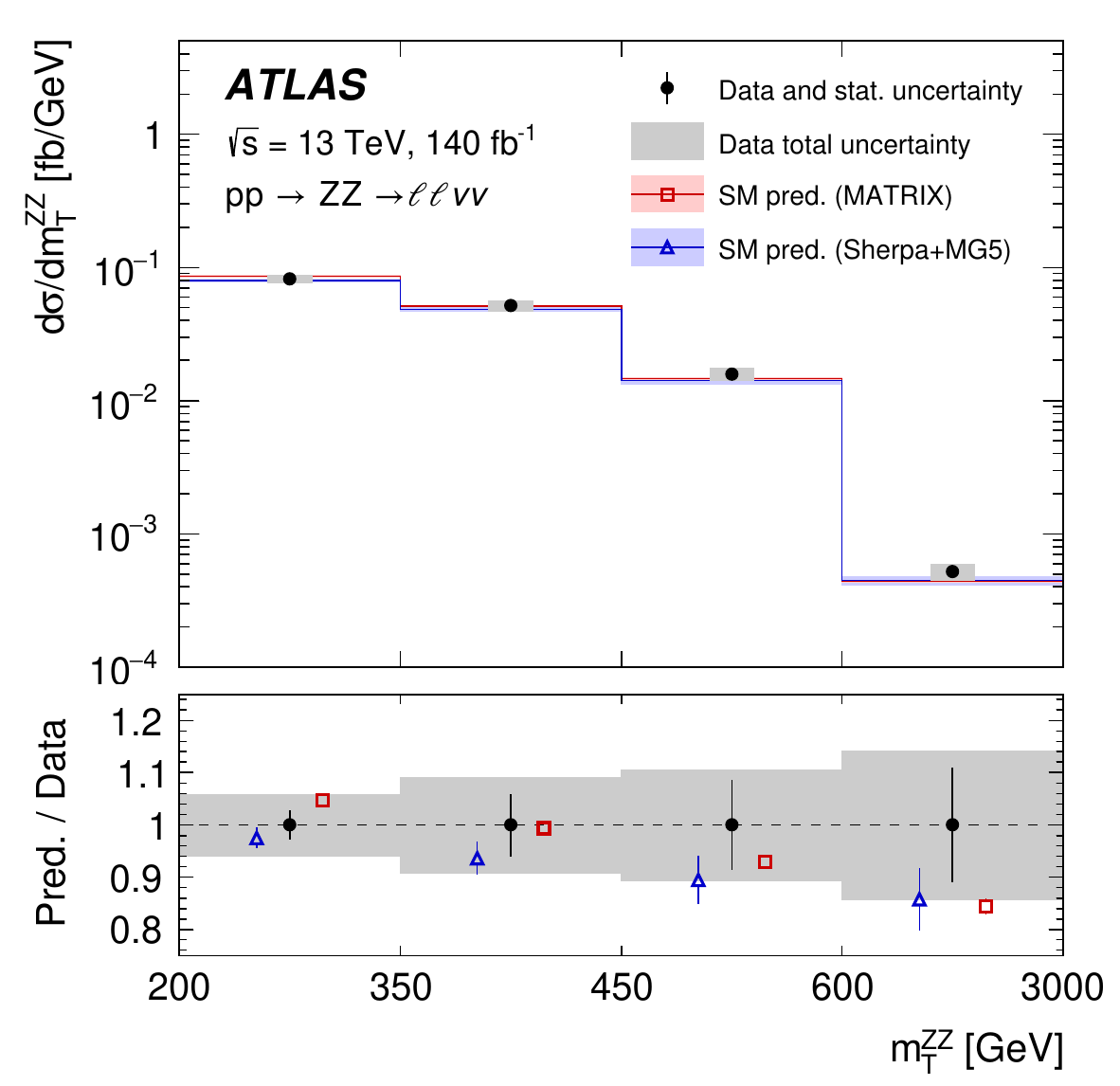}
}\\
\subfloat[]{
\includegraphics[width=0.32\linewidth]{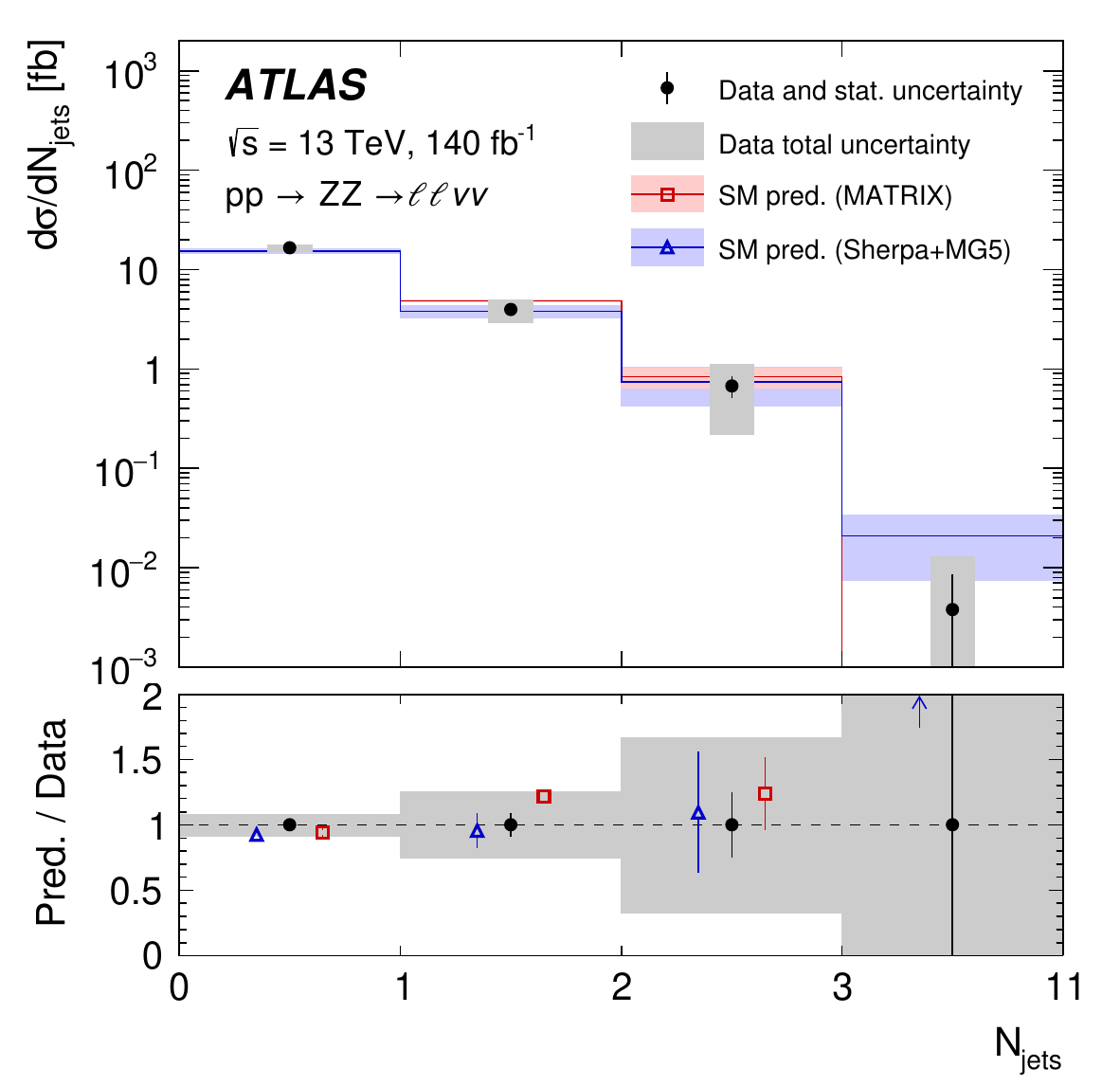}
}
\subfloat[]{
\includegraphics[width=0.32\linewidth]{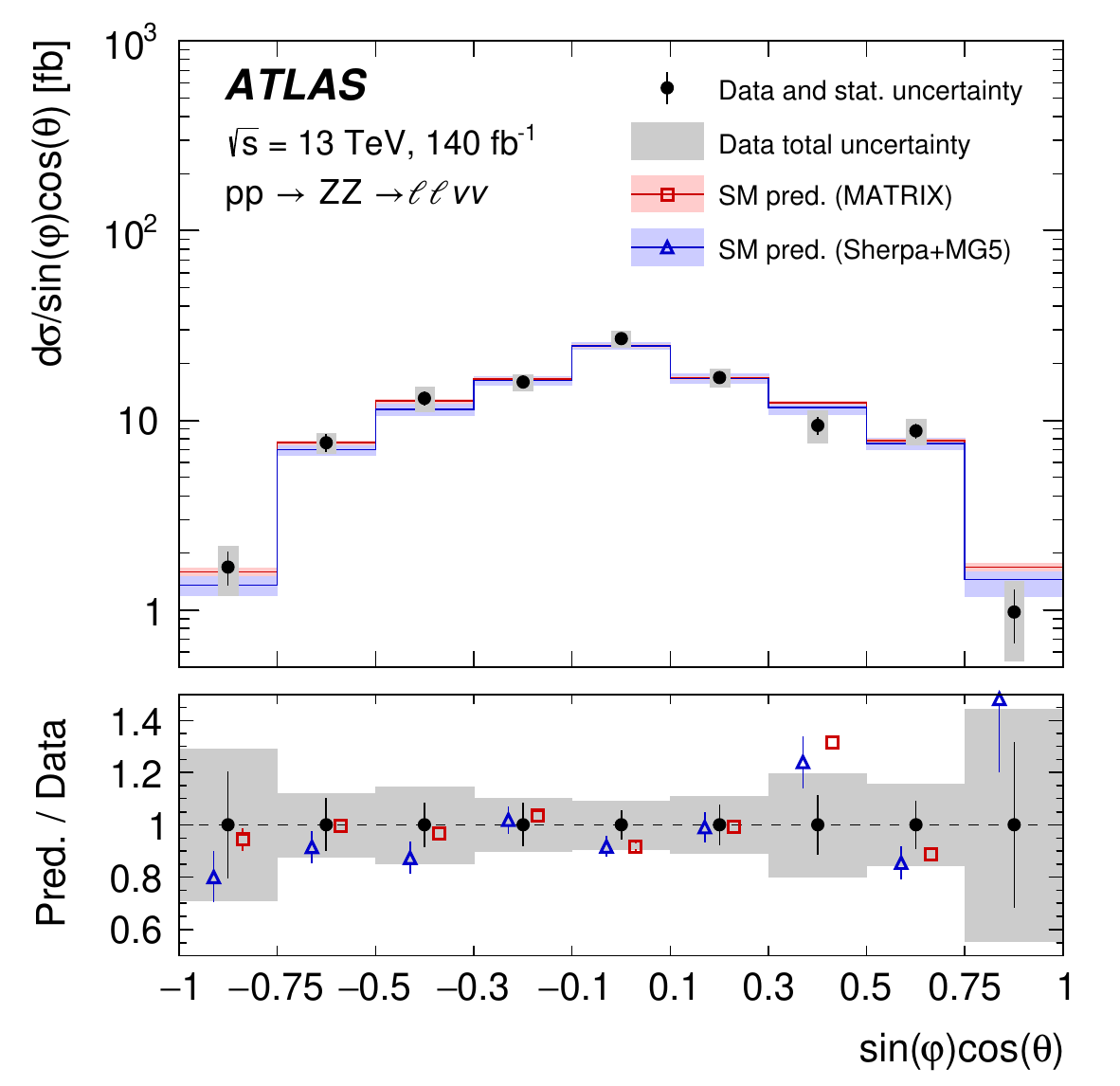}
}
\caption{Differential cross-section measurements in the inclusive $ZZ\rightarrow \ell\ell\nu\nu$ fiducial phase space, compared with MC predictions calculated with the nominal \textsc{Sherpa} simulation %
and with \textsc{Matrix} for a set of kinematic variables:
(a) $p_\text{T}^\text{leading lepton}~,
\text{(b)}~ p_\text{T}^\text{Z}~,
\text{(c)}~\Delta\phi(\ell, \ell),~
\text{(d)}~|y^Z|,~
\text{(e)}~ p_\text{T}^\text{ZZ},~
\text{(f)}~ m_\text{T}^\text{ZZ},~
\text{(g)}~ N_{\text{jets}}$,
and (h) a CP-sensitive angular variable, $\sin(\varphi)\cos(\theta)$.
The statistical and total uncertainties in the data points are displayed in both the differential cross-sections and the corresponding ratios. The total theory uncertainties in the \textsc{Sherpa} and \textsc{Matrix} predictions are represented as shaded bands in the cross-section panels and as error bars in the ratio panels.
}
\label{fig:diff-xs-ZZ}
\end{figure}

\begin{figure}[!hbtp]
\centering

\subfloat[]{
\includegraphics[width=0.32\linewidth]{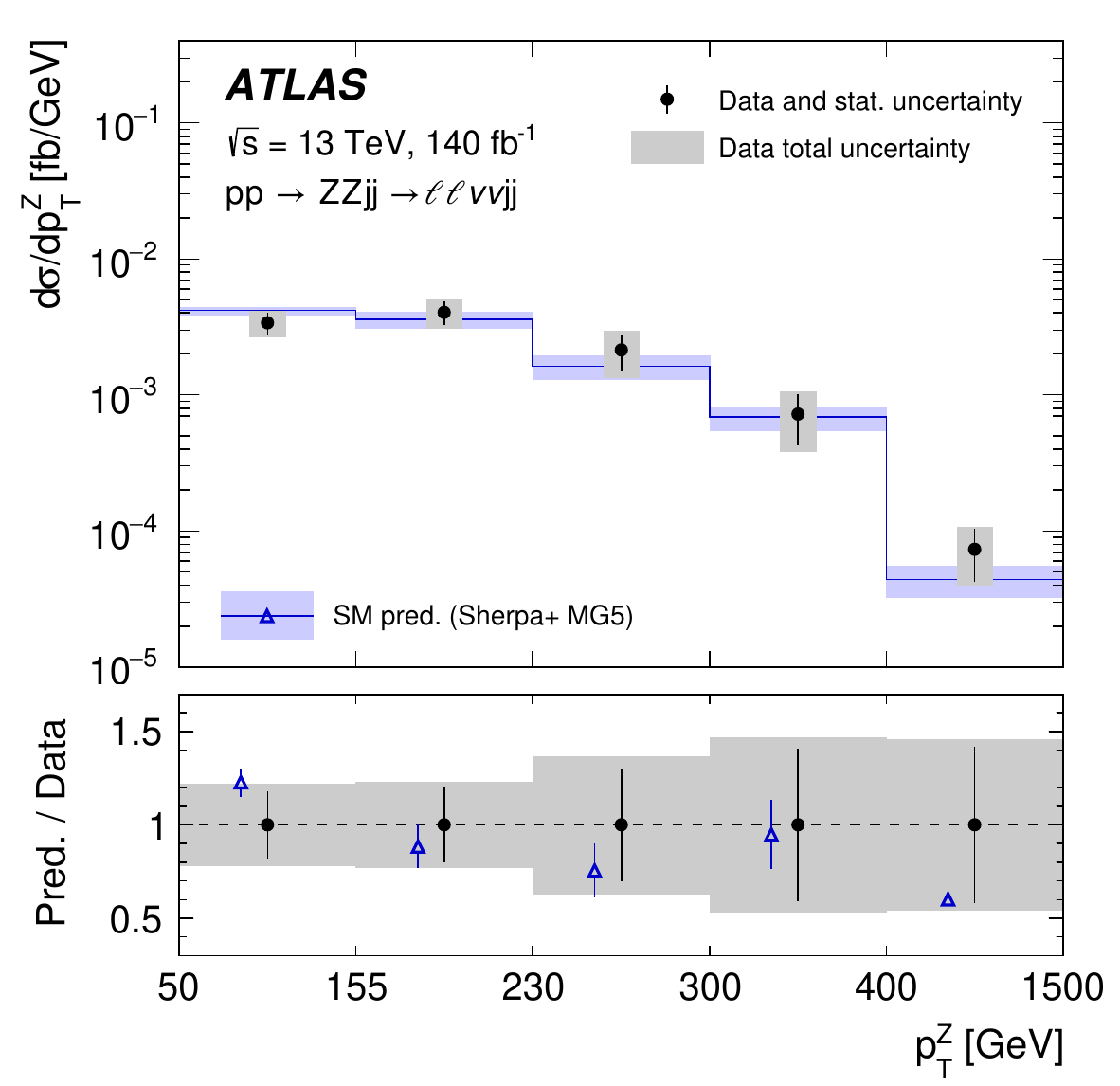}
}
\subfloat[]{
\includegraphics[width=0.32\linewidth]{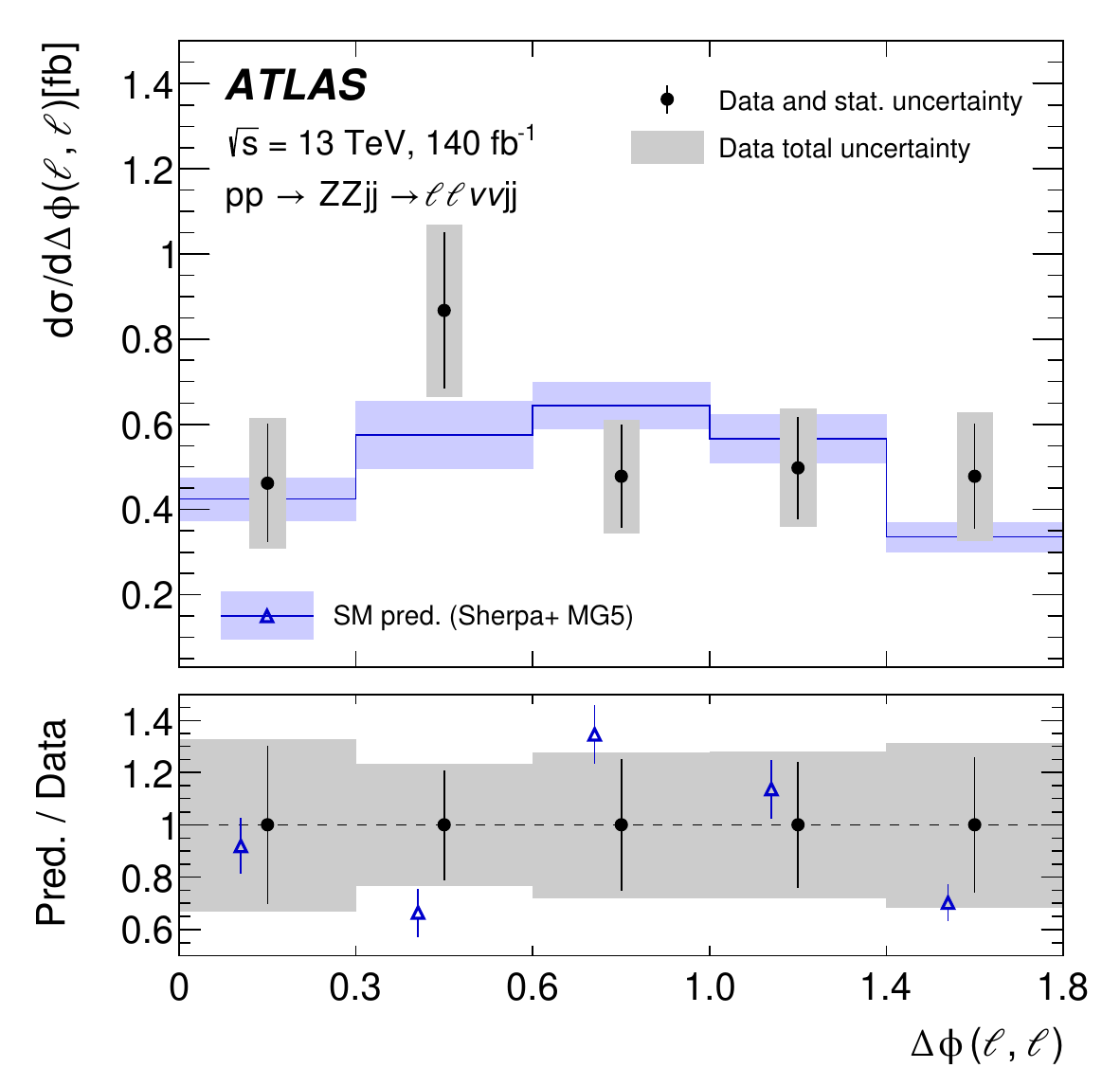}
}
\subfloat[]{
\includegraphics[width=0.32\linewidth]{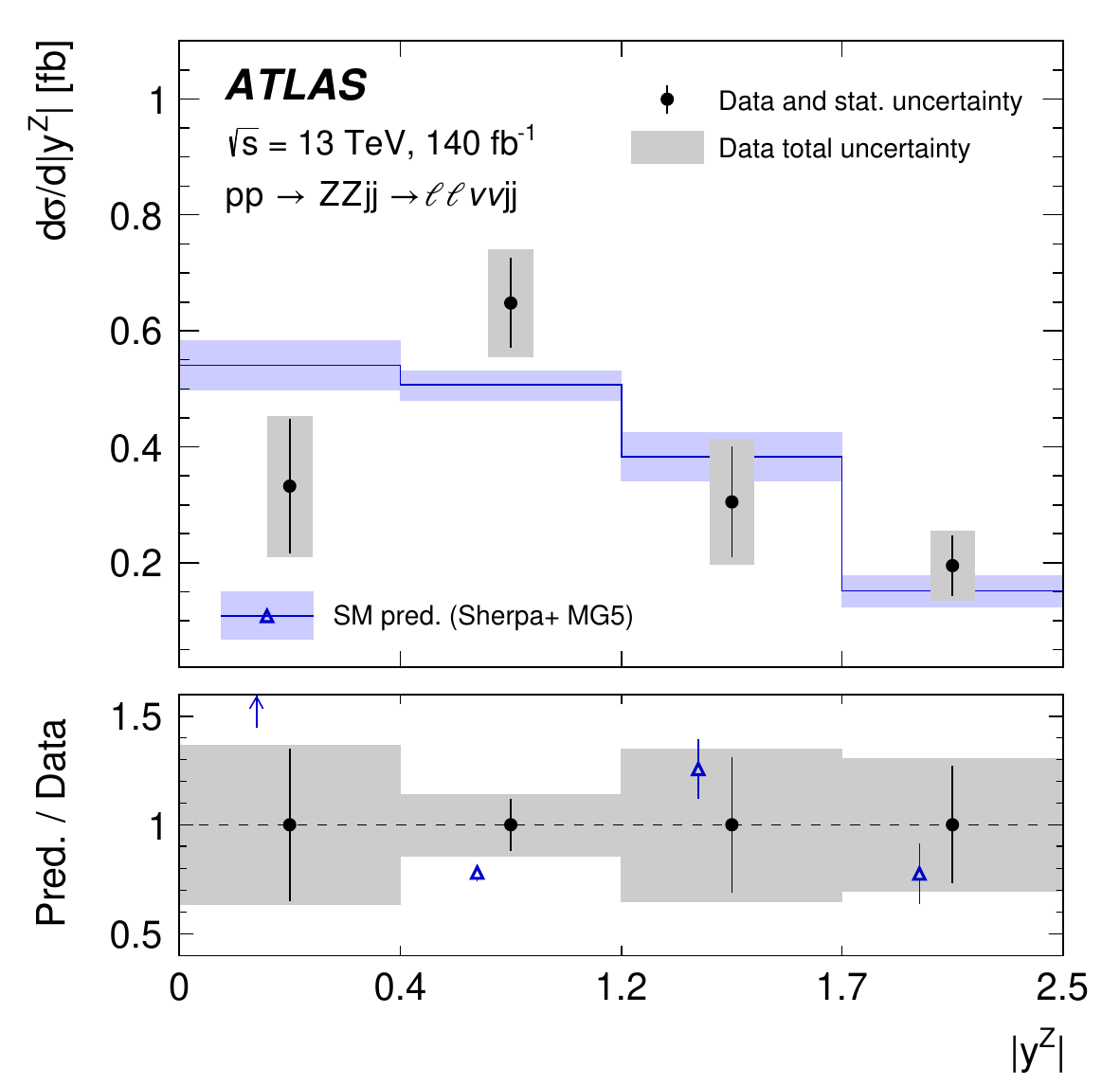}
}\\
\subfloat[]{
\includegraphics[width=0.35\linewidth]{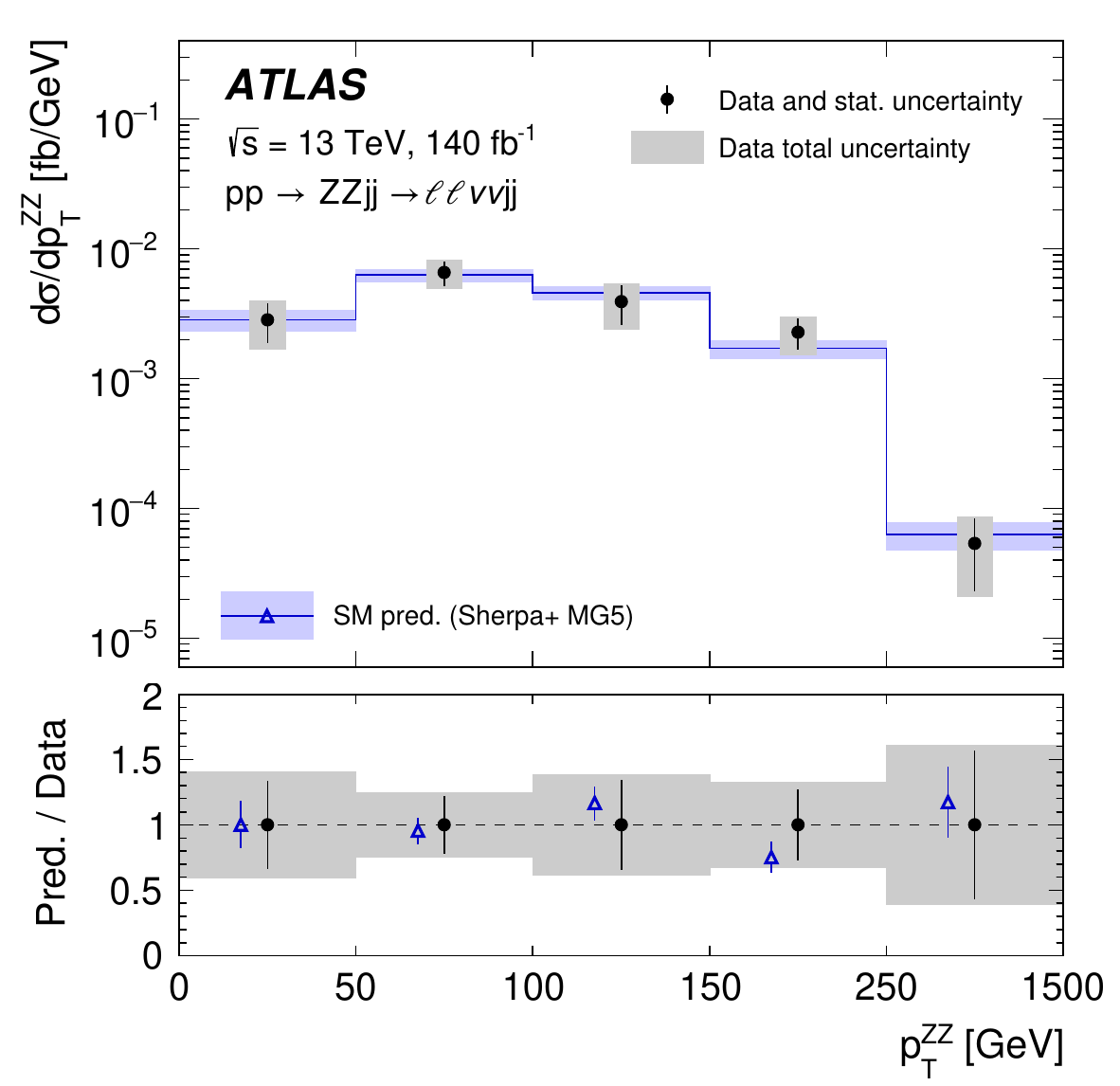}
}
\subfloat[]{
\includegraphics[width=0.35\linewidth]{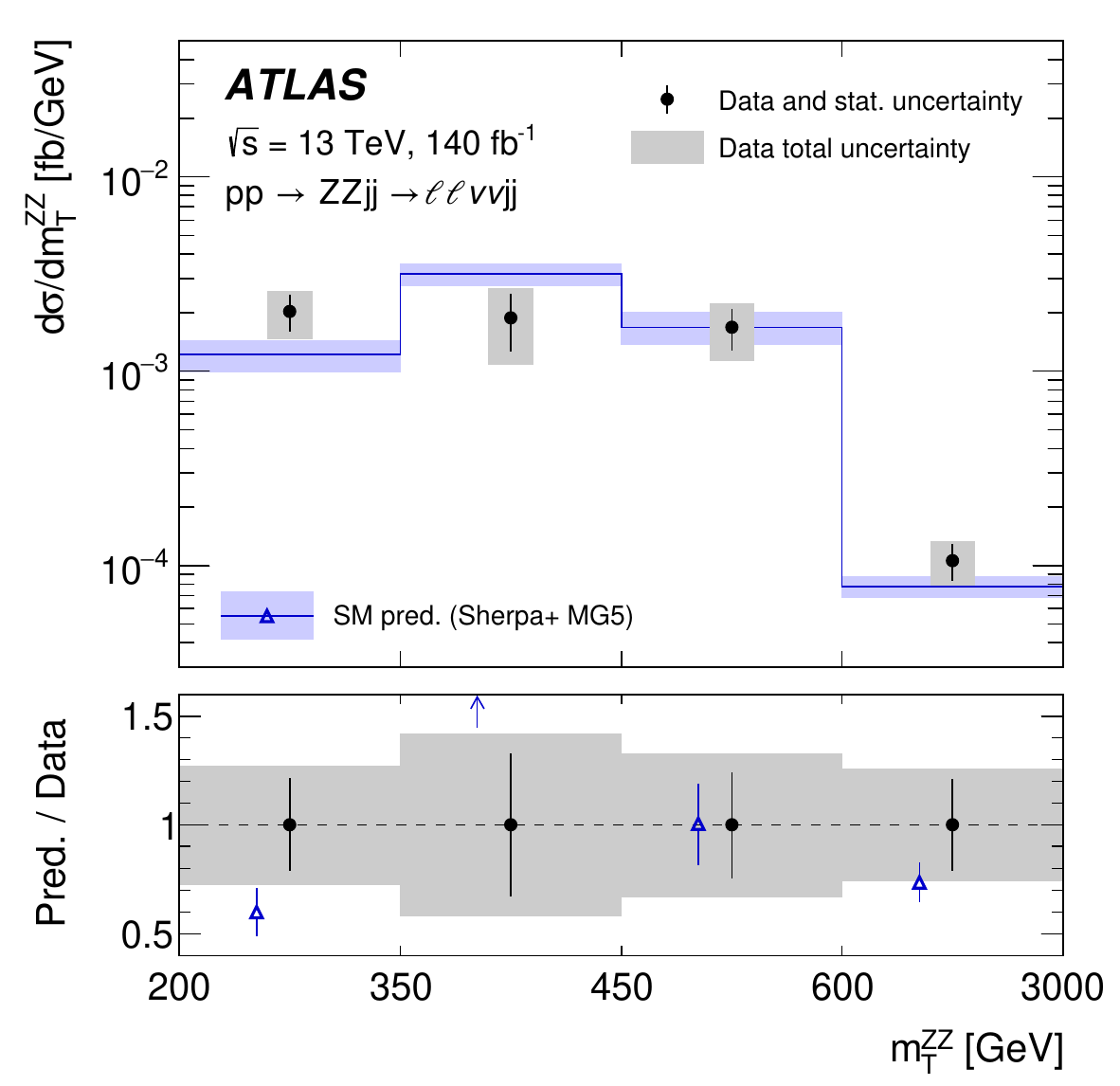}
}\\
\subfloat[]{
\includegraphics[width=0.35\linewidth]{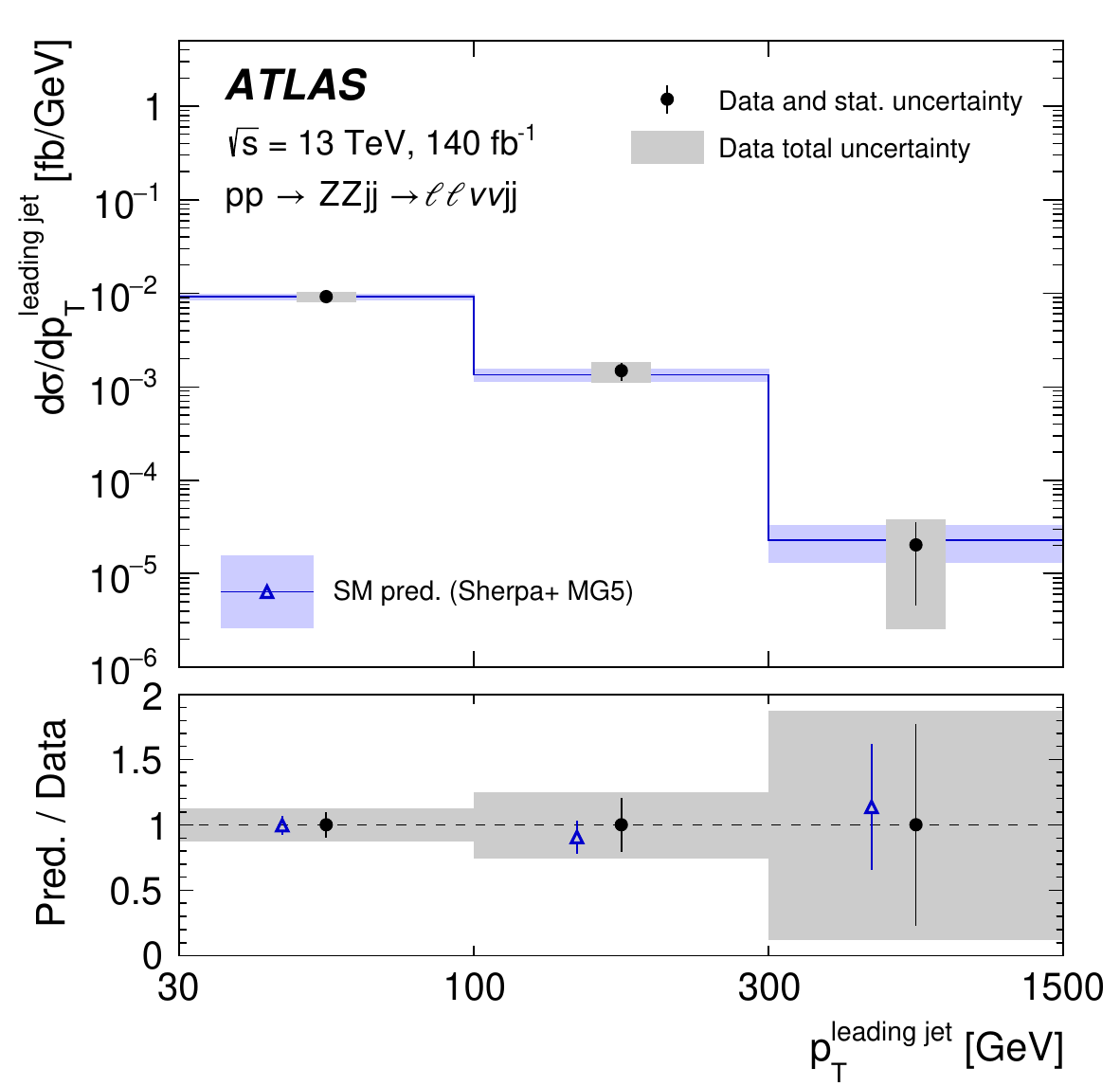}
}
\subfloat[]{
\includegraphics[width=0.35\linewidth]{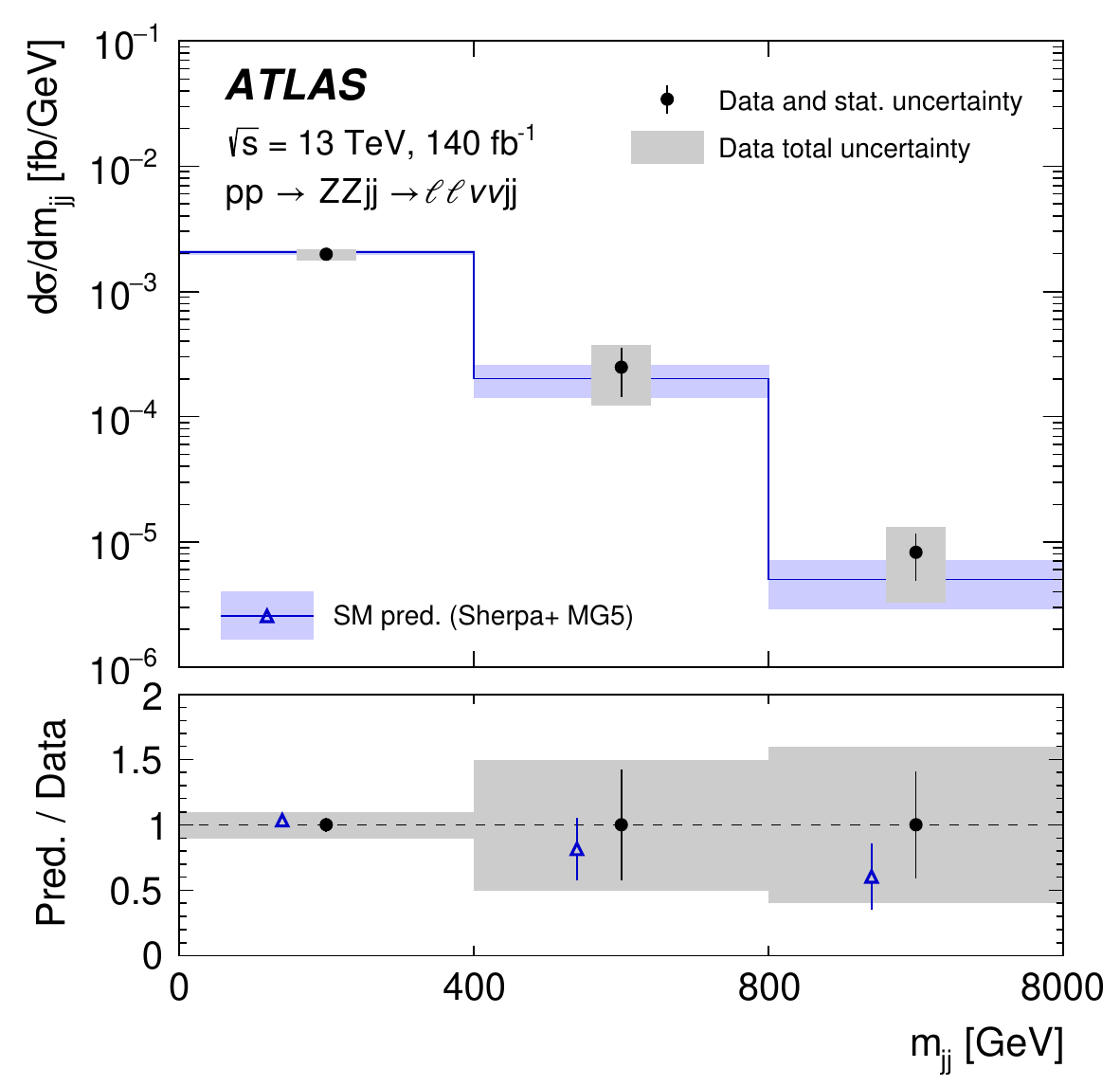}
}
\caption{Differential cross-section measurements in the $ZZjj \rightarrow \ell\ell\nu\nu jj$ fiducial phase space, compared with SM predictions for a set of kinematic variables:
$ \text{(a)}~p_\text{T}^\text{Z},~
\text{(b)}~\Delta\phi(\ell, \ell),~
\text{(c)}~|y^Z|,~
\text{(d)}~p_\text{T}^\text{ZZ},~
\text{(e)}~m_\text{T}^\text{ZZ},~
\text{(f)}~p_\text{T}^\text{leading jet},~
\text{and} ~\text{(g)}~m_\text{jj}.
$
The SM predictions are obtained using \textsc{Sherpa} for strong production and \textsc{MadGraph5} for electroweak production.
The statistical and total uncertainties in the data points are displayed in both the differential cross-sections and the corresponding ratios. The total theory uncertainties in the  MC predictions are represented as shaded bands in the cross-section panels and as error bars in the ratio panels.
}
\label{fig:diff-xs-ZZjj}
\end{figure}


\section{Probing anomalous gauge boson self-interaction couplings }
\label{sec:aTGC-aQGC}

The measured fiducial and total cross-sections are found to be in good agreement with the SM predictions within the quoted uncertainties.
The inclusive $ZZ$ and $ZZjj$ production processes in $pp$ collisions are sensitive to deviations from SM predictions, particularly in the self-interactions between gauge bosons.
Among all the contributing Feynman diagrams, shown in Figures~\ref{fig:feynmanZZinclusive} and \ref{fig:feynmanZZvbs}, those involving aTGCs, specifically the $ZZZ$ and $ZZ\gamma$ vertices and the triple gauge boson $WWZ$ and the quartic gauge boson $WWZZ$ couplings, are of particular interest. These diagrams are directly connected to two fundamental aspects of the SM electroweak theory, namely the non-Abelian gauge group structure and spontaneous electroweak symmetry breaking. The $ZZ$ production process serves as a key channel for probing aTGCs~\cite{Gounaris_sig} while the $ZZjj$ production, which involves both triple and quartic gauge-boson self-interactions, enables searches for both aTGCs and aQGCs~\cite{Eboli2006}.

Since no unified framework simultaneously describing aTGC and aQGC is currently available, the inclusive $ZZ$ measurement is used to constrain aTGC, while the $ZZjj$ process is used to search for aQGC.

Simulation samples  for aTGC and aQGC are produced using the decomposition method, which expresses the matrix element for a process as:
\begin{equation}
|A_{\text{SM}} + \sum_i c_i A_i|^2 = |A_{\text{SM}}|^2 + \sum_i c_i\, 2\, \text{Re}(A_{\text{SM}} A_i^*) + \sum_i c_i^2 |A_i|^2 + \sum_{i \ne j} c_i c_j\, 2\, \text{Re}(A_i A_j^*),
\end{equation}
where $A_{\text{SM}}$ is the SM amplitude, $A_i$ the BSM (i.e., anomalous) contributions, and $c_i$ the corresponding coupling coefficients.
Apart from the pure SM contribution in the first term, the second term captures the interference between the SM amplitude and each BSM contribution, referred to as the linear term, the third term is the pure contribution from each BSM amplitude, referred to as the quadratic term, while the final term describes the interference between different BSM amplitudes. For both the aTGC and aQGC interpretations, all BSM contributions in Eq.~(1) are taken into account, including the linear interference terms ($c_i$), the pure BSM quadratic terms ($c_i^2$), and the cross terms ($c_i c_j$).

The MC samples used for the interpretation of the aTGCs were generated at LO in QCD using \textsc{MadGraph5}  with the \NNPDF[3.0nnlo] PDF set, %
employing Universal FeynRules Output (UFO) models~\cite{UFO-Degrande2012}, which provide the
matrix‑element implementation of the effective operators used for the signal generation. Parton showering, hadronisation, and modelling of the underlying event were performed using \textsc{Pythia~8.2} with the A14 tune.

The MC predictions of the aQGCs in the final state $ZZjj \to \ell^+\ell^- \nu\bar\nu jj$ are generated using \textsc{MadGraph5} with the NNPDF3.0 LO ~\cite{Ball:2014uwa} PDF set. The events were interfaced with \textsc{Pythia~8.2} %
with the A14 tune for modelling the parton shower, hadronisation, and underlying events.

Limits are derived under the linear+quadratic assumption, in which both the interference between the SM and dimension-eight SMEFT operators, and the quadratic contributions of those operators, are included in the cross-section. For both the aTGC and aQGC interpretations, detector-level distributions are used as inputs to the fit, as they allow optimised binning in the most sensitive regions while avoiding uncertainties associated with unfolding.

\subsection{Constraints on anomalous neutral triple gauge couplings}
\label{sec:antgcs}

As discussed in Section~\ref{sec:intro}, aTGCs are absent at tree level in the SM. Consequently, any measurable deviation in the kinematic tails of \(ZZ\) production could be an evidence for aTGC. This subsection presents the corresponding limits, probing both CP-even and CP-odd couplings within two theoretical frameworks: a VF parameterisation~\cite{Baur2000,Biekotter2021} and the SMEFT~\cite{Degrande2014,Ellis2023}.

The most general anomalous $ZZV$ (where $V=Z,\gamma$) VF for on-shell $Z$ bosons is given by~\cite{Hagiwara:1986vm}:
\begin{equation*}
g_{ZZV}\Gamma_{ZZV}^{\alpha\beta\mu}=e\frac{P^2-m_V^2}{m_Z^2}\left[if_4^V(P^{\alpha}g^{\mu\beta}+P^{\beta}g^{\mu\alpha})+if_5^V\epsilon^{\mu\alpha\beta\rho}(q_1-q_2)_\rho\right],
\end{equation*}
where $g_{ZZV}$ represents the $ZZV$ coupling, $\Gamma_{ZZV}^{\alpha\beta\mu}$ is the $ZZV$ vertex function, $e$ is the proton charge, $P$, $q_1$ and $q_2$ indicate the momenta of the $V$ and two on-shell $Z$ bosons, respectively. The anomalous triple gauge coupling parameters are $f_4^V$, which are CP-violating, and $f_5^V$, which are CP-conserving.

In this analysis, both anomalous triple and quartic gauge couplings are modelled via dimension-eight SMEFT operators; hence, the same extended Lagrangian is used for both cases. The SMEFT Lagrangian is extended as
\begin{equation}
\mathcal{L}_\text{SMEFT} = \mathcal{L}_\text{SM} + \sum_i \frac{C^{(8)}_i}{\Lambda^4} \mathcal{O}_i^{(8)},
\label{eqn:eft}
\end{equation}

where \(\mathcal{O}_i\) are dimension-eight operators and \(C_i\) are their Wilson coefficients. In this analysis, the SMEFT formulation incorporates the dimension-eight operators \(\mathcal{O}_{G+}, \mathcal{O}_{G-}, \mathcal{O}_{\widetilde{B}W}, \mathcal{O}_{BB}, \mathcal{O}_{BW}\), and \(\mathcal{O}_{WW}\), each entering the Lagrangian with the corresponding Wilson coefficients \(C_{G+}/\Lambda^4\), \(C_{G-}/\Lambda^4\), \(C_{\widetilde{B}W}/\Lambda^4\), \(C_{BB}/\Lambda^4\), \(C_{BW}/\Lambda^4\), and \(C_{WW}/\Lambda^4\)~\cite{Degrande2014,Ellis2023}. The energy scale \(\Lambda\) denotes the mass scale of heavy new physics that is integrated out and is set to \(\Lambda = 1\)~\TeV throughout this analysis.

The sensitivity to both CP-even and CP-odd aTGCs arises from an enhancement in the high-energy tail of the $ZZ$ production spectrum. Such a tail contains a low number of events and, therefore, the reconstructed variable distribution is more effective compared to the unfolded one. The distribution of the transverse momentum of the dilepton system, $p_\text{T}^{Z}$, is identified as the most sensitive observable and is used to extract limits on %
aTGCs. A binned likelihood fit is performed on the $p_\text{T}^{Z}$  differential distribution shown in Figure~\ref{fig:atgc-ptll-and-cp}, using the same phase space as defined for the $ZZ$ cross‑section measurement, to extract 95\% confidence level intervals for the anomalous parameters. Systematic uncertainties are incorporated as nuisance parameters in the fit, and the results are shown in Table~\ref{tab:limits}, where limits are derived on one coupling coefficient given all other anomalous couplings set to zero. Using the same distribution, limits on two parameters at a time are provided and presented in Figures~\ref{fig:2d-f-limit} and~\ref{fig:atgc_2D_EFT}.

Compared to other results, the limits shown in Table~\ref{tab:limits} on the VF parameters are better by about a factor of 1.7  with respect to those from the previous ATLAS $pp \to ZZ \to \ell\ell\nu\nu$ analysis based on a subset (36.1~fb$^{-1}$) of the dataset~\cite{STDM-2017-03}.

\begin{figure}[!htbp]
\centering
\includegraphics[width=0.45\textwidth]{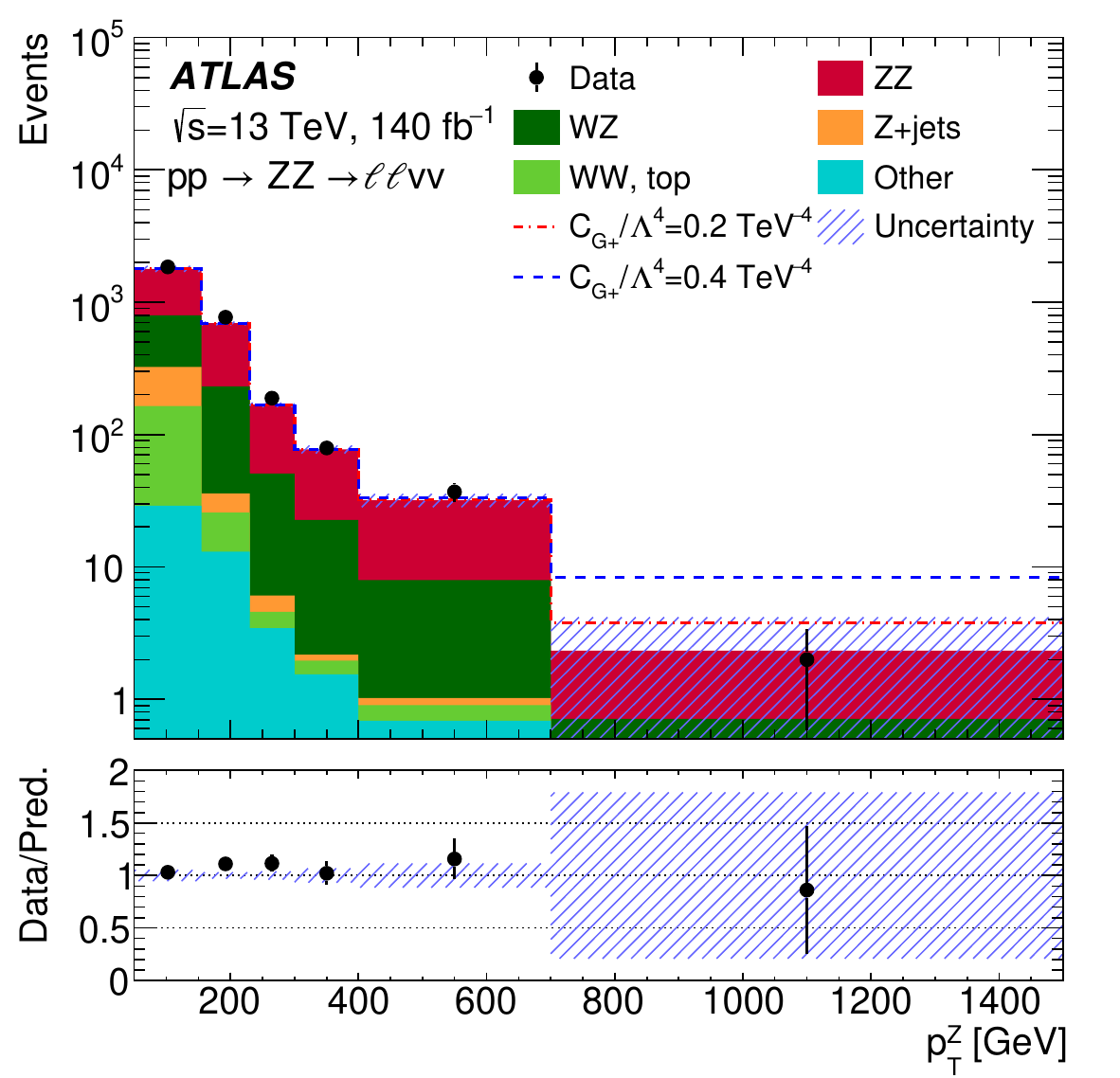}
\caption{Distribution of the dilepton transverse momentum, $p_\mathrm{T}^{Z}$, at detector level. The dashed band represents the total uncertainty in the pre-fit SM prediction, with no normalization factors applied from the signal extraction fit. The expected effects of an aTGC induced by the operator $O_{G+}$, for two different values of the Wilson coefficient, are also shown.}
\label{fig:atgc-ptll-and-cp}
\end{figure}

\begin{table}[htbp]
\caption{Expected and observed 95\% CL intervals on the parameters of the VF and the SMEFT formalisms in the inclusive ZZ channel, obtained without applying any unitarization procedure. %
The results are derived by setting all other aTGC parameters to zero.}
\centering
\scriptsize
\begin{tabular}{lcc}
\hline
\textbf{Coefficient} &\textbf{Expected} & \textbf{Observed} \\
\hline
$f_{4}^{\gamma}$ & $[-7.7,\ 7.7] \times 10^{-4}$ & $[-6.4,\ 6.4] \times 10^{-4}$ \\
$f_{4}^{Z}$       & $[-6.8,\ 6.8] \times 10^{-4}$ & $[-5.7,\ 5.7] \times 10^{-4}$ \\
$f_{5}^{\gamma}$ & $[-7.8,\ 7.7] \times 10^{-4}$ & $[-6.4,\ 6.5] \times 10^{-4}$ \\
$f_{5}^{Z}$       & $[-6.9,\ 6.9] \times 10^{-4}$ & $[-5.8,\ 5.8] \times 10^{-4}$ \\
\hline
$C_{G+}/\Lambda^4$ $[\mathrm{TeV}^{-4}]$       & $[-0.36,\ 0.36]$ & $[-0.29,\ 0.30]$ \\
$C_{G-}/\Lambda^4$ $[\mathrm{TeV}^{-4}]$      & $[-1.1,\ 1.2]$ & $[-0.95,\ 0.95]$ \\
$C_{\tilde{B}W}/\Lambda^4$ $[\mathrm{TeV}^{-4}]$ & $[-1.3,\ 1.3]$ & $[-1.1,\ 1.1]$ \\
$C_{BB}/\Lambda^4$ $[\mathrm{TeV}^{-4}]$       & $[-1.6,\ 1.6]$ & $[-1.3,\ 1.3]$ \\
$C_{BW}/\Lambda^4$ $[\mathrm{TeV}^{-4}]$     & $[-2.9,\ 2.9]$ & $[-2.4,\ 2.4]$ \\
$C_{WW}/\Lambda^4$ $[\mathrm{TeV}^{-4}]$    & $[-2.4,\ 2.4]$ & $[-2.0,\ 2.0]$ \\
\hline
\end{tabular}
\label{tab:limits}
\end{table}

\begin{figure}[htb]
\centering
\subfloat[]{
\includegraphics[width=.40\textwidth]{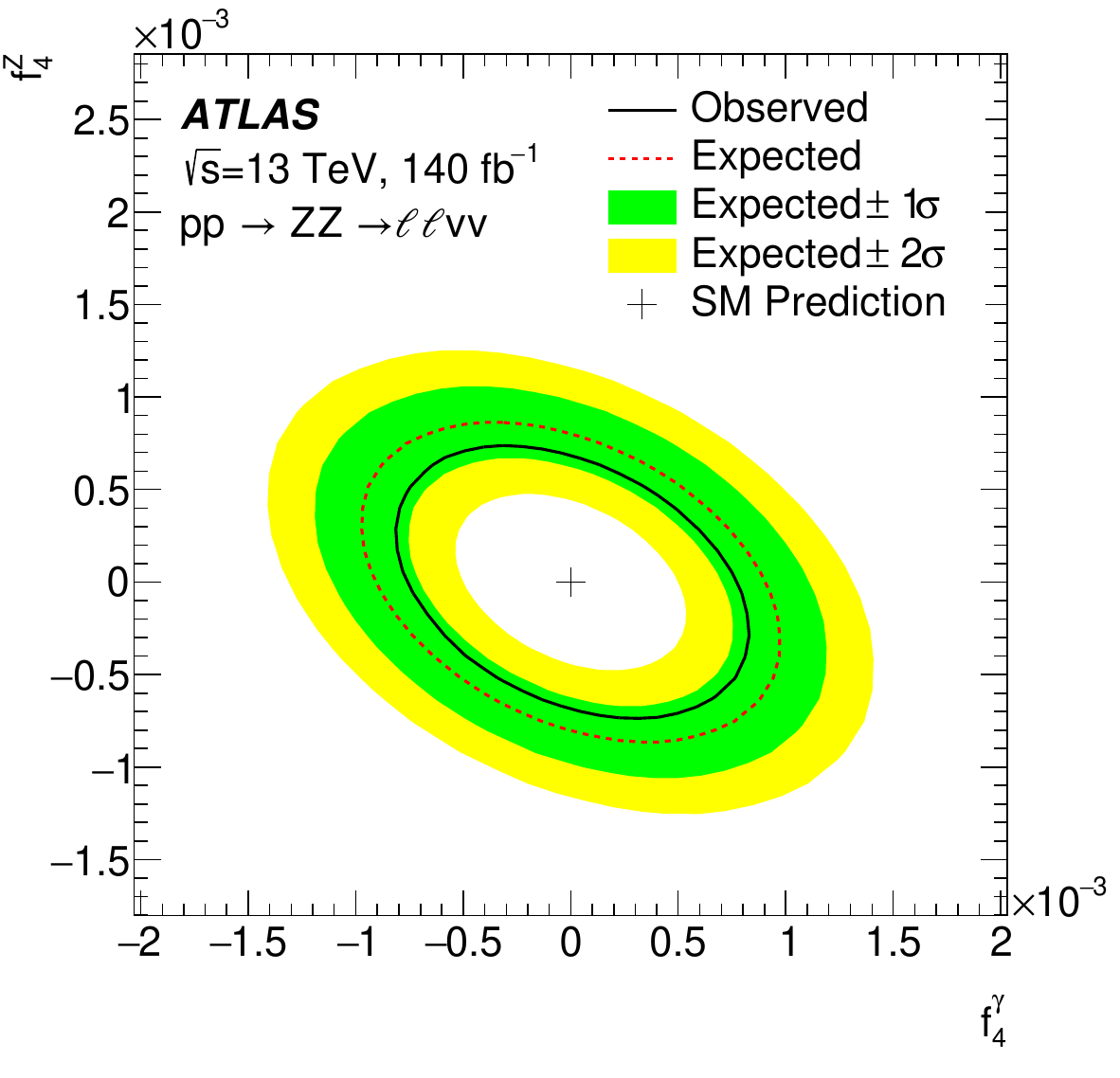}
}
\subfloat[]{
\includegraphics[width=.40\textwidth]{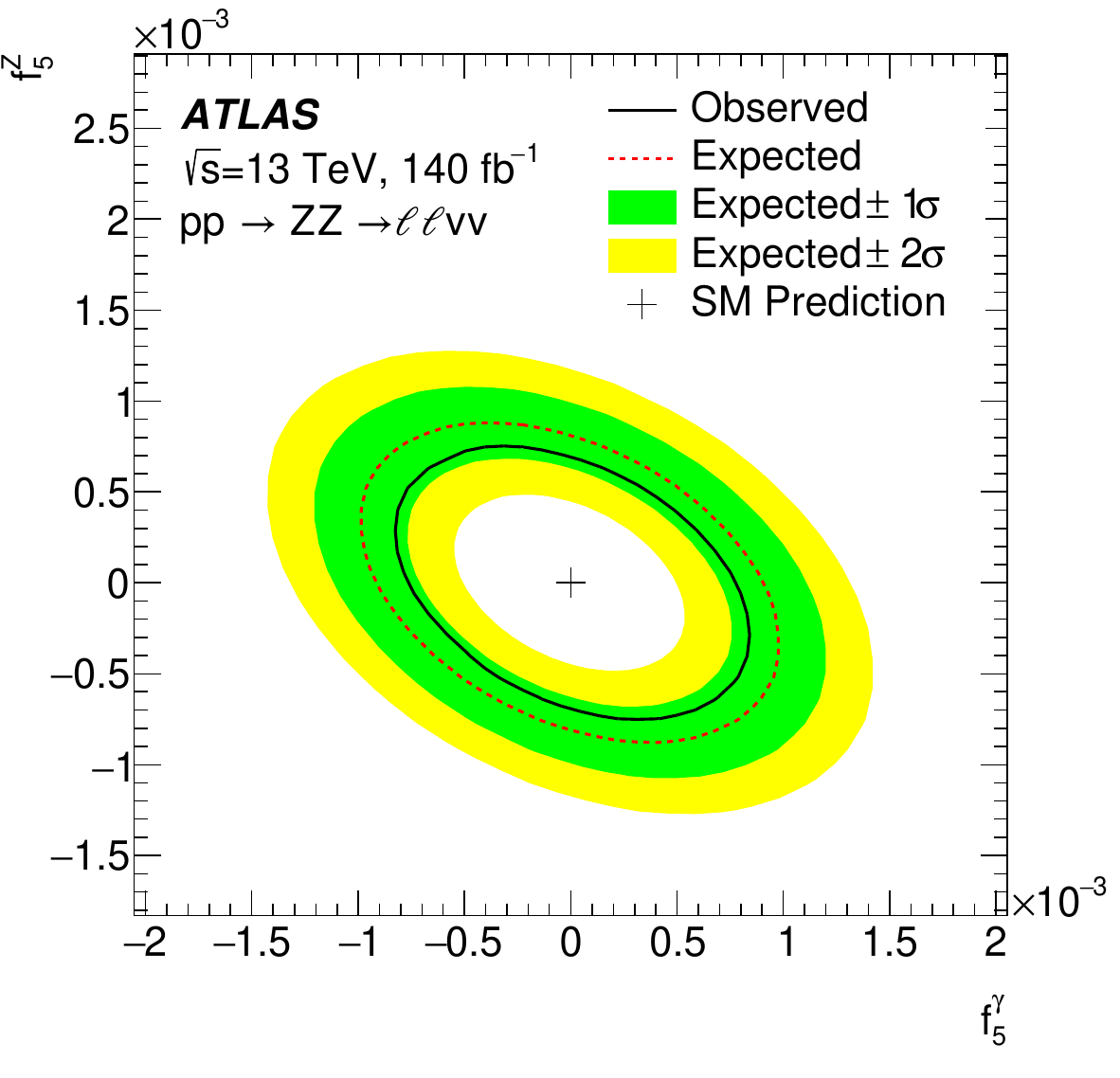}
}
\caption{
Two-dimensional 95\% confidence level limit contours on the aTGC parameters, (a) $f_4^\gamma $ vs. $f_4^Z$ and (b) $f_5^\gamma $ vs. $f_5^Z$ of the vertex function formalism. The inner (outer) bands show the expected $\pm 1\sigma$ ($\pm 2\sigma$) regions around the expected limit contours, corresponding to the range of fluctuations from statistical and systematic uncertainties under the Standard Model hypothesis. The results were obtained without applying any unitarization procedure and by setting the other parameters to zero.
}
\label{fig:2d-f-limit}
\end{figure}

\begin{figure} [!htbp]
\begin{center}
\subfloat[]{
\includegraphics[width=0.38\textwidth]{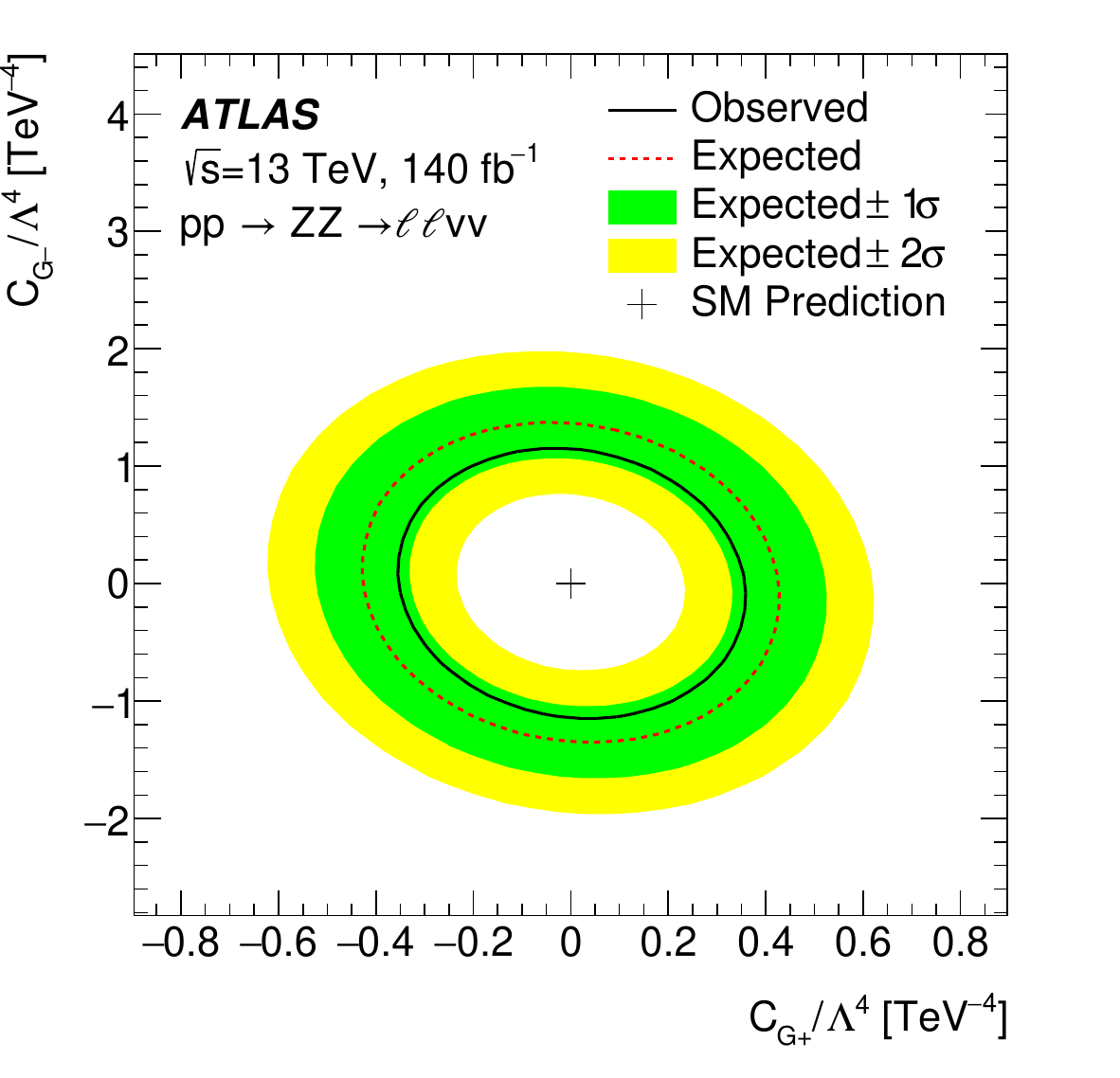}
}
\subfloat[]{
\includegraphics[width=0.38\textwidth]{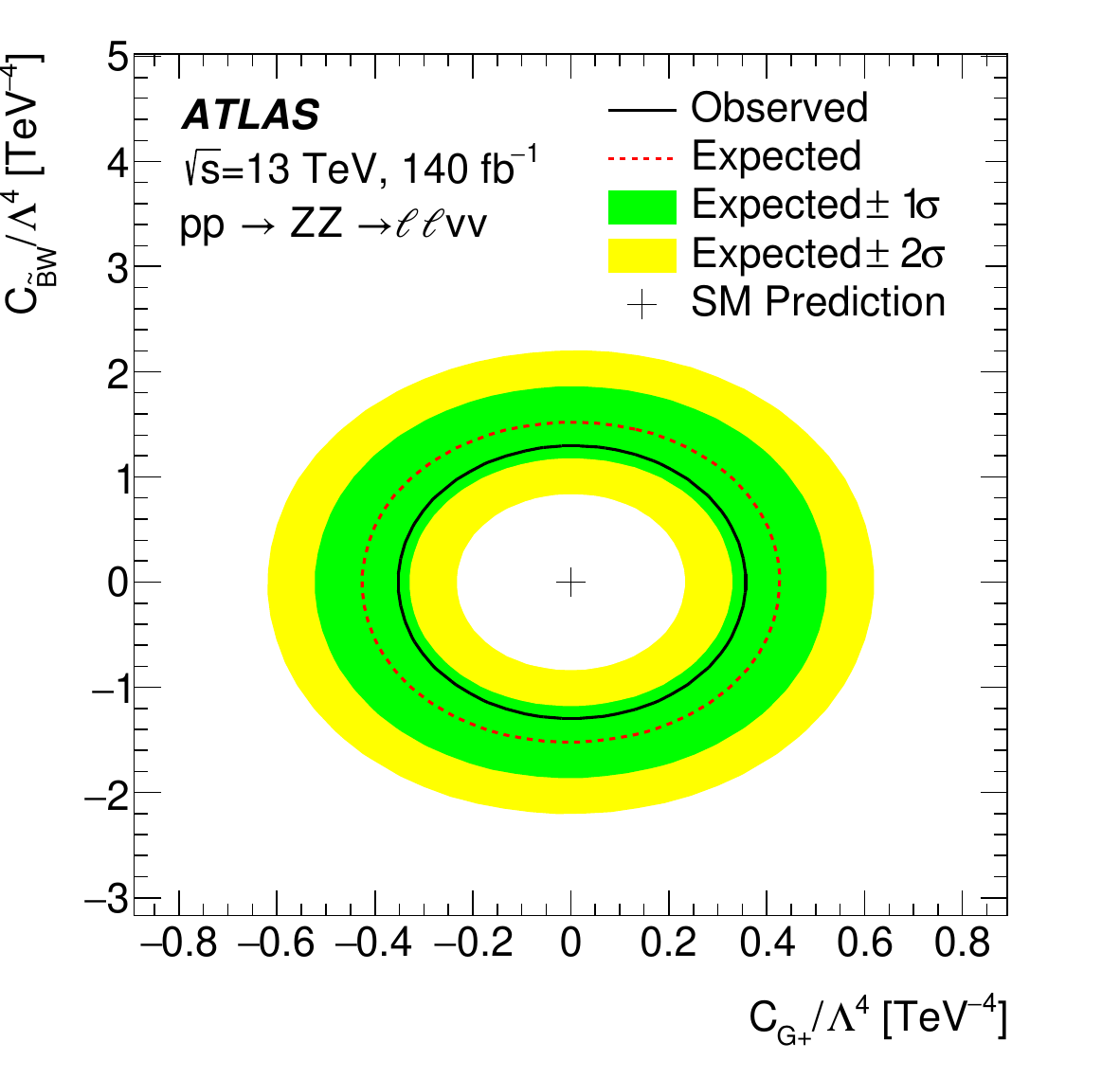}
}\\
\subfloat[]{
\includegraphics[width=0.38\textwidth]{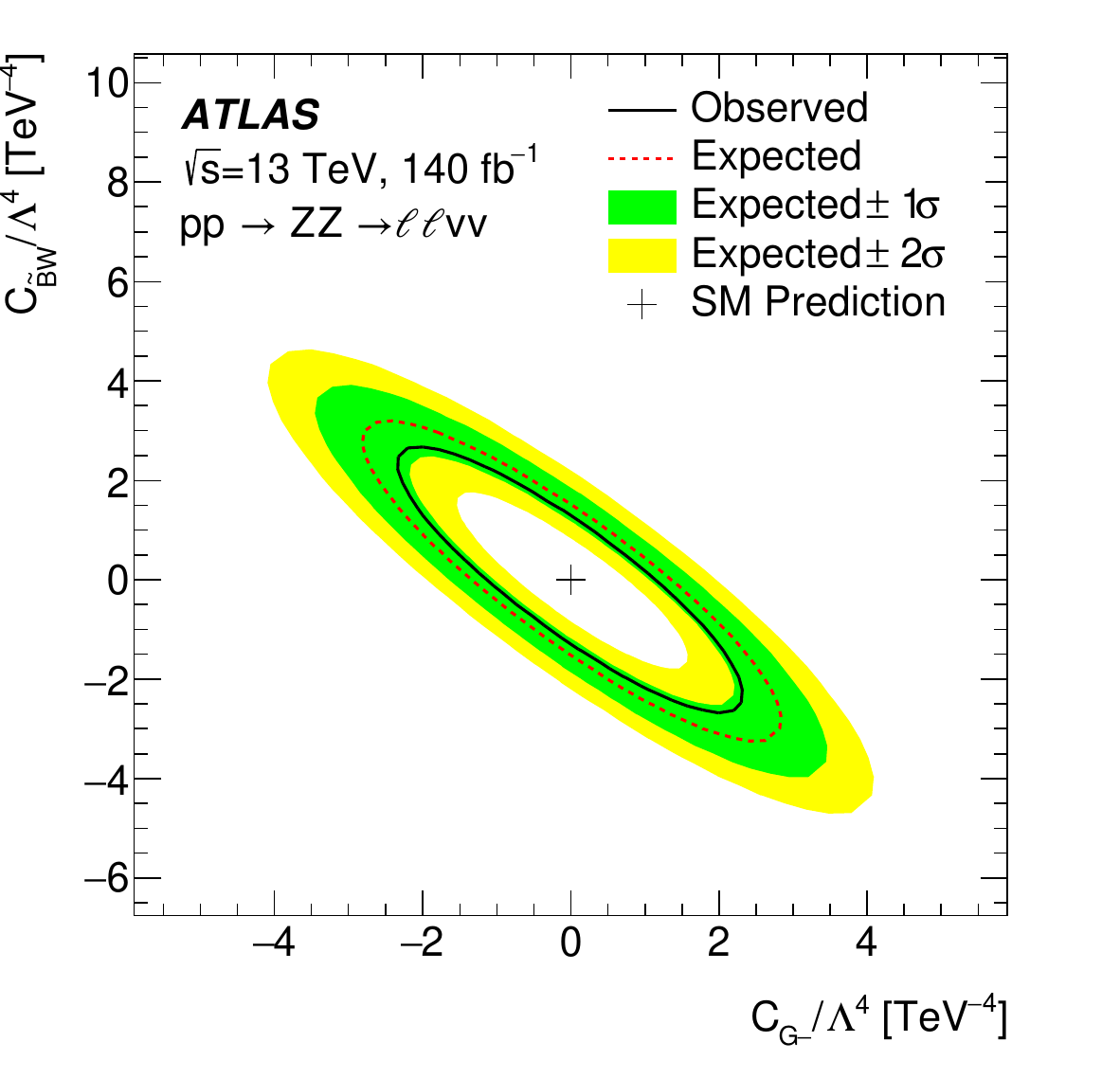}
}
\subfloat[]{
\includegraphics[width=0.38\textwidth]{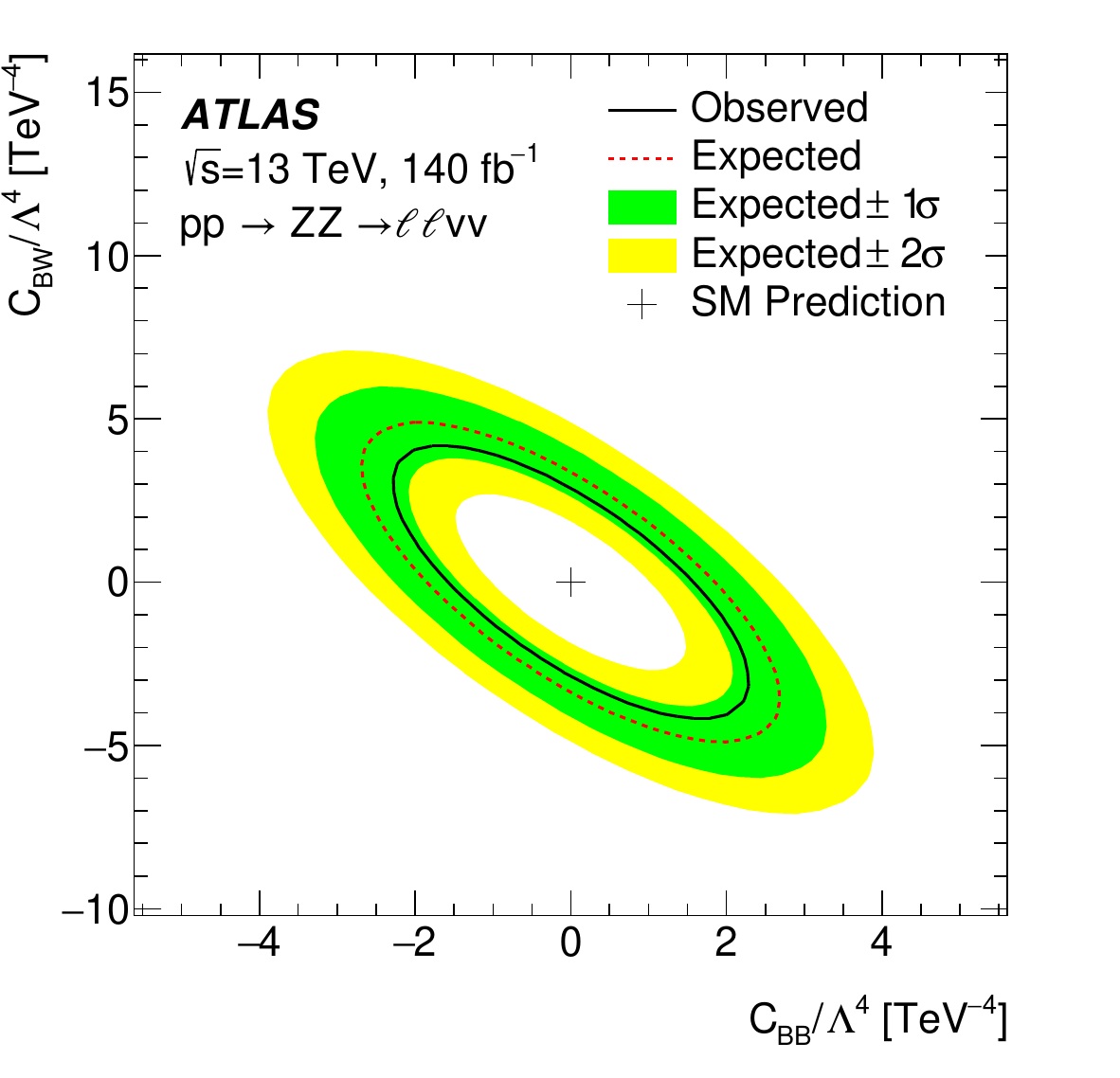}
}\\
\subfloat[]{
\includegraphics[width=0.38\textwidth]{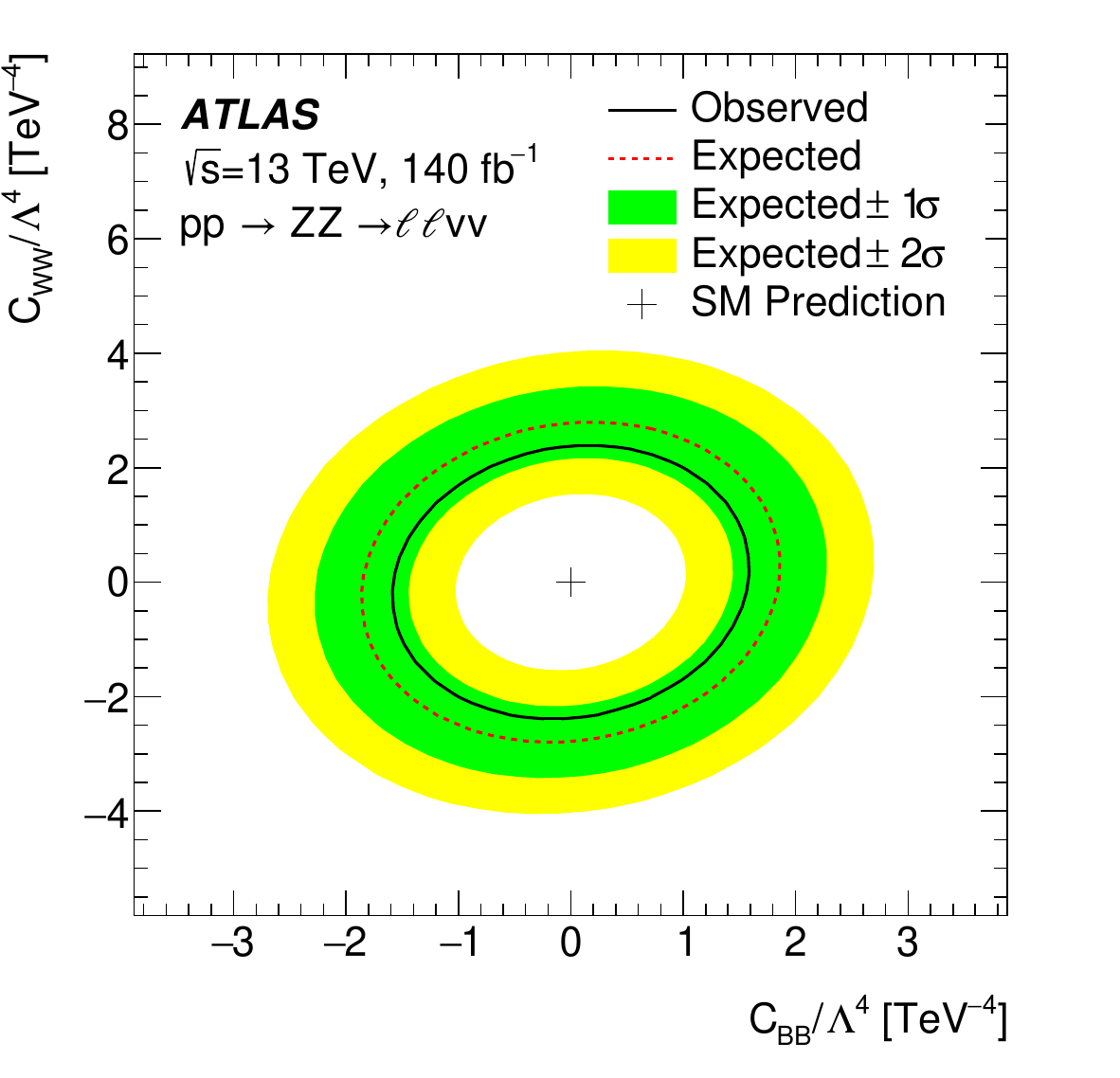}
}
\subfloat[]{
\includegraphics[width=0.38\textwidth]{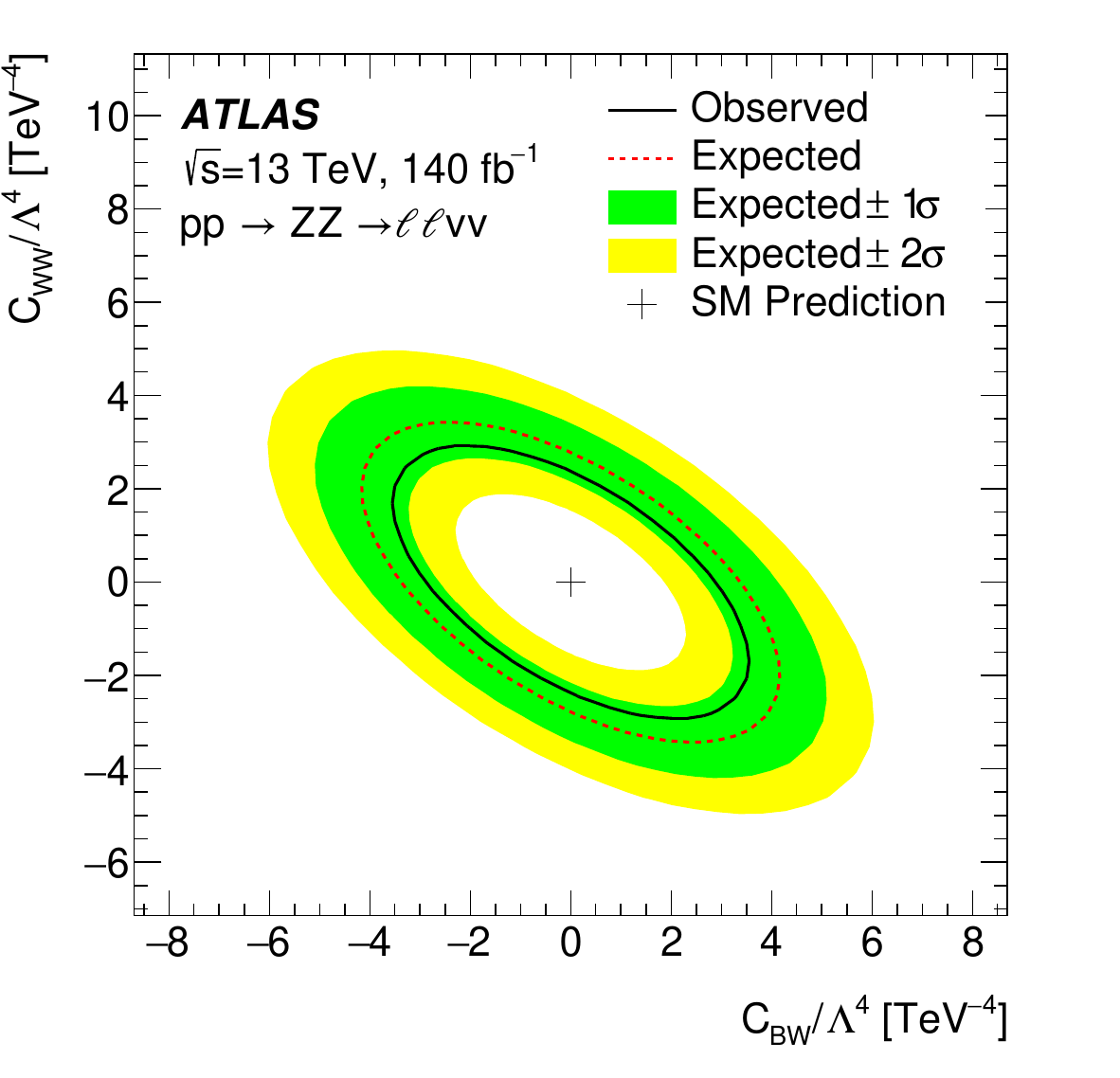}
}
\caption{Two-dimensional 95\% confidence level limit contours on aTGC Wilson coefficients with the EFT formalism. The inner (outer) bands show the expected $\pm 1\sigma$ ($\pm 2\sigma$) regions around the expected limit contours, corresponding to the range of fluctuations from statistical and systematic uncertainties under the Standard Model hypothesis. The results were obtained without applying any unitarization procedure and by setting the other Wilson coefficients to zero.}
\label{fig:atgc_2D_EFT}
\end{center}
\end{figure}

\FloatBarrier


\subsection{Constraints on anomalous quartic gauge couplings}

The measurement of $ZZjj$ production is particularly sensitive to VBS processes, which offer a unique %
probe of quartic electroweak gauge boson interactions. In the context of the SMEFT framework, such effects are parameterised by extending the SM Lagrangian with higher-dimensional operators. Dimension-8 operators are considered here, as they directly contribute to quartic gauge couplings, which do not arise at lower dimensions~\cite{Eboli2016}.

As in Eq.~\eqref{eqn:eft}, the SMEFT Lagrangian includes dimension-eight operators \(\mathcal{O}_{T,i}\), each associated to a Wilson coefficient \(f_{\text{T,i}}/ \Lambda^4\), where \(i = 0, 1, \ldots, 9\),
to distinguish them from the \( C_i \) coefficients used for aTGCs in Section ~\ref{sec:antgcs}.

The reconstructed transverse mass of the $ZZ$ system, $m_{\text{T}}^{ZZ}$, seen in Figure~\ref{fig:zzjj-mTZZ}, is identified as the most sensitive observable and is used to extract limits on these dimension-8 SMEFT operators. A binned likelihood fit is performed on the $m_{\text{T}}^{ZZ}$ differential distribution to derive 95\% CL intervals for the Wilson coefficients. Systematic uncertainties are included as nuisance parameters in the fit, and the resulting limits are summarised in Table~\ref{tab:EFT_limits}, where limits are derived by setting all Wilson coefficients to zero but the one tested.
To ensure the validity of the SMEFT expansion, unitarity-preserving cut-offs are applied to the BSM contributions by restricting \(m_{ZZ} < E_c\), where \(E_c\) is the energy cut-off scale. All coefficients are found to be consistent with zero. The most stringent constraints are obtained for $f_{\text{T0}}$ and $f_{\text{T1}}$. The present limits are better by more than a factor of two relative to previous ATLAS results obtained with $ZZ \to 4\ell$ events~\cite{STDM-2020-02}.

\begin{figure}[!hbtp]
\centering
\includegraphics[width=0.5\textwidth]{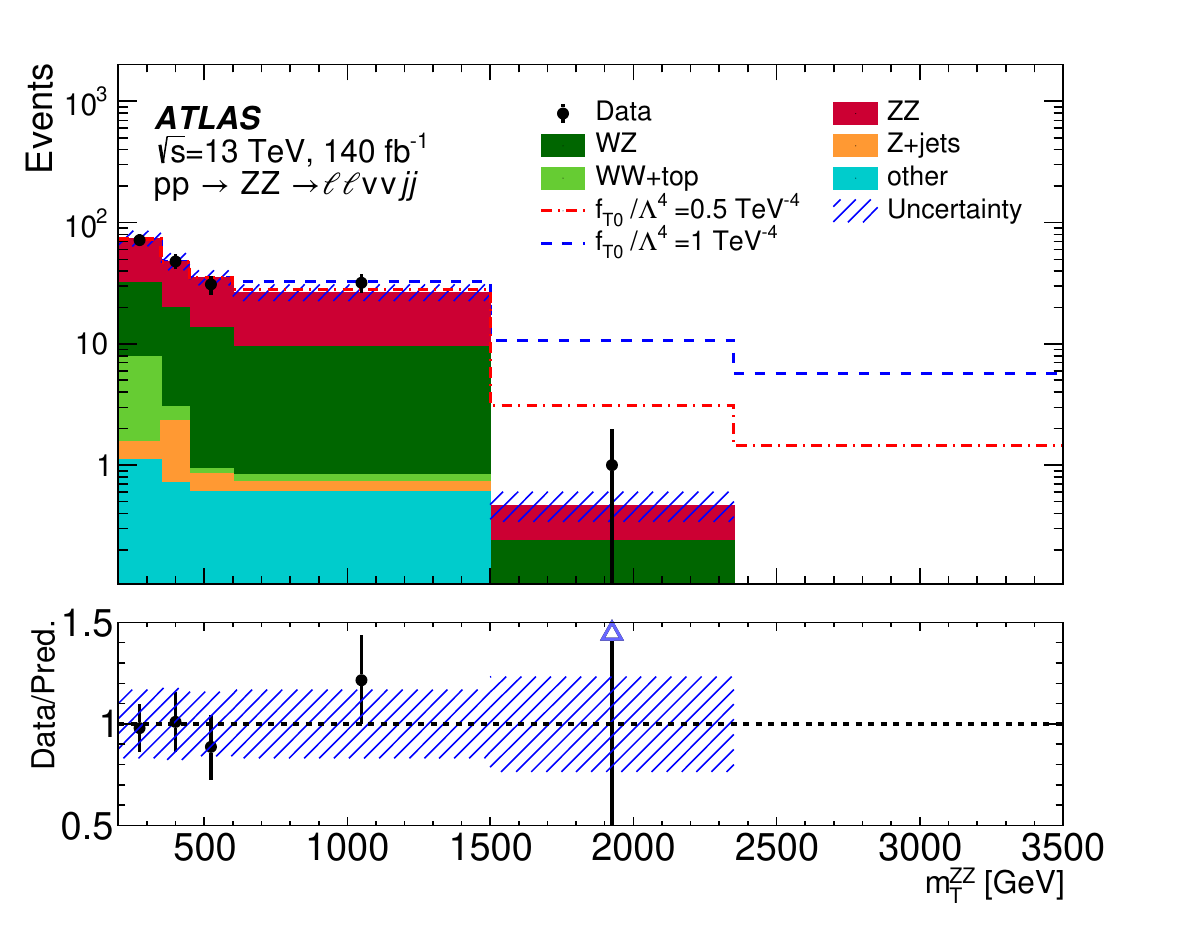}
\caption{
Distribution of the transverse mass of the $ZZ$ system, $m_\mathrm{T}^{ZZ}$, at detector level. The dashed band represents the total uncertainty in the pre-fit SM prediction, with no normalization factors applied from the signal extraction fit. The expected effects of aQGC induced by the SMEFT operator $f_{\mathrm{T0}}$, for two different values of the Wilson coefficient, are also shown.}
\label{fig:zzjj-mTZZ}
\end{figure}

\begin{table}[htbp]
\centering
\caption{Expected and observed 95\% CL intervals on the dimension-8 Wilson coefficients, quoted in units of $\text{TeV}^{-4}$,  for different cut-off scales.}
\label{tab:EFT_limits}
\renewcommand{\arraystretch}{1.1}
\begin{tabular}{lcccccc}
\hline
\textbf{Wilson Coefficient} & \multicolumn{2}{c}{\textbf{No cut-off}} &
\multicolumn{2}{c}{$E_c = 1~\text{TeV}$} &
\multicolumn{2}{c}{$E_c = 1.5~\text{TeV}$} \\
\cline{2-7}
$[\mathrm{TeV}^{-4}]$ & Expected & Observed & Expected & Observed & Expected & Observed \\
\hline
$f_{T0}/\Lambda^4$ & [$-0.42$, 0.40] & [$-0.50$, 0.49] & [$-3.9$, 3.8] & [$-4.9$, 4.9] & [$-2.8$, 2.6] & [$-4.0$, 3.4] \\
$f_{T1}/\Lambda^4$ & [$-0.46$, 0.46] & [$-0.56$, 0.56] & [$-3.7$, 3.6] & [$-4.6$, 4.6] & [$-2.9$, 2.8] & [$-3.8$, 3.8] \\
$f_{T2}/\Lambda^4$ & [$-0.95$, 0.93] & [$-1.1$, 1.1] & [$-7.7$, 7.6] & [$-9.8$, 9.7] & [$-5.9$, 5.6] & [$-7.7$, 7.4] \\
$f_{T3}/\Lambda^4$ & [$-1.3$, 1.3] & [$-1.6$, 1.6] & [$-9.9$, 9.9] & [$-13$, 13] & [$-7.7$, 7.4] & [$-10$, 9.7] \\
$f_{T4}/\Lambda^4$ & [$-8.8$, 8.5] & [$-11$, 11] & [$-55$, 54] & [$-69$, 69] & [$-43$, 41] & [$-56$, 53] \\
$f_{T5}/\Lambda^4$ & [$-0.75$, 0.71] & [$-0.89$, 0.87] & [$-6.9$, 6.7] & [$-8.7$, 8.6] & [$-4.8$, 4.4] & [$-6.2$, 5.8] \\
$f_{T6}/\Lambda^4$ & [$-1.1$, 1.1] & [$-1.4$, 1.4] & [$-6.9$, 6.9] & [$-8.9$, 8.9] & [$-5.4$, 5.3] & [$-7.1$, 7.0] \\
$f_{T7}/\Lambda^4$ & [$-2.5$, 2.4] & [$-3.0$, 3.0] & [$-14$, 14] & [$-18$, 18] & [$-12$, 12] & [$-17$, 16] \\
$f_{T8}/\Lambda^4$ & [$-0.60$, 0.60] & [$-0.74$, 0.74] & [$-6.7$, 6.7] & [$-8.7$, 8.7] & [$-3.8$, 3.8] & [$-4.9$, 4.9] \\
$f_{T9}/\Lambda^4$ & [$-1.3$, 1.3] & [$-1.6$, 1.6] & [$-14$, 14] & [$-18$, 18] & [$-7.9$, 7.9] & [$-10$, 10] \\
\hline
\end{tabular}
\end{table}

Figures~\ref{fig:fT0-fT1_zzjj_EFT} and~\ref{fig:fT2-fT9_zzjj_EFT} show the evolution of the limits on the Wilson coefficients $f_{\text{T,i}}$,
as a function of $E_c$. The unitarity bounds shown in Figures~\ref{fig:fT0-fT1_zzjj_EFT} and~\ref{fig:fT2-fT9_zzjj_EFT}  indicate the parameter values above which the SMEFT description breaks down due to violation of perturbative unitarity. They are computed following the prescription in Ref.~\cite{Almeida:2020ylr}, based on the energy scale at which the partonic scattering amplitude exceeds the unitarity limit.

\begin{figure}[!htbp]
\centering

\begin{subfigure}{0.4\linewidth}
\centering
\includegraphics[width=\linewidth]{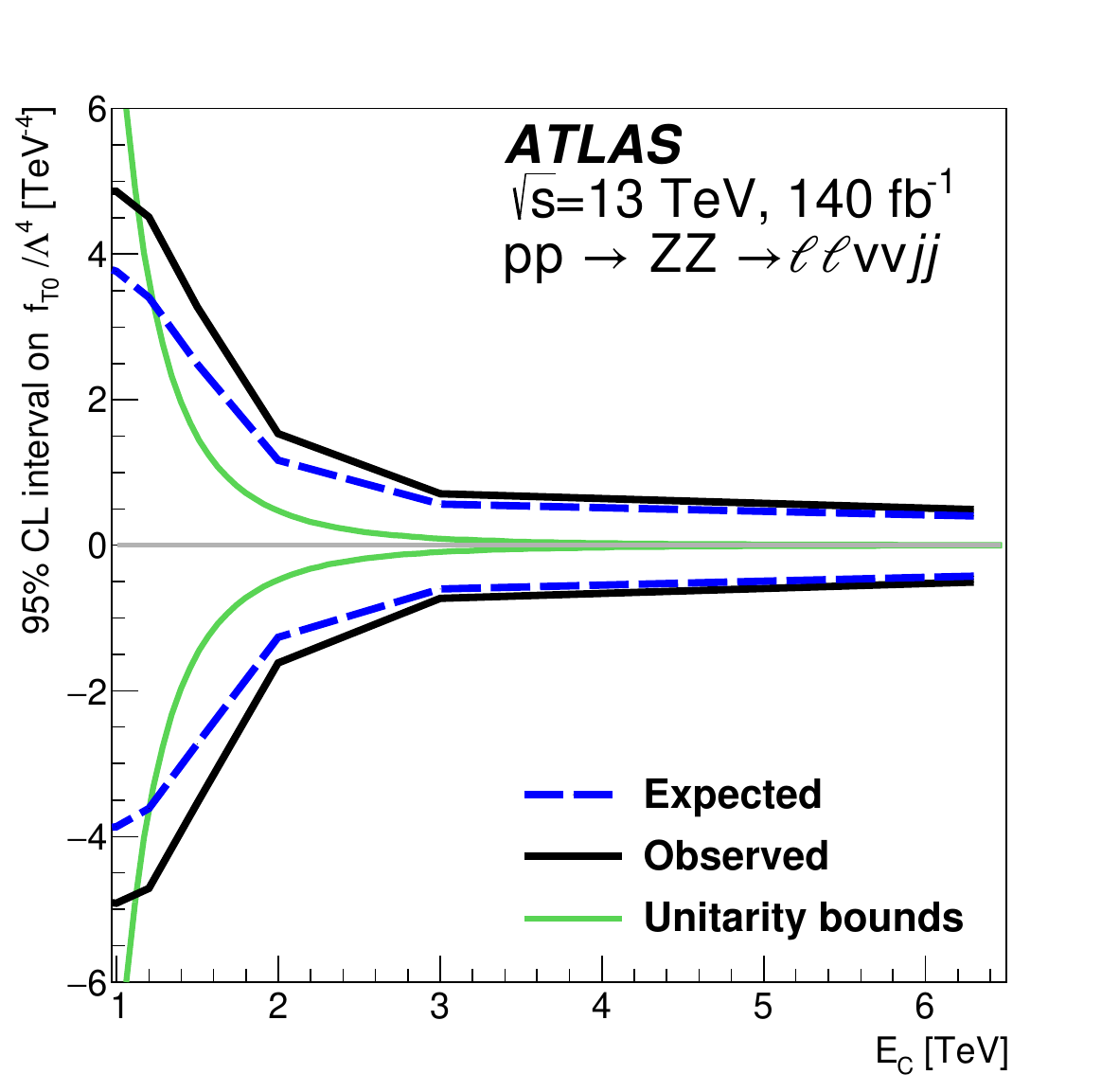}
\caption{}
\end{subfigure}
\begin{subfigure}{0.4\linewidth}
\centering
\includegraphics[width=\linewidth]{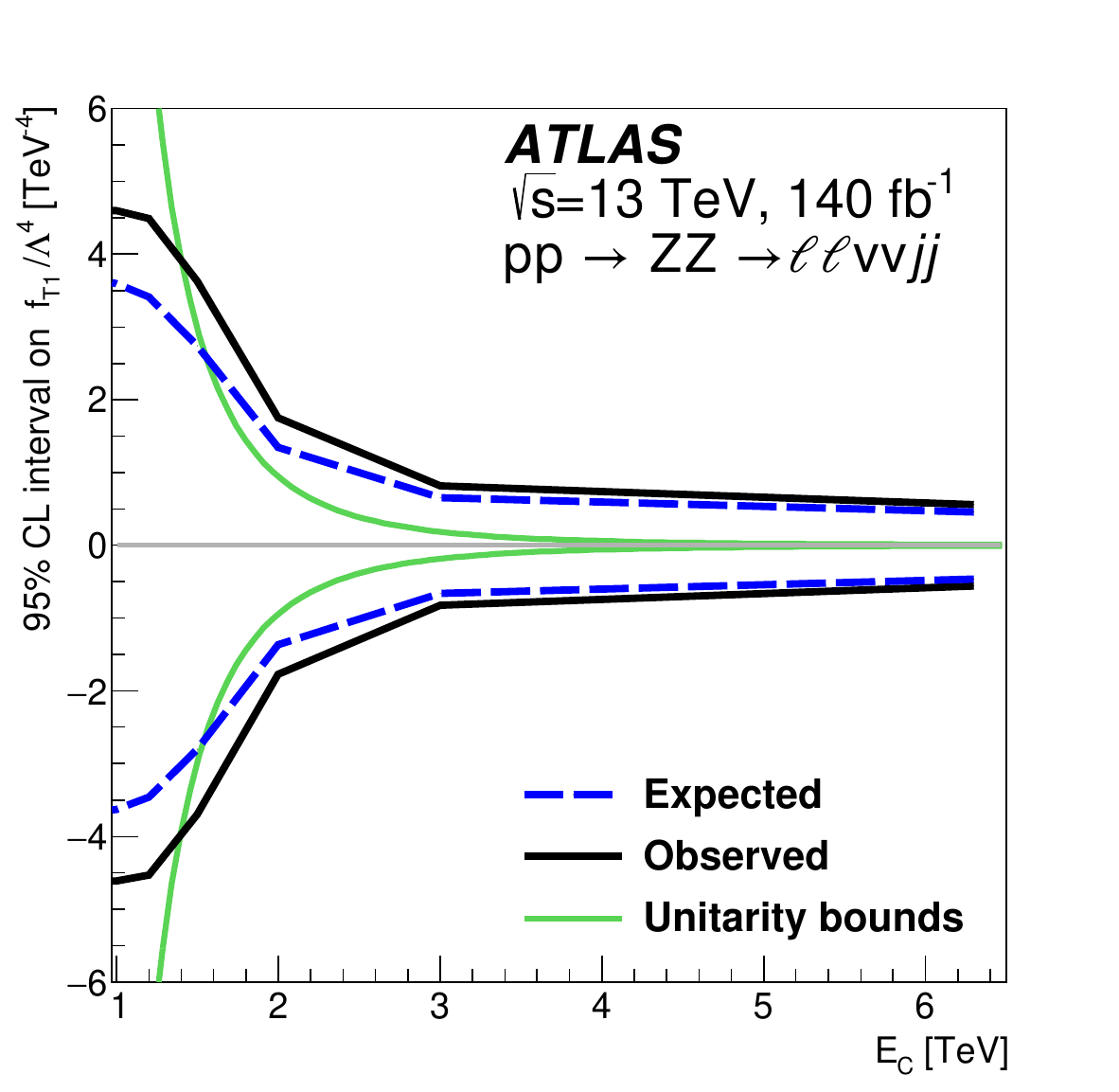}
\caption{}
\end{subfigure}

\begin{subfigure}{0.4\linewidth}
\centering
\includegraphics[width=\linewidth]{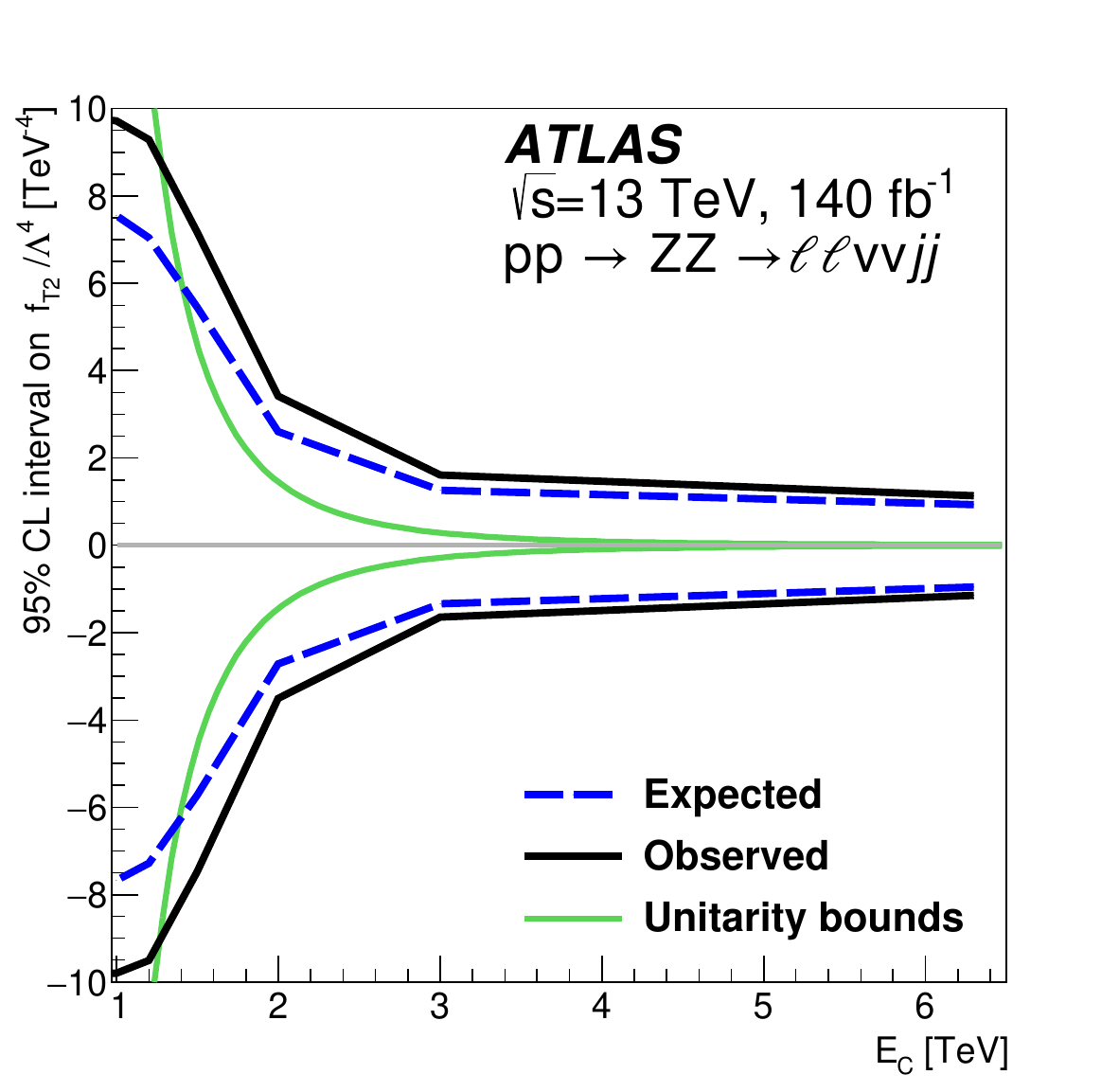}
\caption{}
\end{subfigure}
\begin{subfigure}{0.4\linewidth}
\centering
\includegraphics[width=\linewidth]{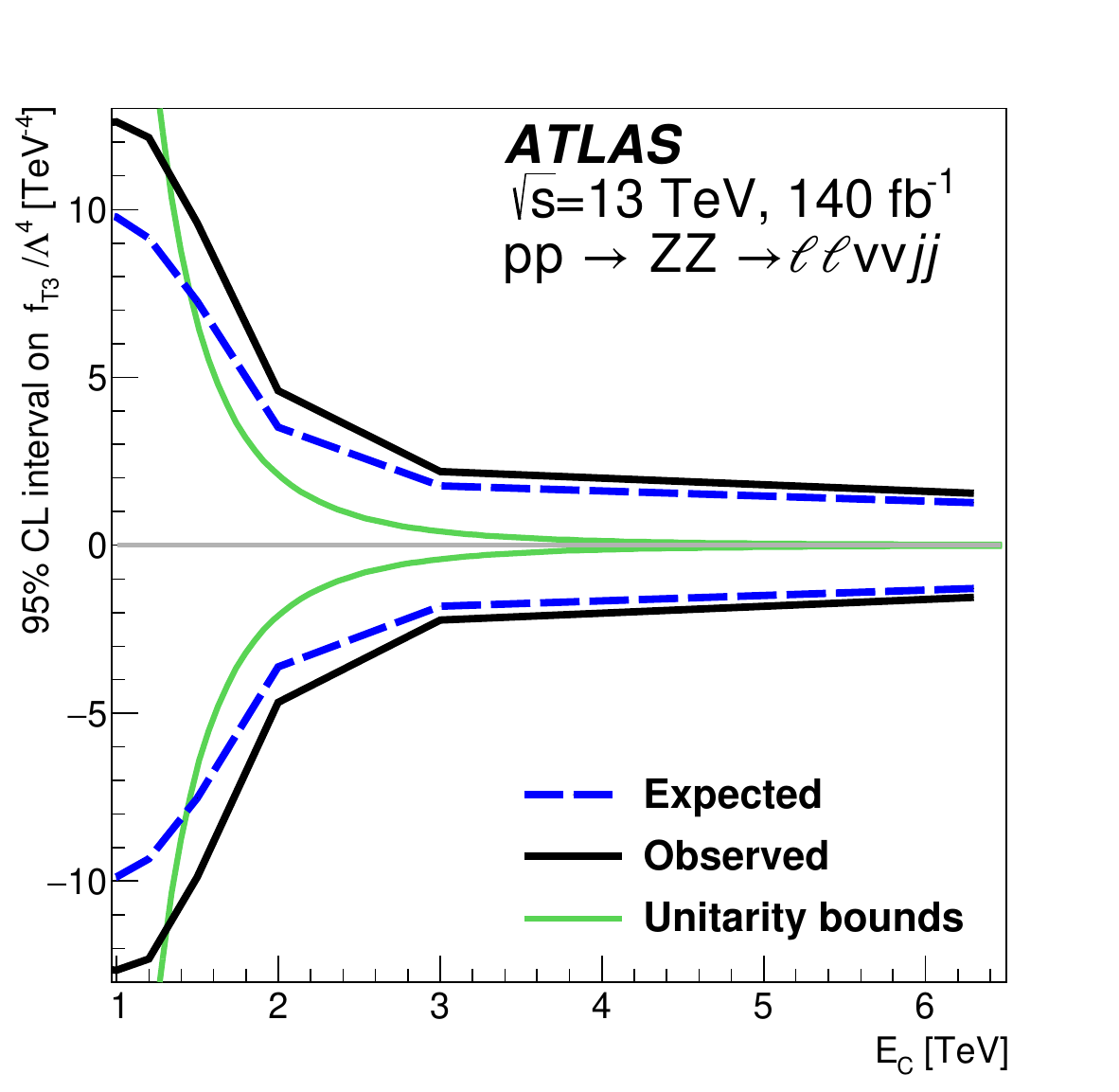}
\caption{}
\end{subfigure}

\caption{The expected and observed 95\% CL intervals for the (a)$f_\text{T0}$, (b)$f_\text{T1}$, (c)$f_\text{T2}$ and (d)$f_\text{T3}$ Wilson coefficients are derived as a function of the cut-off scale, $E_\text{c}$, which imposes a restriction such that the BSM amplitudes of interference and pure dimension-eight are set to zero when $m_\text{ZZ} < E_\text{c}$.
These constraints are determined by performing a one-dimensional fit to the $m_\text{T}^\text{ZZ}$ distribution in the $\ell \ell \nu \nu jj$ signal region.
The dashed (solid) lines show the expected (observed) 95\% CL limits, and the solid curve corresponds to the unitarity bounds~\cite{Almeida:2020ylr}.}
\label{fig:fT0-fT1_zzjj_EFT}
\end{figure}

\begin{figure}[!htbp]
\centering

\begin{subfigure}{0.40\linewidth}
\centering
\includegraphics[width=\linewidth]{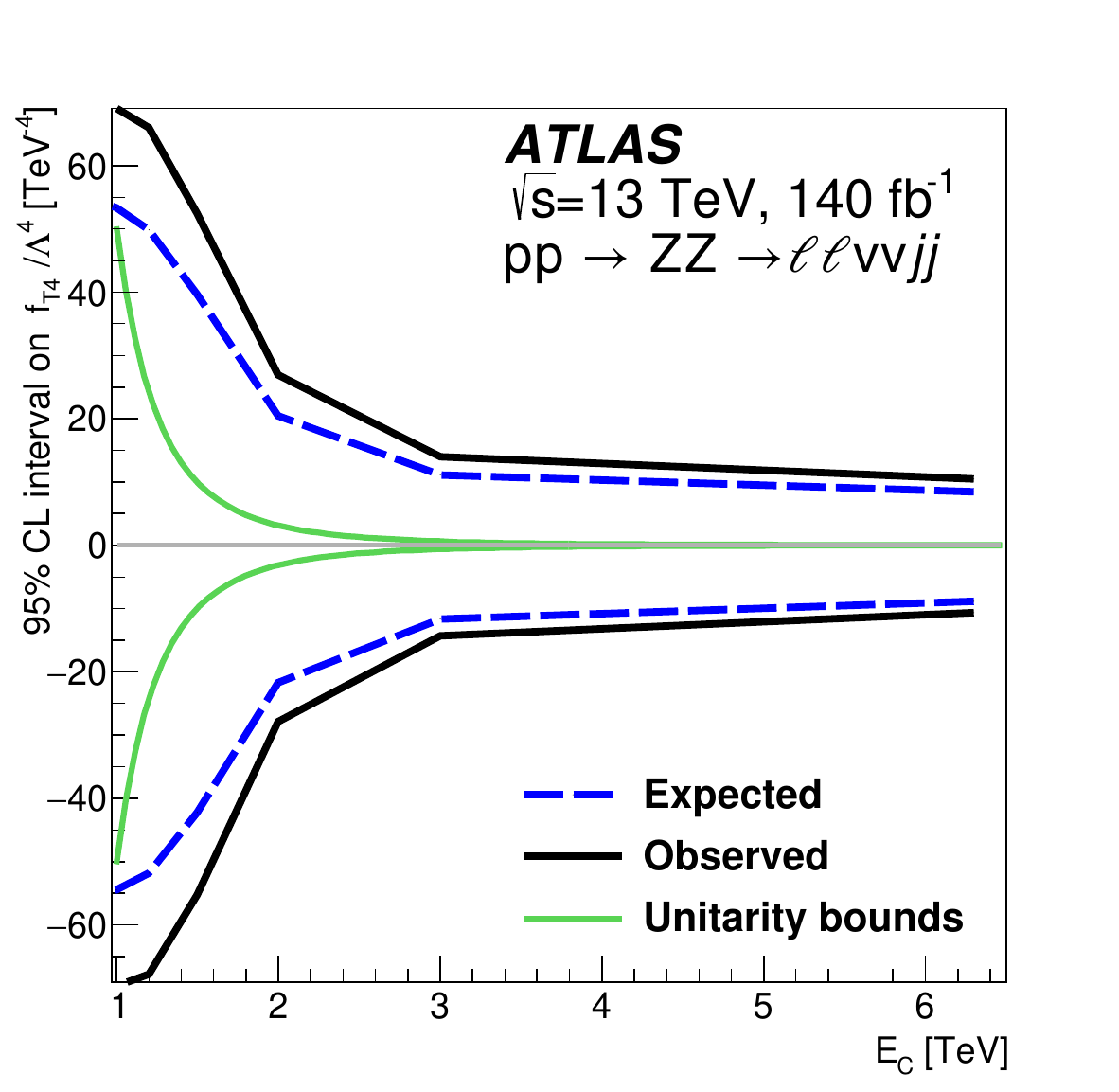}
\caption{}
\end{subfigure}
\begin{subfigure}{0.40\linewidth}
\centering
\includegraphics[width=\linewidth]{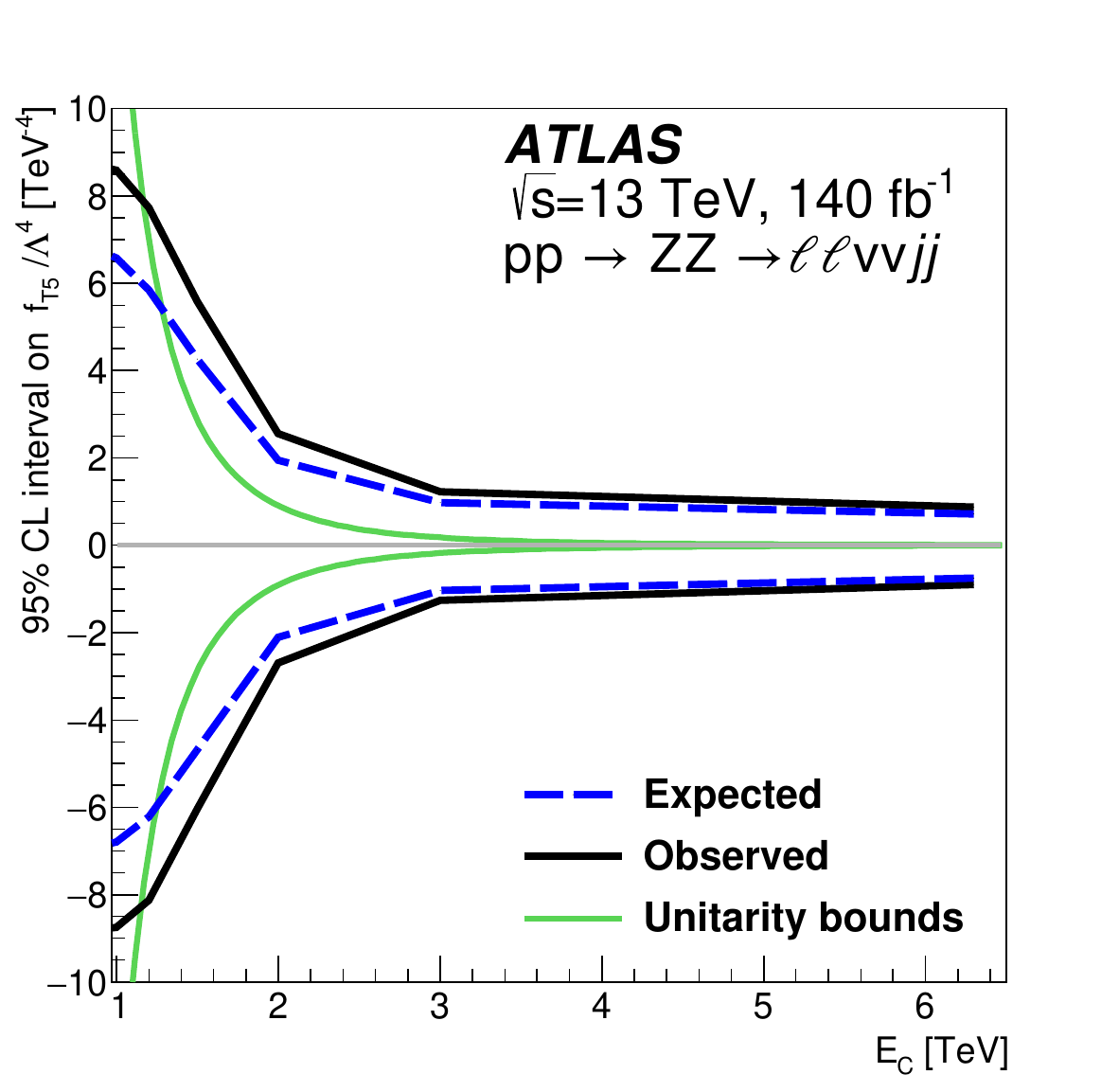}
\caption{}
\end{subfigure}

\begin{subfigure}{0.40\linewidth}
\centering
\includegraphics[width=\linewidth]{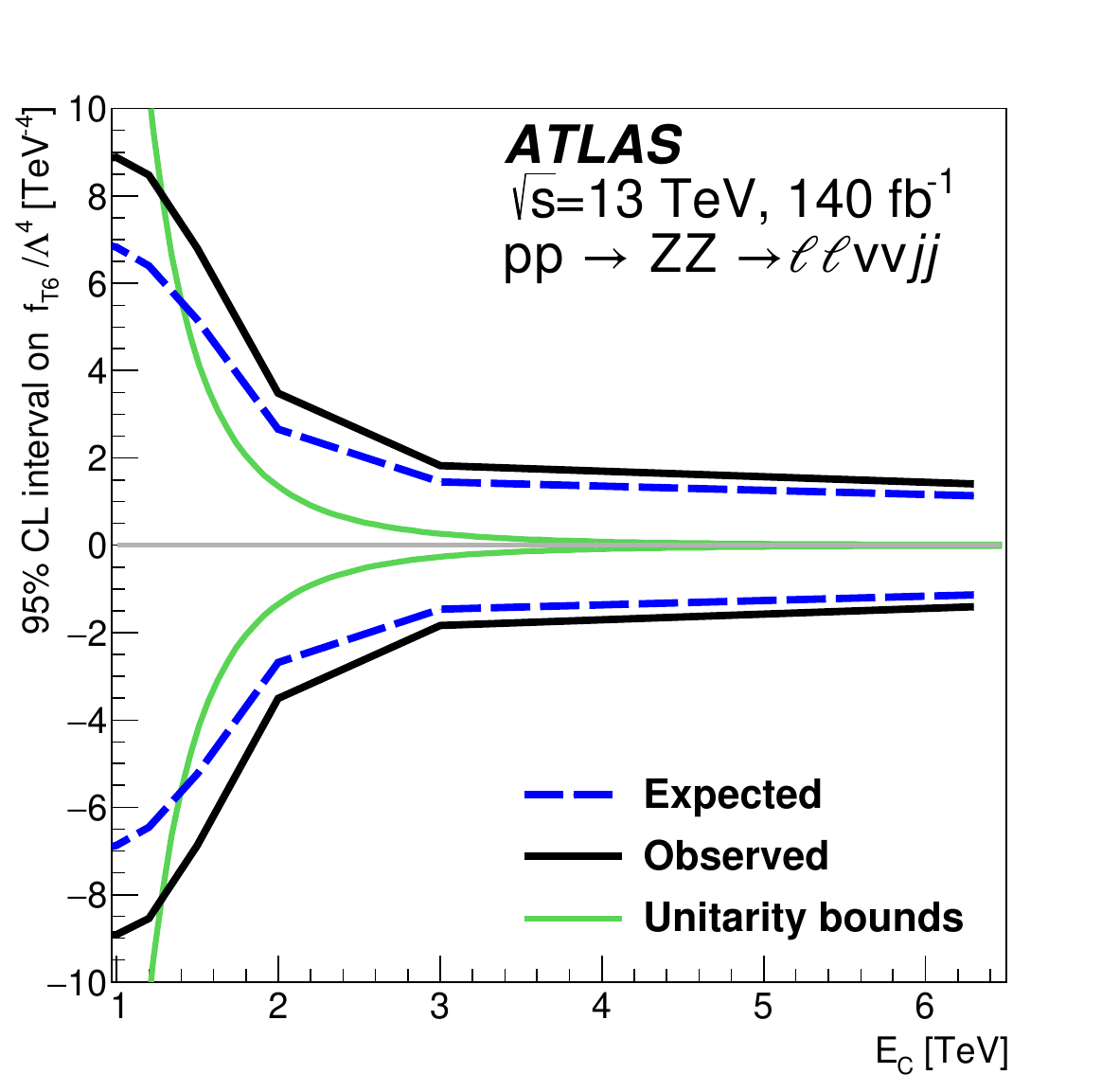}
\caption{}
\end{subfigure}
\begin{subfigure}{0.40\linewidth}
\centering
\includegraphics[width=\linewidth]{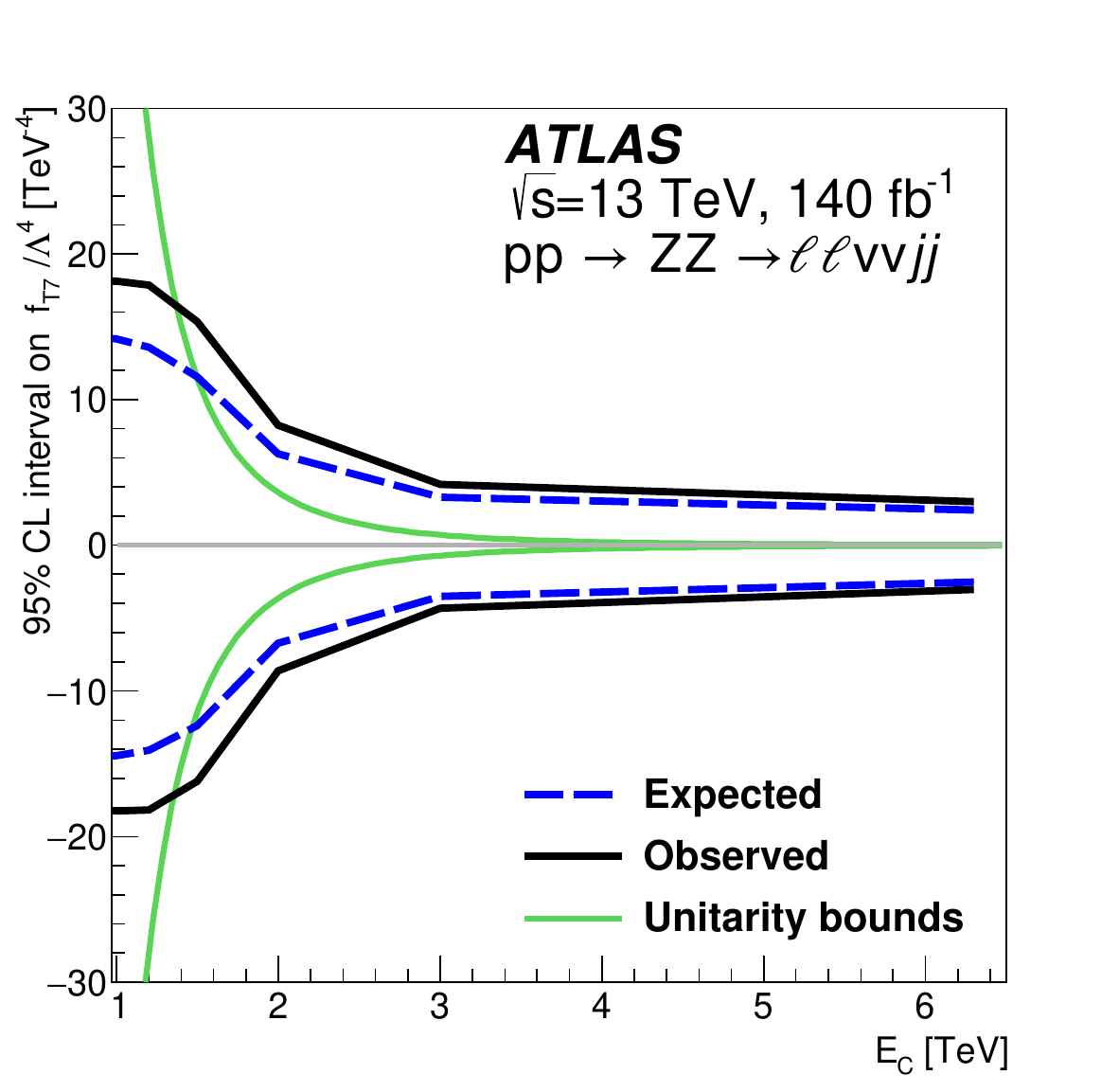}
\caption{}
\end{subfigure}

\begin{subfigure}{0.40\linewidth}
\centering
\includegraphics[width=\linewidth]{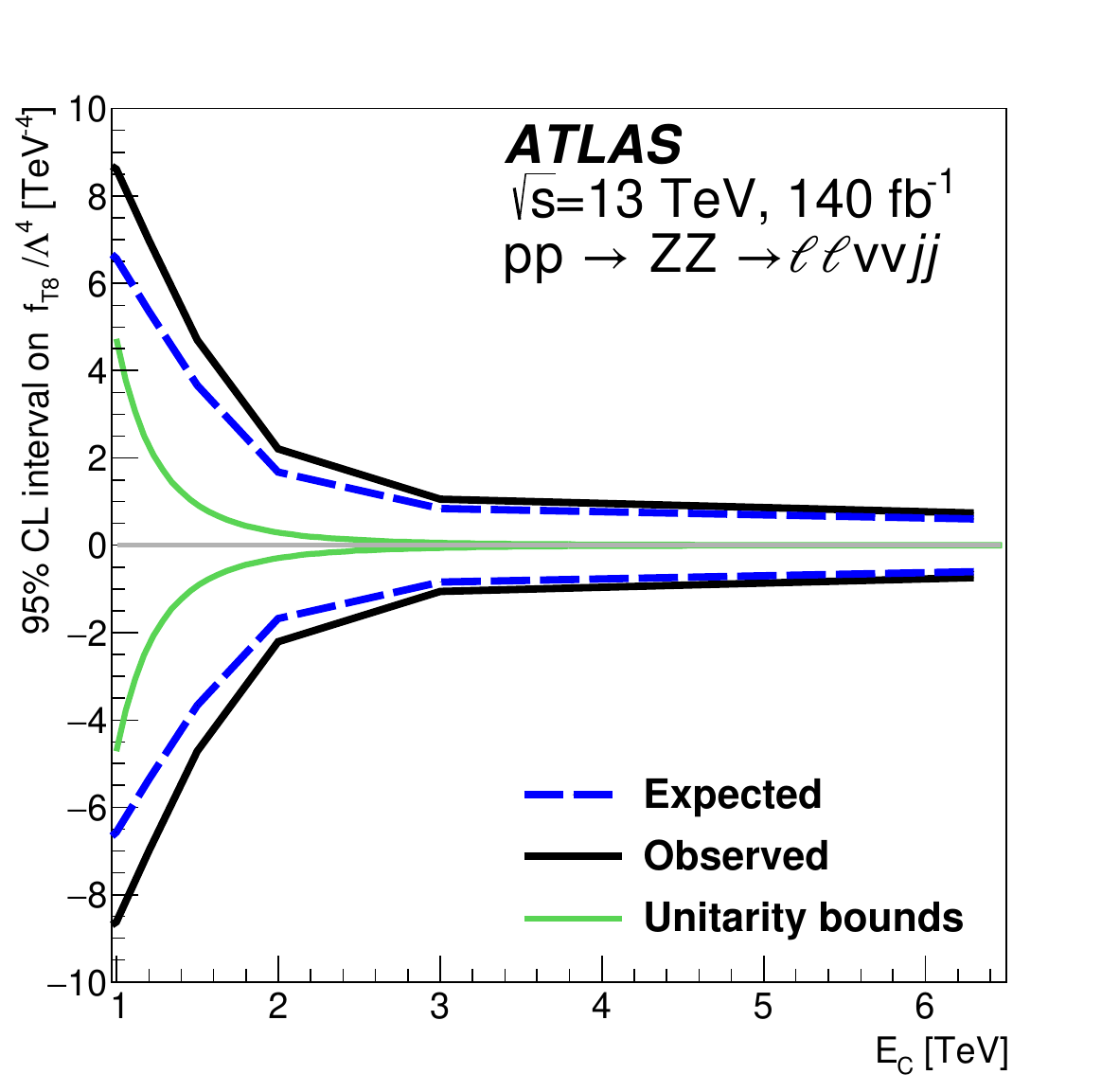}
\caption{}
\end{subfigure}
\begin{subfigure}{0.40\linewidth}
\centering
\includegraphics[width=\linewidth]{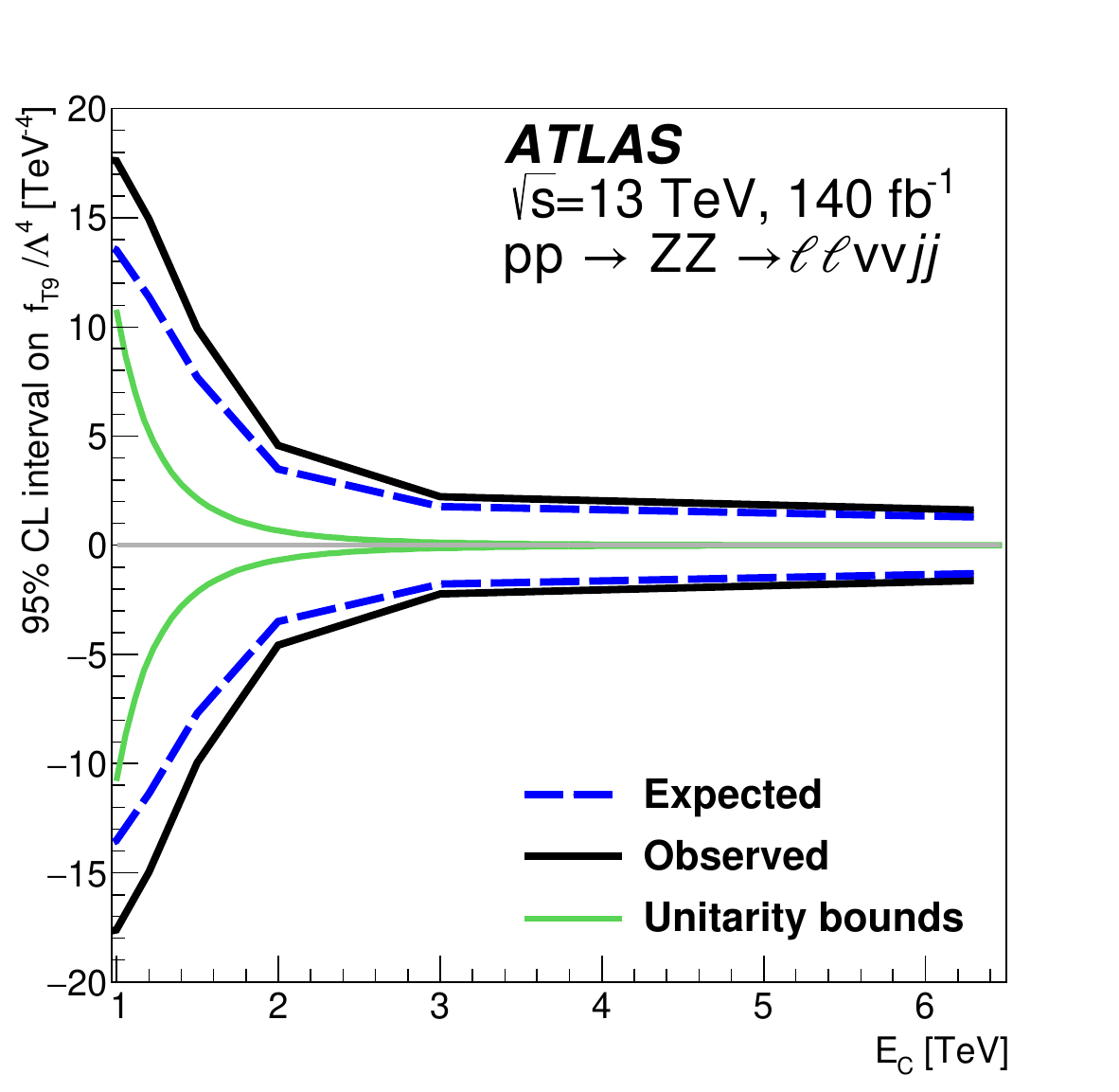}
\caption{}
\end{subfigure}

\caption{The expected and observed 95\% CL intervals for the (a)$f_\text{T4}$, (b)$f_\text{T5}$, (c)$f_\text{T6}$, (d)$f_\text{T7}$, (e)$f_\text{T8}$ and (f)$f_\text{T9}$ Wilson coefficients are derived as a function of the cut-off scale, $E_\text{c}$, which imposes a restriction such that the BSM amplitudes of interference and pure dimension-eight are set to zero when $m_\text{ZZ} < E_\text{c}$.
These constraints are determined by performing a one-dimensional fit to the $m_\text{T}^\text{ZZ}$ distribution in the $\ell \ell \nu \nu jj$ signal region.
The dashed (solid) lines show the expected (observed) 95\% CL limits, and the solid curve corresponds to the unitarity bounds~\cite{Almeida:2020ylr}.}
\label{fig:fT2-fT9_zzjj_EFT}
\end{figure}

\FloatBarrier


%

%
%

%
%
%
%

%
%

%
%
%
%

%

%
%
%

%


\section{Conclusion}
\label{sec:conclusion}

Integrated and differential cross-sections for $Z$ boson pair production in the
$ZZ \to \ell \ell \nu\nu$ final state, where $\ell=e,\mu$, are measured in both inclusive
$ZZ$ and $ZZjj$ production modes at $\sqrt{s}=13$~\TeV using 140~fb$^{-1}$ of ATLAS data collected at the LHC.
The $ZZjj$ measurement requires at least two jets.
The inclusive $ZZ \to \ell\ell\nu\nu$ and $ZZjj \to \ell\ell\nu\nu jj$ fiducial cross-sections are measured as
$21.0 \pm 1.0$~fb and $0.96^{+0.18}_{-0.16}$~fb, respectively.
Compared with earlier results using a partial Run-2 dataset, the total uncertainty in the inclusive $ZZ$ fiducial measurement improves from 7.0\% to 4.6\% and is limited by the statistical uncertainty.
The measured values are in agreement with the corresponding SM predictions of $20.1 \pm 1.1$~fb and $0.94 \pm 0.20$~fb from \textsc{Sherpa}+\textsc{MadGraph5}.
The \textsc{Matrix} prediction for the inclusive $ZZ \to \ell\ell\nu\nu$ fiducial cross-section is $21.03 \pm 0.30$~fb, demonstrating that the theory uncertainty in the prediction is reduced from 5.8\% to 1.5\% thanks to calculations incorporating NNLO QCD and NLO EW corrections.
The inclusive $ZZ \to \ell\ell\nu\nu$ cross-section is extrapolated to the total phase space, yielding a value of $15.38 \pm 0.81$~pb, in agreement with the SM prediction of $15.40 \pm 0.38$~pb from \textsc{Matrix}.
Differential cross-sections for both $ZZ$ and $ZZjj$ productions are measured as functions of observables that provide high-precision tests of the SM and input for modelling studies.
The transverse momentum of the dilepton system is used to constrain anomalous neutral triple gauge couplings via both the effective vertex approach and the SMEFT framework.
In the vertex‑function formalism, limits are set on $f_{4}^{Z}$, $f_{4}^{\gamma}$, $f_{5}^{\gamma}$ and $f_{5}^{Z}$, improving previous results in this final state by a factor of 1.7.
In the SMEFT interpretation, limits are derived on the Wilson coefficients $C_{G+}/\Lambda^{4}$, $C_{G-}/\Lambda^{4}$, $C_{BW}/\Lambda^{4}$, $C_{BB}/\Lambda^{4}$, $C_{\tilde{B}W}/\Lambda^{4}$ and $C_{WW}/\Lambda^{4}$.
In the $ZZjj$ channel, the transverse mass of the $ZZ$ system is used to constrain dimension-eight operators contributing to aQGCs, with limits improving sensitivity by a factor of 2 relative to the $4\ell jj$ channel due to the larger branching ratio of the $\ell\ell\nu\nu$ final state.
Overall, the results show excellent agreement with SM predictions and provide significantly improved sensitivity to new physics in the electroweak sector.


%

%

%
%
%
%
%
\section*{Acknowledgements}
%

%

%
%

%
%

We thank CERN for the very successful operation of the LHC and its injectors, as well as the support staff at
CERN and at our institutions worldwide without whom ATLAS could not be operated efficiently.

The crucial computing support from all WLCG partners is acknowledged gratefully, in particular from CERN, the ATLAS Tier-1 facilities at TRIUMF/SFU (Canada), NDGF (Denmark, Norway, Sweden), CC-IN2P3 (France), KIT/GridKA (Germany), INFN-CNAF (Italy), NL-T1 (Netherlands), PIC (Spain), RAL (UK) and BNL (USA), the Tier-2 facilities worldwide and large non-WLCG resource providers. Major contributors of computing resources are listed in Ref.~\cite{ATL-SOFT-PUB-2025-001}.

We gratefully acknowledge the support of ANPCyT, Argentina; YerPhI, Armenia; ARC, Australia; BMWFW and FWF, Austria; ANAS, Azerbaijan; CNPq and FAPESP, Brazil; NSERC, NRC and CFI, Canada; CERN; ANID, Chile; CAS, MOST and NSFC, China; Minciencias, Colombia; MEYS CR, Czech Republic; DNRF and DNSRC, Denmark; IN2P3-CNRS and CEA-DRF/IRFU, France; SRNSFG, Georgia; BMFTR, HGF and MPG, Germany; GSRI, Greece; RGC and Hong Kong SAR, China; ICHEP and Academy of Sciences and Humanities, Israel; INFN, Italy; MEXT and JSPS, Japan; CNRST, Morocco; NWO, Netherlands; RCN, Norway; MNiSW, Poland; FCT, Portugal; MNE/IFA, Romania; MSTDI, Serbia; MSSR, Slovakia; ARIS and MVZI, Slovenia; DSI/NRF, South Africa; MICIU/AEI, Spain; SRC and Wallenberg Foundation, Sweden; SERI, SNSF and Cantons of Bern and Geneva, Switzerland; NSTC, Taipei; TENMAK, T\"urkiye; STFC/UKRI, United Kingdom; DOE and NSF, United States of America.

Individual groups and members have received support from BCKDF, CANARIE, CRC and DRAC, Canada; CERN-CZ, FORTE and PRIMUS, Czech Republic; COST, ERC, ERDF, Horizon 2020, ICSC-NextGenerationEU and Marie Sk{\l}odowska-Curie Actions, European Union; Investissements d'Avenir Labex, Investissements d'Avenir Idex and ANR, France; DFG and AvH Foundation, Germany; Herakleitos, Thales and Aristeia programmes co-financed by EU-ESF and the Greek NSRF, Greece; BSF-NSF and MINERVA, Israel; NCN and NAWA, Poland; La Caixa Banking Foundation, CERCA Programme Generalitat de Catalunya and PROMETEO and GenT Programmes Generalitat Valenciana, Spain; G\"{o}ran Gustafssons Stiftelse, Sweden; The Royal Society and Leverhulme Trust, United Kingdom.

In addition, individual members wish to acknowledge support from CERN: European Organization for Nuclear Research (CERN DOCT); Chile: Agencia Nacional de Investigaci\'on y Desarrollo (FONDECYT 1230812, FONDECYT 1240864, Fondecyt 3240661, Fondecyt Regular 1240721); China: Chinese Ministry of Science and Technology (MOST-2023YFA1605700, MOST-2023YFA1609300), National Natural Science Foundation of China (NSFC - 12175119, NSFC 12275265); Czech Republic: Czech Science Foundation (GACR - 24-11373S), Ministry of Education Youth and Sports (ERC-CZ-LL2327, FORTE CZ.02.01.01/00/22\_008/0004632), PRIMUS Research Programme (PRIMUS/21/SCI/017); EU: H2020 European Research Council (ERC - 101002463); European Union: European Research Council (BARD No. 101116429, ERC - 948254, ERC 101089007), European Regional Development Fund (SMASH COFUND 101081355, SLO ERDF), European Union, Future Artificial Intelligence Research (FAIR-NextGenerationEU PE00000013), Italian Center for High Performance Computing, Big Data and Quantum Computing (ICSC, NextGenerationEU); France: Agence Nationale de la Recherche (ANR-21-CE31-0022, ANR-22-EDIR-0002, ANR-24-CE31-0504-01); Germany: Baden-Württemberg Stiftung (BW Stiftung-Postdoc Eliteprogramme), Deutsche Forschungsgemeinschaft (DFG - 469666862, DFG - CR 312/5-2); China: Research Grants Council (GRF); Italy: Istituto Nazionale di Fisica Nucleare (ICSC, NextGenerationEU), Ministero dell'Università e della Ricerca (NextGenEU 153D23001490006 M4C2.1.1, NextGenEU I53D23000820006 M4C2.1.1, NextGenEU I53D23001490006 M4C2.1.1, SOE2024\_0000023); Japan: Japan Society for the Promotion of Science (JSPS KAKENHI JP22H01227, JSPS KAKENHI JP22H04944, JSPS KAKENHI JP22KK0227, JSPS KAKENHI JP24K23939, JSPS KAKENHI JP24KK0251, JSPS KAKENHI JP25H00650, JSPS KAKENHI JP25H01291, JSPS KAKENHI JP25K01023); Norway: Research Council of Norway (RCN-314472); Poland: Ministry of Science and Higher Education (IDUB AGH, POB8, D4 no 9722), Polish National Science Centre (NCN 2021/42/E/ST2/00350, NCN OPUS 2023/51/B/ST2/02507, NCN UMO-2019/34/E/ST2/00393, UMO-2022/47/O/ST2/00148, UMO-2023/49/B/ST2/04085, UMO-2023/51/B/ST2/00920, UMO-2024/53/N/ST2/00869); Portugal: Foundation for Science and Technology (FCT); Spain: Generalitat Valenciana (ASFAE/2022/008), Ministry of Science and Innovation (MCIN \& NextGenEU PCI2022-135018-2, MICIN \& FEDER PID2021-125273NB, RYC2019-028510-I, RYC2020-030254-I, RYC2021-031273-I, RYC2022-038164-I), Ministerio de Ciencia, Innovación y Universidades/Agencia Estatal de Investigaci\'on (PID2022-142604OB-C22); Sweden: Carl Trygger Foundation (Carl Trygger Foundation CTS 22:2312), Swedish Research Council (Swedish Research Council 2023-04654, VR 2021-03651, VR 2022-03845, VR 2022-04683, VR 2023-03403, VR 2024-05451), Knut and Alice Wallenberg Foundation (KAW 2018.0458, KAW 2022.0358, KAW 2023.0366); Switzerland: Swiss National Science Foundation (SNSF - PCEFP2\_194658); United Kingdom: Royal Society (NIF-R1-231091); United States of America: U.S. Department of Energy (ECA DE-AC02-76SF00515), Neubauer Family Foundation.

%
%


\newpage
\printbibliography
\newpage
 
\begin{flushleft}
\hypersetup{urlcolor=black}
{\Large The ATLAS Collaboration}

\bigskip

\AtlasOrcid[0000-0002-6665-4934]{G.~Aad}$^\textrm{\scriptsize 104}$,
\AtlasOrcid[0000-0001-7616-1554]{E.~Aakvaag}$^\textrm{\scriptsize 17}$,
\AtlasOrcid[0000-0002-5888-2734]{B.~Abbott}$^\textrm{\scriptsize 123}$,
\AtlasOrcid[0000-0002-0287-5869]{S.~Abdelhameed}$^\textrm{\scriptsize 119a}$,
\AtlasOrcid[0000-0002-1002-1652]{K.~Abeling}$^\textrm{\scriptsize 55}$,
\AtlasOrcid[0000-0001-5763-2760]{N.J.~Abicht}$^\textrm{\scriptsize 49}$,
\AtlasOrcid[0000-0002-8496-9294]{S.H.~Abidi}$^\textrm{\scriptsize 30}$,
\AtlasOrcid[0009-0003-6578-220X]{M.~Aboelela}$^\textrm{\scriptsize 45}$,
\AtlasOrcid[0000-0002-9987-2292]{A.~Aboulhorma}$^\textrm{\scriptsize 36e}$,
\AtlasOrcid[0000-0001-5329-6640]{H.~Abramowicz}$^\textrm{\scriptsize 157}$
% ...
\bigskip
\\

% ...
$^{17}$Department for Physics and Technology, University of Bergen, Bergen; Norway.\\
% ...
$^{30}$Physics Department, Brookhaven National Laboratory, Upton NY; United States of America.\\
% ...
$^{36}$$^{(a)}$Facult\'e des Sciences Ain Chock, Universit\'e Hassan II de Casablanca;$^{(b)}$Facult\'{e} des Sciences, Universit\'{e} Ibn-Tofail, K\'{e}nitra;$^{(c)}$Facult\'e des Sciences Semlalia, Universit\'e Cadi Ayyad, LPHEA-Marrakech;$^{(d)}$LPMR, Facult\'e des Sciences, Universit\'e Mohamed Premier, Oujda;$^{(e)}$Facult\'e des sciences, Universit\'e Mohammed V, Rabat;$^{(f)}$Institute of Applied Physics, Mohammed VI Polytechnic University, Ben Guerir; Morocco.\\
% ...
$^{45}$Physics Department, Southern Methodist University, Dallas TX; United States of America.\\
% ...
$^{49}$Fakult\"{a}t Physik, Technische Universit{\"a}t Dortmund, Dortmund; Germany.\\
% ...
$^{55}$II. Physikalisches Institut, Georg-August-Universit\"{a}t G\"ottingen, G\"ottingen; Germany.\\
% ...
$^{104}$CPPM, Aix-Marseille Universit\'e, CNRS/IN2P3, Marseille; France.\\
% ...
$^{119}$$^{(a)}$New York University Abu Dhabi, Abu Dhabi;$^{(b)}$United Arab Emirates University, Al Ain; United Arab Emirates.\\
% ...
$^{123}$Homer L. Dodge Department of Physics and Astronomy, University of Oklahoma, Norman OK; United States of America.\\
% ...
$^{157}$Raymond and Beverly Sackler School of Physics and Astronomy, Tel Aviv University, Tel Aviv; Israel.\\
% ...
$^{a}$ Also at Affiliated with an institute formerly covered by a cooperation agreement with CERN.\\
$^{b}$ Also at An-Najah National University, Nablus; Palestine.\\
$^{c}$ Also at Borough of Manhattan Community College, City University of New York, New York NY; United States of America.\\
$^{d}$ Also at Center for Interdisciplinary Research and Innovation (CIRI-AUTH), Thessaloniki; Greece.\\
$^{e}$ Also at Centre of Physics of the Universities of Minho and Porto (CF-UM-UP); Portugal.\\
$^{f}$ Also at CERN, Geneva; Switzerland.\\
% ...

\end{flushleft}


%
\clearpage

\end{document}